\documentclass[a4paper,10pt]{article}

\pdfoutput=1 

\usepackage{jheppub} 

\usepackage{slashed}

\usepackage{floatrow}
\newfloatcommand{capbtabbox}{table}[][\FBwidth]

\usepackage{blindtext}

\usepackage[dvipsnames,table]{xcolor}

\usepackage{hyperref} 
\usepackage{xspace}
\usepackage[tight]{subfigure}
\usepackage{amsmath}
\usepackage{amssymb}
\usepackage{amsfonts}
\usepackage{mathrsfs}
\usepackage{comment}
\usepackage{afterpage}
\usepackage{booktabs}
\usepackage{array}

\usepackage[title,titletoc]{appendix}

\usepackage{tabularx}
\newcolumntype{C}[1]{>{\centering\arraybackslash}p{#1}}

\newcommand{\GeV}{\,{\rm GeV}\,}

\newcommand{\beq}{\begin{eqnarray}}
\newcommand{\eeq}{\end{eqnarray}}

\newcommand{\eqn}[1]{Eq.(\ref{#1})}

\newcommand{\eqns}[2]{Eqs.(\ref{#1}--\ref{#2})}

\newcommand{\tbn}[1]{Tab.~\ref{#1}}

\newcommand{\fig}[1]{Fig.~\ref{#1}}
\newcommand{\Fig}[1]{Fig.~\ref{#1}}

\newcommand{\sect}[1]{Sect.~\ref{#1}}
\newcommand{\sects}[2]{Sect.~\ref{#1}--\ref{#2}}

\newcommand{\rf}[1]{Ref.~\cite{#1}}

\newcommand{\iu}{{i\mkern1mu}}
\newcommand{\ab}{ab}

\newcommand{\mc}{\mathcal}

\newcommand{\nnb}{\nonumber}

\preprint{VBSCAN-PUB-06-19}

\title{\sffamily Polarized vector boson scattering in the fully leptonic WZ and ZZ channels at the LHC}
\author[a]{A.~Ballestrero}
\author[a,b]{E.~Maina}
\author[a,b]{and G.~Pelliccioli}

\affiliation[a]{INFN, Sezione di Torino, Via P. Giuria 1, I-10125 Torino, Italy}
\affiliation[b]{Dipartimento di Fisica, Universit\`a di Torino, Via P. Giuria 1, I-10125 Torino, Italy}

\emailAdd{ballestr@to.infn.it}
\emailAdd{maina@to.infn.it}
\emailAdd{gpellicc@to.infn.it}

\abstract{Isolating the scattering of longitudinal weak bosons at the LHC 
is an important tool to probe the ElectroWeak Symmetry Breaking mechanism. Separating 
polarizations of $W$ and $Z$  bosons is complicated, because of non resonant 
contributions and interference effects.
Additional care is necessary when considering $Z$ bosons, due to the $\gamma/Z$ mixing in the 
coupling to charged leptons.

We propose a method to define polarized signals in $Z\!Z$ and $W^+\!Z$ 
scattering at the LHC, which relies on the separation of weak boson 
polarizations at the amplitude level in Monte Carlo simulations.

After validation in the absence of lepton cuts, we investigate how polarized 
distributions are affected by a realistic set of kinematic cuts (and neutrino reconstruction, 
when needed). The total and differential polarized cross sections computed at the amplitude 
level are well defined, and their sum reproduces the full results, up to non negligible {but computable} 
interference effects which should be included in experimental analyses.
We show that polarized cross sections computed using the reweighting method are inaccurate, particularly at large energies.
We also present two procedures which address the model independent extraction of polarized 
components from LHC data, using Standard Model angular distribution templates.}

\begin{document} 
\maketitle
\flushbottom

\section{Introduction}
\label{intro}
Vector Boson Scattering (VBS) of longitudinally polarized, on shell,  $W$'s and $Z$'s is the perfect 
exemplification of the interplay of gauge invariance and unitarity in ElectroWeak Symmetry Breaking (EWSB).
The delicate cancellations which occur in the set of diagrams which involve only the exchange of vector bosons and 
between them and the Higgs exchange diagrams  are crucial for Unitarity. These cancellations are not needed when
the scattering involves transverse polarized vector bosons. 
Hence a polarization analysis of VBS nicely complements the study of Higgs boson properties in the effort to fully 
characterize the details of the EWSB mechanism.

Since vector boson lifetimes are too short to allow for stable beams or direct observation before decay, 
we can only access VBS as a subprocess of more complicated reactions which include
the emission of weak bosons from initial state quarks and their decay to stable particles. 

In Run 2 CMS and ATLAS have finally produced convincing evidence that VBS actually takes place in the complex 
environment of the LHC \cite{Sirunyan:2017fvv,Sirunyan:2017ret,ATLAS:2018ogo,Aaboud:2018ddq,Sirunyan:2019ksz}.
Unfortunately, the statistics is still too small for any attempt to analyze vector boson 
polarizations.
Hopefully the higher rates which will be available after the Long Stop in 2019 and 2020 and later in the High 
Luminosity phase of the LHC will allow polarization studies \cite{CMS:2018mbt, Azzi:2019yne}.

Vector boson polarizations at the LHC have been studied in a number of papers.
$W+$ jets processes, without cuts on the charged leptons, have been
studied in \rf{Bern:2011ie}. The modifications introduced by selection cuts have been examined in 
\rf{Stirling:2012zt}, where, in addition to $W +$ jets, several other $W$ and $Z$ production mechanisms
have been discussed.
The interplay between interference among polarizations and selection cuts has also been analyzed in
\rf{Belyaev:2013nla}.
Recently, the vector boson polarizations in $ p p \rightarrow W\!Z$ have been studied, taking into account both QCD and
EW NLO corrections \cite{Baglio:2018rcu}.

Both CMS  and ATLAS have measured the $W$ polarization fractions 
in the $W+\,$jets \cite{Chatrchyan:2011ig,ATLAS:2012au} channel and in $t\,\bar{t}$ events \cite{Aaboud:2016hsq, Khachatryan:2016fky}.

In the Feynman amplitudes which describe VBS in a realistic accelerator framework, all information about the vector 
boson polarization is confined to the polarization sum in the corresponding propagators. Therefore,
the individual polarizations interfere among themselves. These interference contributions cancel 
exactly only when an integration over the full azimuth of the decay products is performed. 
Acceptance cuts, however, inhibit collecting data over the full angular range and the cancellation cannot be 
complete. 

In addition, electroweak boson production processes are typically described by amplitudes including non 
resonant diagrams, which cannot be interpreted as production times decay of any vector boson. These diagrams 
are essential for gauge invariance and cannot be ignored. For them, separating polarizations is simply unfeasible. 

In a previous paper \cite{Ballestrero:2017bxn}, we have shown that it is possible to define, in a simple and 
natural way, cross sections corresponding to vector bosons of definite polarization.

We have further demonstrated that the sum of cross sections with definite polarization, 
even in the presence of cuts on the final state leptons, describes reasonably well the full total 
cross section and most of the differential distributions. Therefore, it is possible to fit 
the data using single polarized templates and the interference, to extract polarization fractions.

In \rf{Ballestrero:2017bxn} we have focused on the 
$jj W^+ W^-$ final state, with both $W$'s decaying leptonically,
as a proof of concept of our method to separate the different polarizations, without 
worrying too much about its practical observability.

In this paper we study $jj \, e^+e^- \mu^+\mu^-$ and 
$jj \,\mu^+{\nu}_\mu e^+e^- $. The cross section for the first reaction is small, but the 
decay angles for each $Z$ and the invariant mass of the $Z\!Z$ pair can be determined with high precision.
The $W\!Z$ channel has a  much larger cross section but, as any reaction involving a $W$ decaying
leptonically, is affected by the need to reconstruct the unknown component along the beam direction of the
neutrino momentum.
This reconstruction, from which the decay distribution of the $W$ and the total mass of the $W\!Z$ system are 
inferred, can only be approximate.

In \sect{sec:poldef} we recall the basic features of unstable vector boson polarizations and  their relationship 
with the angular distribution of the charged fermions produced in the boson decay. In \sect{sec:separating}
we present our proposal for a definition of polarized amplitudes which entails dropping non resonant diagrams 
and projecting the resulting amplitude on the vector boson mass shell. In sections \ref{sec:ZZ} and \ref{sec:WZ} we 
study the production of $Z\!Z$ and $W\!Z$ pairs in VBS, first in the absence of cuts on the leptonic variables and 
then in the realistic case in which acceptance cuts are imposed. In both cases we focus on how well the sum of 
single polarized cross sections reproduces the full result.
In \sect{subsec:reweight} we compare our approach to the reweighting 
procedure which has been so far adopted in experimental analyses.
In \sect{subsec:extracting_ZW_ZZ} we discuss, using the $M_h \rightarrow \infty$ toy model, to what extent
the angular distribution obtained, in the presence of lepton cuts, in the SM can be employed for extracting 
polarization fractions from the data even in case the underlying dynamics goes beyond the Standard Model.
Finally, in \sect{sec:conclusions} we summarize our findings.

\section{Vector bosons polarization and angular distributions of their decay products}
\label{sec:poldef}
Let us consider an amplitude in which a weak vector boson decays to a final state fermion pair.
In the Unitary Gauge, it can be expressed as

\begin{equation}\label{eq:Mlep}
\mathcal{M} = \mathcal{M}_{\mu} \frac{i}{k^2 - M^2 + i \Gamma M}\left(-g^{\mu\nu}+\frac{k^{\mu}
k^{\nu}}{M^2}\right)
\left[ {-i\,g \, \bar{\psi}_f \,\, \gamma_{\nu}
\left(c_L \frac{1 - \gamma^5}{2} + c_R\frac{1 + \gamma^5}{2} \right)\,
\psi_{f^\prime} } \right]\,,
\end{equation}
where $M$ and $\Gamma$ are the vector boson mass and width, respectively. $c_R$ and $c_L$ are the right and left handed couplings of the fermions to the $W^+(Z)$, as shown in \tbn{table:couplings}. .

The polarization tensor can be expressed in terms of four polarization vectors \cite{Kadeer:2005aq}
\begin{equation}
-g^{\mu\nu} + \frac{k^{\mu}k^{\nu}}{M^2} = \sum_{\lambda = 1}^4 \varepsilon^{\mu}_\lambda(k)
\varepsilon^{\nu^*}_{\lambda}(k)\,\,.
\label{eq:polexpansion}
\end{equation}
In the following we call single polarized amplitude with polarization
$\lambda$ an amplitude in which the sum on the left hand side of \eqn{eq:polexpansion}
is substituted by one of the terms on the right hand side,
$\sum_{\lambda'} \varepsilon_{\lambda'}^{\mu}\varepsilon_{\lambda'}^{\nu *}\,\,\,\rightarrow\,\,\,
\varepsilon_{\lambda}^{\mu}\varepsilon_{\lambda}^{\nu *}$.

In a frame in which the off shell vector boson propagates along the $(\theta_V,\phi_V)$ axis,
with three momentum $\kappa$,
energy $E$ and invariant mass $\sqrt{Q^2}=\sqrt{E^2-\kappa^2}$, the polarizations vectors read:
\begin{align}
\varepsilon^{\mu}_{L} &= \frac{1}{\sqrt 2}
(0,\cos\theta_V\cos\phi_V +  \iu\sin\phi_V , \cos\theta_V\sin\phi_V - \iu\cos\phi_V, - \sin\theta_V)\,\, \textrm{(left)} \,,\nonumber \\
\varepsilon^{\mu}_{R} &= \frac{1}{\sqrt 2}
(0,- \cos\theta_V\cos\phi_V +  \iu\sin\phi_V , - \cos\theta_V\sin\phi_V - \iu\cos\phi_V, \sin\theta_V)\,\, \textrm{ (right)} \,,\\
\varepsilon^{\mu}_{0} &= 
(\kappa,E \sin\theta_V\cos\phi_V,E \sin\theta_V\sin\phi,E \cos\theta_V)
/\sqrt{Q^2}\,\, \textrm{(longitudinal)}\,, \nonumber \\
\varepsilon^{\mu}_{A} &= \sqrt{\frac{Q^2 - M^2}{Q^2\,M^2}}
(E, \kappa \sin\theta_V\cos\phi_V,\kappa \sin\theta_V\sin\phi_V,\kappa \cos\theta_V)\,\, \textrm{(auxiliary)}\,.
\nonumber
\end{align}
In this paper, they are computed in the lab frame.

The auxiliary polarization in \eqn{eq:polexpansion} does not contribute if the decay fermions are massless.
Therefore, the normalized cross section, after integration over the azimuthal angle of the decay products,
can be expressed, in the absence of cuts on decay leptons, as follows:
\begin{eqnarray}\nonumber
 \frac{1}{\frac{d\sigma(X)}{d X}} \,\,\frac{d\sigma(\theta,X)}{d\cos\theta\, d X}
 &=&\frac{3}{8}  f_L(X) \bigg(1+\cos^2{\theta}-
\frac{2(c_L^2-c_R^2)}{(c_L^2+c_R^2)}\cos\theta\bigg)\\
&+&\frac{3}{8} f_R(X) \bigg(1+\cos^2{\theta}
+\frac{2(c_L^2-c_R^2)}{(c_L^2-c_R^2)}\cos\theta\bigg)
+\frac{3}{4} f_0(X) \sin^2\theta,
\label{eq:diffeqV}
\end{eqnarray}
$X$ stands for all additional phase space variables in addition to the decay angle $\theta$. 
The three polarization fractions $ f_L,\, f_0,\, f_R$ sum to one. 
For $W^+$ leptonic decays, $\theta$ is the angle measured in the $W^+$ rest frame between the charged 
particle
and the $W^+$ direction of flight in the lab frame. For $Z$ decays, $\theta$ is the angle measured in the $Z$
rest frame between the antifermion 
and the $Z$ direction of flight in the lab frame. For $W^-$, $\cos\theta \rightarrow - \cos\theta $.

\begingroup
\setlength{\tabcolsep}{10pt} 
\renewcommand{\arraystretch}{1.5} 

\begin{table}[th]
\begin{center}
\begin{tabular}{|C{1.5cm}||C{3cm}|C{3cm}|}
\hline
     & \bf $c_L$    & \bf $c_R$ \\ 
\hline
\hline
W    &  $1/(s \,\sqrt{2})$  & 0   \\
\hline
Z    &  $(I^3_{W,f} - s^2\, Q_f)/(s \, c)$  &  $ - s \, Q_f/c$   \\
\hline
\end{tabular}
\end{center}
\caption{
Weak couplings. $c =  \cos\theta_W = M_W/M_Z$ , $s  = \sin\theta_W $}
\label{table:couplings}
\end{table}

\endgroup
 
 Hence, each physical polarization is uniquely associated with a specific angular distribution of the charged
 lepton, even when the vector boson is off mass shell.

Defining polarized production and decay amplitudes,
\begin{equation}\label{eq:Mprod}
\mathcal{M^P}_{\lambda} = \mathcal{M}_{\mu} \varepsilon^{\mu}_{\lambda}\,,\qquad
\mathcal{M^D}_{\lambda} = 
\varepsilon^{\nu^*}_{\lambda}(k) \left[ {-i\,g \, \bar{\psi}_f \,\, \gamma_{\nu}
\left(c_L \frac{1 - \gamma^5}{2} + c_R\frac{1 + \gamma^5}{2} \right)\,
\psi_{f^\prime} } \right]\,,
\end{equation}
the full amplitude can be written as:
\begin{equation}\label{eq:Msum}
\mathcal{M} = \sum_{\lambda = 1}^3\mathcal{M^P}_{\lambda} \,\, \frac{i}
{k^2 - M^2 + i \Gamma M} \,\mathcal{M^D}_\lambda = \sum_{\lambda = 1}^3
\mathcal{M^F}_\lambda\,,
\end{equation}
where $\mathcal{M^F}_\lambda$ is the amplitude with a single polarization for the intermediate vector boson.
Notice that in each $\mathcal{M^F}_\lambda$ all correlations between production and decay are exact.

The squared amplitude becomes:
\begin{equation}\label{eq:interfpol}
\underbrace{\left|\mathcal{M}\right|^2}_{\textrm{coherent sum}} = \underbrace{\sum_{\lambda}\left|
\mathcal{M^F}_{\lambda}\right|^2}_{\textrm{incoherent sum}} + \underbrace{\sum_{\lambda \neq \lambda'}
\mathcal{M^F}_{\lambda}^{ *}\mathcal{M^F}_{\lambda'}}_{\textrm{interference terms}}\,.
\end{equation}

The interference terms in \eqn{eq:interfpol} are not, in general, zero. They cancel only when the squared
amplitude is integrated over the full range of the angle $\phi$. Acceptance cuts on the charged leptons and 
on the transverse missing momentum, unavoidable in practice, prevent from full $\phi$ integration. They
break the factorization of the angular dependence of the decay from that on the 
remaining kinematic variables $X$ which is embodied in \eqn{eq:diffeqV}. Cuts affect differently the different
single polarized angular decay distributions. The effect depends on the kinematics of the intermediate
vector bosons which in turn is determined by the underlying physics model.

\begin{figure}[bth]
\includegraphics[scale=0.45]{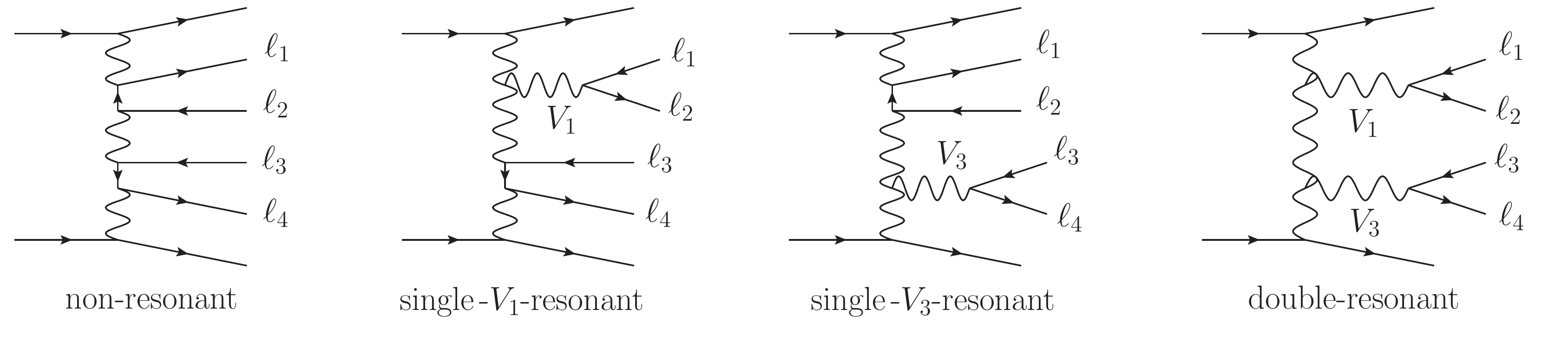}
\caption{Sample tree level diagrams for VBS at the LHC.
Scattering diagrams (like the rightmost one) are only a subset of double resonant diagrams.}
\label{fig:FeynRes}
\end{figure}

We now turn to VBS processes which feature two lepton pairs in the final state
produced in association with two quarks.
Here and in all the following we consider pure electroweak contributions at tree level,
$\mathcal{O}(\alpha^6)$.
Let us assume for simplicity that the final state lepton flavours
are chosen such that $\ell_1,\ell_2$ can be decay products of a vector boson $V_1$ and $\ell_3,\ell_4$ can
be decay products of a boson $V_3$. In such a situation, non resonant diagrams, single-$V_1$-resonant,
single-$V_3$-resonant and double resonant diagrams contribute to the tree level amplitude. Sample diagrams are shown in Fig.~\ref{fig:FeynRes}.
In order to separate the polarizations of $V_1$, non resonant and single-$V_3$-resonant contributions must be dropped.
Analogously, non resonant and single-$V_1$-resonant contributions must be dropped to separate $V_3$ polarizations.
Dropping a set of diagrams may violate gauge invariance. If this results in large numerical discrepancies
we need a procedure to produce a reliable prediction.
If both $V_1$ and $V_3$ are $W$ bosons, we have shown \cite{Ballestrero:2017bxn} that the
selection of double resonant diagrams can give physical predictions
(\emph{i.e.} approximate the full computation), provided that double On Shell projections (OSP)
are performed on the two $W$'s. Such procedure is known in the literature as Double Pole Approximation
\cite{Aeppli:1993cb, Aeppli:1993rs, Denner:2000bj, Billoni:2013aba, Biedermann:2016guo}, and
preserves the gauge invariance of squared electroweak amplitudes that can be written as the
production of two massive vector bosons times their leptonic decay. In the $WW$ final state, an additional
advantage of this
method consists in avoiding any invariant mass cut on decay products of $W$ bosons
(lepton-neutrino pairs) to reproduce accurately the results of the full calculation.

 The pole approximation has an intrinsic uncertainty of $\Gamma/M$ and thus of a few percent for weak 
 bosons. 
 Since polarizations of intermediate unstable particles have no sound theoretical basis and their definition 
 necessarily depends on conventions and approximations, predictions beyond this accuracy should not,
in any case,  be expected.

\section{Separating $Z$ resonant contributions}
\label{sec:separating}
As an introduction to separating polarizations in processes with $Z$ bosons
which decay leptonically, we briefly recall the main issues which affect isolating polarizations in
 the $WW$ channel in VBS \cite{Ballestrero:2017bxn}.
 In Tab.~\ref{table:sigma4V} we show cross sections for two partonic channels contributing to
$pp\rightarrow jje^- \bar{\nu}_{e}\mu^+\nu_{\mu}$.
\begin{table}[th]
\begin{center}
\begin{tabular}{|C{1.8cm}||C{2.5cm}|C{2.5cm}|C{2.7cm}|}
\hline
\bf $\Delta_M$      & \bf FULL     & \bf RES OSP   & \bf RES NO OSP\\
\hline
\hline
\multicolumn{4}{|c|}{\cellcolor{blue!9}$ u\ u \rightarrow u\ u \ e^-\ \bar{\nu}_{e}\ \mu^+\ \nu_{\mu} $  (2Z2W)}\\
\hline
$\infty$ (no cut)    &   403.3 (4)  &   398.1(4)  &  1025.9(9)   \\
\hline
\multicolumn{4}{|c|}{\cellcolor{blue!9}$ u\ s \rightarrow d\ c \ e^-\ \bar{\nu}_{e}\ \mu^+\ \nu_{\mu} $  (4W)}\\
\hline
$\infty$ (no cut)   &   23.80(3)   &   23.62(2)   &  29.11(4)   \\
\hline
\end{tabular}
\end{center}
\caption{
Cross sections in attobarns ($\ab$) for some VBS processes, for $\vert M_{\ell\nu_\ell} - M_W \vert < \Delta_M$ ($\ell=e,\mu$).
All diagrams are taken into account for the full  calculation (FULL), while only
double resonant diagrams contribute to the two rightmost columns, with (RES OSP) and without
(RES NO OSP) projection.
Selection cuts are:  $p_t^j>20\GeV,\,|\eta_j|<5.5,\,M_{jj}>600\GeV,\,|\Delta\eta_{jj}|>3.6,\,M_{4\ell}>300 \GeV$.}
\label{table:sigma4V}
\end{table}
The first process receives contributions only from 2Z2W amplitudes, \emph{i.e.} it includes
$Z\!Z\rightarrow W^+W^-$ scattering diagrams. The second process receives contributions only
from 4W amplitudes, as it includes $W^+W^-\rightarrow W^+W^-$ scattering diagrams.
In both cases the full result (FULL) is reproduced at the 1\% level by
the On Shell projected one (RES OSP). If double On Shell projections are not applied (RES NO OSP),
double resonant diagrams fail to reproduce the full result, since gauge invariance is violated and no cut
on $M_{\ell\nu_{\ell}}$ is imposed.
We note that unprojected resonant diagrams overestimate the cross section much more
in the 2Z2W process (+150\%) than in 4W one (+25\%).

\subsection{$Z\!Z$ processes}
\begin{figure}[h]
\includegraphics[scale=0.45]{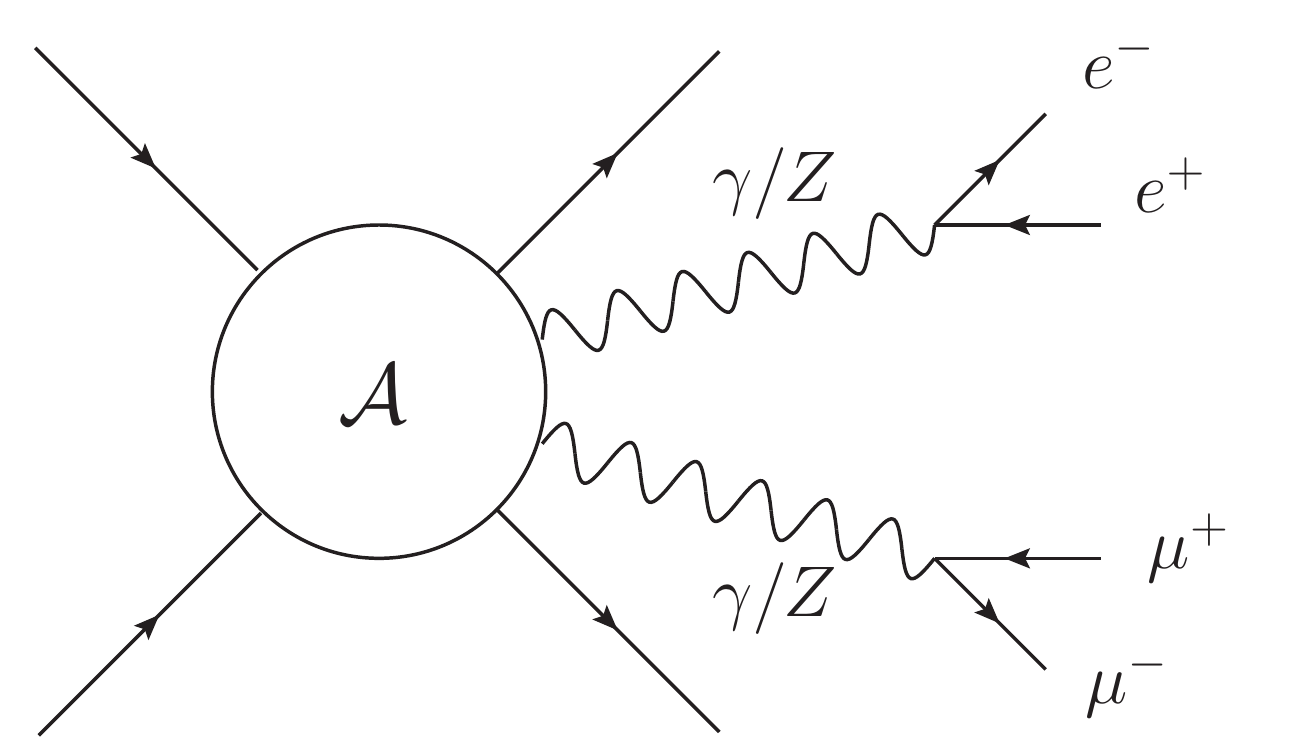}
\caption{$\gamma/Z$ resonant diagrams in $pp\rightarrow jje^+e^-\mu^+\mu^-$.}
\label{fig:zgammares}
\end{figure}

We now consider VBS $Z\!Z$ production in the four charged leptons decay channel
($pp\rightarrow jj\, e^- e^+ {\mu}^- \mu^+$), which contains both $Z\!Z\rightarrow Z\!Z$
and $W^+W^-\rightarrow Z\!Z$ scattering. Therefore, some VBS processes receive contributions
only from 4Z amplitudes, others only from 2W2Z amplitudes. Some processes, which we label mixed,
receive contributions from both of them.
In the Standard Model, charged leptons couple both to the $Z$ boson and to the
photon: this issue, when studying $Z$ bosons phenomenology, is usually treated by
selecting lepton pair invariant masses close to the $Z$ pole mass ($M_Z$).
In VBS, $\gamma\gamma$ and $\gamma Z$ resonant diagrams interfere with $Z\!Z$ resonant ones,
as shown in Fig.~\ref{fig:zgammares}.

The OSP treatment of $Z\!Z$ resonant diagrams,
defined in complete analogy with the one applied for $W\!W$ in \rf{Ballestrero:2017bxn},
is a gauge invariant procedure, \emph{i.e.}
preserves $SU(2)$ and $U(1)$ Ward identities.
However, it leads to results which are not sufficiently close to the full ones.
We have computed a 4Z process contributing to $Z\!Z$ VBS production,
with different cuts on $| M_{\ell^+\ell^-} - M_Z |$ of both lepton pairs, either
including all diagrams, or selecting only $Z\!Z$ resonant diagrams
(with or without OSP). We have imposed standard cuts on single jet kinematics
($p_t^j>20\,\GeV,\,|\eta_j|<5.5$), strong VBS cuts on the jet pair
($M_{jj}>600\GeV,\,|\Delta\eta_{jj}|>3.6$), a minimum invariant mass of the
four leptons system ($M_{4\ell}>300$ GeV), and a minimum invariant mass cut
on same flavour opposite sign lepton pairs  ($M_{\ell^+\ell^-}>40\,\GeV$), to avoid
infrared singularities due to $\gamma^*\rightarrow \ell^+\ell^-$ diagrams.

Numerical results for the total cross sections
are shown in Tab.~\ref{table:sigma4Zid_ZZ}.
\begin{table}[th]
\begin{center}
\begin{tabular}{|C{1.8cm}||C{2.7cm}|C{2.7cm}|C{2.7cm}|}
\hline
\bf $\Delta_M$      & \bf FULL     & \bf RES OSP   & \bf RES NO OSP\\
\hline
\hline
\multicolumn{4}{|c|}{\cellcolor{blue!9} $u\ u \rightarrow u\ u \ e^-\ e^+\ \mu^-\ \mu^+ $  (4Z)}\\
\hline
$\infty$ (no cut)   &  0.4479(5)  &  0.1302(2)    &   0.1318(2)  \\
\hline
30 GeV   &  0.1776(5)  &  0.1264(2)    &   0.1266(2)  \\
\hline
5  GeV   &  0.1009(1)  &   0.0955(2)   &    0.0953(1)  \\
\hline
\end{tabular}
\end{center}
\caption{
Cross sections ($\ab$) for $u\,u \rightarrow u\,u\,  e^-e^+ \mu^- \mu^+ $, for $| M_{\ell^+\ell^-} - M_Z |<\Delta_M$ ($\ell=e,\mu$). All diagrams are taken into account for the full  calculation (FULL), while only
double resonant diagrams contribute to the two rightmost columns, with (RES OSP) and without
(RES NO OSP) projection. Selection cuts are: $p_t^j>20\GeV,\,|\eta_j|<5.5,\,M_{jj}>600\GeV,\,|\Delta\eta_{jj}|>3.6,\,M_{4\ell}>300 \GeV$, $M_{\ell^+\ell^-}>40\,\GeV$.}
\label{table:sigma4Zid_ZZ}
\end{table}
If no cut is imposed on $|M_{\ell^+\ell^-}-M_Z|$, double resonant $Z\!Z$ diagrams
fail badly to describe the full result (either with or without OSP):
this is the effect of neglecting resonant $\gamma Z$ and $\gamma\gamma$
diagrams, which interfere with $Z\!Z$ ones giving large contributions in the
low $M_{\ell^+\ell^-}$ region, despite the $M_{\ell^+\ell^-}>40\,\GeV$ cut.
The situation slightly improves when imposing the cut
$|M_{\ell^+\ell^-}-M_Z|<30$ GeV, but the resonant predictions are still unreliable as they
underestimate by 30\% the full result. Imposing a sharper cut on $|M_{\ell^+\ell^-}-M_Z|$
(5 GeV), the $Z\!Z$ resonant diagrams still underestimate by 5\% the full cross section.
It is evident that the discrepancy between the resonant and full calculations is due to
the $\gamma/Z$ mixing, not only at the level of the decay into charged leptons, but more
generally at the level of the complete amplitude. In any case, the application of On Shell
projections doesn't change substantially the resonant calculation.

We have investigated further this effect, by simulating the same process with final state
neutrinos, instead of charged leptons, \emph{i.e.} $u\,u \rightarrow u\, u \,\nu_e \bar{\nu}_e \nu_{\mu} \bar{\nu}_{\mu}$.
Numerical results are shown in Tab.~\ref{table:sigma4VZ_vvvv}.
The presence of neutrinos implies that there are no contributions in which the photon couples to
the final state leptons.
In fact, the large discrepancy between full and resonant results obtained in the four charged
leptons case (first line of Tab.~\ref{table:sigma4Zid_ZZ}) is much reduced, even without any
cut on $M_{\nu_\ell\bar{\nu}_\ell}$.
Nevertheless, the resonant calculation, either with or without OSP, gives a
cross section which is 8\% smaller than the full one, at variance with the case
of final state $W$'s \cite{Ballestrero:2017bxn}.
%
%

\begin{table}[th]
\begin{center}
\begin{tabular}{|C{1.8cm}||C{2.5cm}|C{2.5cm}|C{2.7cm}|}
\hline
\bf $\Delta_M$     & \bf FULL     & \bf RES OSP   & \bf RES NO OSP\\
\hline
\hline
\multicolumn{4}{|c|}
{\cellcolor{blue!9}$ u\ u \rightarrow u\ u \ \nu_e\ \bar{\nu}_e\ \nu_{\mu}\ \bar{\nu}_{\mu}$  (4Z)}\\
\hline
$\infty$ (no cut)    &   0.5580(1)   &   0.5113(2)   &  0.5165(2)   \\
\hline
\end{tabular}
\end{center}
\caption{
Cross sections ($\ab $) for $u\,u \rightarrow u\, u \,\nu_e \bar{\nu}_e \nu_{\mu} \bar{\nu}_{\mu}$. All diagrams are taken into account for the full  calculation (FULL), while only double resonant diagrams contribute to the two rightmost columns, with (RES OSP) and without (RES NO OSP) projection.
Selection cuts are:  $p_t^j>20\GeV,\,|\eta_j|<5.5,\,M_{jj}>600\GeV,\,|\Delta\eta_{jj}|>3.6,\,M_{4\ell}>300 \GeV$.}
\label{table:sigma4VZ_vvvv}
\end{table}

So far we have considered a partonic process which receives contribution from 4Z amplitudes.
We now perform an analogous study for $u\, s \rightarrow c\, d\, e^- e^+ {\mu}^- \mu^+ $,
which receives contribution from 2W2Z amplitudes. The chosen kinematic cuts are identical
to those detailed above. Numerical results are shown in Tab.~\ref{table:sigma2W2Z_ZZ} for
different choices of the cut $\Delta_M$ on both lepton pair invariant masses around the $Z$
pole mass, \emph{i.e.} $|M_{e^+e^-} - M_Z|<\Delta_M$ and $|M_{\mu^+\mu^-} - M_Z|<\Delta_M$.
\begin{table}[th]
\begin{center}
\begin{tabular}{|C{1.8cm}||C{2.5cm}|C{2.5cm}|C{2.7cm}|}
\hline
\bf $\Delta_M$      & \bf FULL     & \bf RES OSP   & \bf RES NO OSP\\
\hline
\hline
\multicolumn{4}{|c|}{\cellcolor{blue!9} $u\ s \rightarrow c\ d \ e^-\ e^+\ \mu^+\ \mu^- $  (2W2Z)}\\
\hline
$\infty$ (no cut)   & 2.680(3) & 2.483(3) & 3.245(8)\\
\hline
30 GeV   & 2.457(3) & 2.410(3) & 2.432(2)\\
\hline
5  GeV   & 1.824(3) & 1.822(4) & 1.823(6)\\
\hline
\end{tabular}
\end{center}
\caption{
Cross sections ($\ab $) for  $u\, s \rightarrow c\, d\, e^- e^+ \mu^- \mu^+ $,  for
$| M_{\ell^+\ell^-} - M_Z |<\Delta_M$ ($\ell=e,\mu$). All diagrams are taken into account for the full calculation (FULL),
while only double resonant diagrams contribute to the two rightmost columns, with (RES OSP) and without
(RES NO OSP) projection. Selection cuts are: $p_t^j>20\GeV,\,|\eta_j|<5.5,\,M_{jj}>600\GeV,\,|\Delta
\eta_{jj}|>3.6,\,M_{4\ell}>300 \GeV$, $M_{\ell^+\ell^-}>40\,\GeV$.}
\label{table:sigma2W2Z_ZZ}
\end{table}
If no restriction on $|M_{\ell^+\ell^-}-M_Z|$ is imposed, the resonant calculation (RES NO OSP)
overestimates the full cross section by more than 20\%. If OSP are applied, the
cross section becomes 7\% smaller than the full one: OSP play a different role in 2W2Z amplitudes,
as they seem to regularize partially the gauge violating double resonant calculation, and the
remaining discrepancy with respect to the full result is due to the $\gamma$ coupling to charged
leptons which is the main missing contribution in this setup.
This is confirmed by the cross sections obtained for neutrinos in the final state, shown in
Tab.~\ref{table:sigma4VZ_vvvv_W}.
\begin{table}[th]
\begin{center}
\begin{tabular}{|C{1.8cm}||C{2.5cm}|C{2.5cm}|C{2.7cm}|}
\hline
\bf $\Delta_M$     & \bf FULL     & \bf RES OSP   & \bf RES NO OSP\\
\hline
\hline
\multicolumn{4}{|c|}
{\cellcolor{blue!9}$ u\ s \rightarrow d\ c \ \nu_e\ \bar{\nu}_e\ \nu_{\mu}\ \bar{\nu}_{\mu} $  (2W2Z)}\\
\hline
$\infty$ (no cut)    &   9.888(8)   &   9.758(9)  &  12.66(4)   \\
\hline
\end{tabular}
\end{center}
\caption{
Cross sections ($\ab $) for $ u\ s \rightarrow d\ c \ \nu_e\ \bar{\nu}_e\ \nu_{\mu}\ \bar{\nu}_{\mu} $, for $\vert M_{\nu_\ell\bar{\nu}_\ell} - M_Z \vert<\Delta_M$ ($\ell=e,\mu$).
All diagrams are taken into account for the full  calculation (FULL), while only
double resonant diagrams contribute to the two rightmost columns, with (RES OSP) and without
(RES NO OSP) projection.
Selection cuts are:  $p_t^j>20\GeV,\,|\eta_j|<5.5,\,M_{jj}>600\GeV,\,|\Delta\eta_{jj}|>3.6,\,M_{4\ell}>300 \GeV$, $M_{\ell^+\ell^-}>40\,\GeV$.}
\label{table:sigma4VZ_vvvv_W}
\end{table}
In this case, since photons do not couple to neutrinos, the OSP result
provides a good description of the full one (-1\%). The resonant
computation without OSP doesn't provide a trustworthy prediction (+30\%).
Turning back to the four charged leptons case, if a cut on $|M_{\ell^+\ell^-}-M_Z|$ is
applied for both lepton pairs (see last two lines of Tab.~\ref{table:sigma2W2Z_ZZ}),
$Z\!Z$ resonant diagrams reproduce accurately the full result, both with and
without OSP. For a very sharp cut (5 GeV), the full result is reproduced at the per mill level.

Other relevant processes contributing to $Z\!Z$ scattering at the LHC are the mixed ones, which
receive contributions both from 4Z and from 2W2Z amplitudes.
We have checked that for such partonic channels the 2W2Z contribution is dominant,
the 4Z one accounts for a few percent of
the total, and the interference between the two sets of contributions is negligible.
The numerical results in Tab.~\ref{table:sigmamixed_ZZ} show that, in the presence of a reasonable cut on $M_{\ell^+\ell^-}$ around $M_Z$, the double resonant calculation describes accurately the full one.
\begin{table}[th]
\begin{center}
\begin{tabular}{|C{1.8cm}||C{2.5cm}|C{2.5cm}|C{2.7cm}|}
\hline
\bf $\Delta_M$      & \bf FULL     & \bf RES OSP   & \bf RES NO OSP\\
\hline
\hline
\multicolumn{4}{|c|}{\cellcolor{blue!9} $u\ d \rightarrow u\ d \ e^-\ e^+\ \mu^+\ \mu^- $  (mixed)}\\
\hline
$\infty$ (no cut)   & 20.09(2) & 18.62(2) &  24.67(4) \\
\hline
30 GeV   & 18.45(3) & 18.08(3) &  1.826(2)\\
\hline
5  GeV   & 13.69(2) & 13.65(4) &  1.362(2)\\
\hline
\end{tabular}
\end{center}
\caption{
Cross sections ($\ab $) for
$u\, d \rightarrow u\, d \ e^- e^+ \nu_{\mu} \mu^+ $, for $\vert M_{\ell^+\ell^-} - M_Z \vert<\Delta_M$ ($\ell=e,\mu$).
All diagrams are taken into account for the full  calculation (FULL), while only
double resonant diagrams contribute to the two rightmost columns, with (RES OSP) and without
(RES NO OSP) projection.
Selection cuts are: $p_t^j>20\GeV,\,|\eta_j|<5.5,\,M_{jj}>600\GeV,\,|\Delta\eta_{jj}|>3.6,\,M_{4\ell}>300 \GeV$, $M_{\ell^+\ell^-}>40\,\GeV$.}
\label{table:sigmamixed_ZZ}
\end{table}

We remark that, simulating $Z\!Z$ scattering at the LHC including all possible partonic processes, the mixed ones
account for 80\% of the total cross section, pure 2W2Z processes account for more than 18\%, and 4Z processes
give a very small contribution (order 1\%). This gives us confidence that double resonant diagrams provide
reliable
predictions, provided that a reasonable cut is imposed on $|M_{\ell^+\ell^-}-M_Z|$, both with and without OSP.

\subsection{$W\!Z$ processes}
We now investigate how well resonant diagrams can reproduce
full results in the $W\!Z$ scattering channel. For this purpose we consider a single
partonic process which features three charged leptons and a neutrino in the final state,
namely $u \,c \rightarrow u \,s\, e^+e^-\mu^+\nu_\mu $. We apply exactly the same cuts as
those applied for $Z\!Z$.

We show in Tab.~\ref{table:sigma2W2Z_WZ} the total cross sections corresponding to a
few representative cuts, $\Delta_M$, on the difference between the invariant mass of
the decay particles and the vector boson mass. We require $|M_{\mu^+\nu_\mu}-M_W| < \Delta_M$
and $|M_{e^+e^-}-M_Z| < \Delta_M $, both with full matrix elements, and selecting only $W\!Z$
resonant diagrams (with or without OSP).

\begin{table}[th]
\begin{center}
\begin{tabular}{|C{1.8cm}||C{2.5cm}|C{2.5cm}|C{2.7cm}|}
\hline
\bf    $\Delta_M$   & \bf FULL     & \bf RES OSP   & \bf RES NO OSP\\
\hline
\hline
\multicolumn{4}{|c|}{\cellcolor{blue!9} $u\ s \rightarrow d\ s \ e^-\ e^+\ \nu_{\mu}\ \mu^+ $  (WZWZ)}\\
\hline
$\infty$ (no cut)   &  2.897(3)  &  2.746(4)    &   3.776(3)  \\
\hline
30 GeV   &  2.701(2)  &  2.667(2)    &   2.686(2)  \\
\hline
5  GeV   &  2.064(2)  &  2.060(3)    &   2.059(3)  \\
\hline
\end{tabular}
\end{center}
\caption{
Cross sections ($\ab $) for  $u\ s \rightarrow d\ s \ e^-\ e^+\ \nu_{\mu}\ \mu^+ $,  for
$\vert M_{e^+e^-} - M_Z \vert  < \Delta_M $ and $\vert M_{\mu^+\nu_{\mu}} - M_W \vert   < \Delta_M$.
All diagrams are taken into account for the full  calculation (FULL), while only
double resonant diagrams contribute to the two rightmost columns, with (RES OSP) and without
(RES NO OSP) projection.
Selection cuts are: $p_t^j>20\GeV,\,|\eta_j|<5.5,\,M_{jj}>600\GeV,\,|\Delta\eta_{jj}|>3.6,\,M_{4\ell}>300 \GeV$, $M_{\ell^+\ell^-}>40\,\GeV$.}
\label{table:sigma2W2Z_WZ}
\end{table}
In the absence of invariant mass restrictions on lepton pairs, the OSP cross section is
5\% smaller than the full one, as a result of the unconstrained $M_{e^+e^-}$ which can be far from
the $Z$ pole mass (but $M_{\ell^+\ell^-}>40\,\GeV$), where photons play an important role.
The resonant calculation without OSP is far from providing
reliable results in this situation (+30\% w.r.t. the full cross section).
In the presence of the cuts $\vert M_{e^+e^-} - M_Z \vert<30\,\GeV$ and
$\vert M_{\mu^+\nu_{\mu}} - M_W \vert<30\,\GeV$, resonant calculations underestimate the full
result by only 1\%. Making the invariant mass cuts even sharper (5 GeV), the resonant calculation describes
perfectly (0.1\% accuracy) the full result, both with and without On Shell projections.
The conclusions we can draw for $W\!Z$ are very similar to those for 2W2Z processes in $Z\!Z$ scattering, apart from
smaller effects related to $\gamma$ decay, since $W\!Z$ features only one opposite charge, same flavour lepton
pair in the final state.
However, differently from $Z\!Z$, in fully leptonic $W\!Z$ scattering imposing a cut on $M_{\ell\nu_\ell}$
is physically unfeasible, thus we have to work out an alternative procedure to separate resonant contributions,
which allows to avoid any cut on $M_{\ell\nu_\ell}$.

A possible way to avoid cuts on the $\ell^+\nu_\ell$ pair invariant mass consists in performing an On Shell projection on the $W$ boson. This procedure is rather different from the double On Shell projections introduced above.

We select only $W$ resonant diagrams (single-$W$-resonant and
double resonant), dropping all the other contributions, which cannot be interpreted as the production
of a $W$ times its leptonic decay. The single On Shell projection (OSP1, for brevity) procedure consists
in projecting on mass shell the numerator of the $W$ resonant amplitude, leaving the Breit Wigner
modulation untouched. In formulas,
\beq
\mathcal A\,&=&\,\mathcal{A}_{\rm res}\,+\,\mathcal{A}_{\rm nonres}\nnb\\
&=&\,\sum_{\lambda} \left[\frac{\mathcal{A}^{\mathcal{P}}_{\mu}(q_1,q_2;k,\{p_i\})\,
\varepsilon_{\lambda}^{\mu}(k)\,\varepsilon_{\lambda}^{*\nu}(k)\,\mathcal{A}^{\mathcal{D}}_{\nu}(k,\{l_1,l_2\})}
{k^2-M_W^2 + i\Gamma_WM_W}\right]\,+\,\mathcal{A}_{\rm nonres}\nnb\\
&\rightarrow & \,\sum_{\lambda}\left[\frac{ \mathcal{A}^{\mathcal{P}}_{\mu}(\bar{q}_1,\bar{q}_2;\bar{k},\{p_i\})\,
\varepsilon_{\lambda}^{\mu}(\bar{k})\,\varepsilon_{\lambda}^{*\nu}(\bar{k})\,\mathcal{A}^{\mathcal{D}}_{\nu}
(\bar{k},\{\bar{l}_1,\bar{l}_2\})}{k^2-M_W^2 + i\Gamma_WM_W}\right]\,=\,\mathcal{A}_{\rm OSP1}\,,
\eeq

where $q_1,q_2$ are the initial parton momenta, $k$ is the $W$ momentum, $l_1,l_2$ are the $W$ decay
product momenta and $\{p_i\}$ are the momenta of the other final state particles. Barred momenta are the
projected ones, in particular, ${\bar{k}}^2=M_W^2$.

The projection procedure is not uniquely defined, since different sets of physical quantities can be kept unmodified.
Our choice is to preserve:
\begin{enumerate}
\item the space like components of the $W$ boson momentum in the laboratory reference frame,
\item the direction of the leptonic decay products momenta in the $W$ rest frame,
\item the four momenta of all other final state particles (system $X$).
\end{enumerate}
Since $k$, the off shell momentum of the $W$,  is projected to the on shell momentum $\bar{k}$,
while the other final state particles ($X$) are left untouched, the recoil $\Delta k = k-\bar{k}$
must be absorbed by the initial state partons ($q_1,q_2\rightarrow \bar{q}_1,\bar{q}_2$).

The same procedure can be applied when separating $Z$ resonant diagrams. In this case, it amounts to
selecting single-$Z$-resonant and double resonant diagrams, and then projecting on shell the $Z$ boson.

To evaluate how well OSP1 results reproduce the full matrix element ones, we have computed the
same cross sections of Tab.~\ref{table:sigma2W2Z_WZ} without any cut on $M_{\mu^+\nu_\mu}$. We have performed
the calculation in three different ways: including all contributions (FULL), applying OSP1 on $W$ resonant diagrams (OSP1-W),
and applying OSP1 on $Z$ resonant diagrams (OSP1-Z). Numerical results are shown in Tab.~\ref{tableosp1}.
\begin{table}[th]
\begin{center}
\begin{tabular}{|C{1.8cm}||C{2.5cm}|C{2.7cm}|C{2.7cm}|}
\hline
\bf $\Delta_M$      & \bf FULL     & \bf OSP1-W   & \bf OSP1-Z\\
\hline
\hline
\multicolumn{4}{|c|}{\cellcolor{blue!9} $u\ s \rightarrow d\ s \ e^-\ e^+\ \nu_{\mu}\ \mu^+ $  (WZWZ)}\\
\hline
$\infty$ (no cut)   & 2.897(3) &  2.880(4) &  2.752(3) \\
\hline
30 GeV   & 2.749(3) &  2.741(4) &  2.707(3)\\
\hline
5  GeV   & 2.359(3) &  2.357(2)&  2.354(3)\\
\hline
\end{tabular}
\end{center}
\caption{
Cross sections ($\ab $) for  $u\ s \rightarrow d\ s \ e^-\ e^+\ \nu_{\mu}\ \mu^+ $, for $| M_{e^+e^-} - M_Z |<\Delta_M$.
All diagrams are taken into account for the full  calculation (FULL), only
$W(Z)$ resonant diagrams contribute to the approximate calculation with OSP1-W(Z).
Selection cuts are: $p_t^j>20\GeV,\,|\eta_j|<5.5,\,M_{jj}>600\GeV,\,|\Delta\eta_{jj}|>3.6,\,M_{4\ell}>300 \GeV$, $M_{\ell^+\ell^-}>40\,\GeV$.}
\label{tableosp1}
\end{table}

Both in the absence and in the presence of cuts on $|M_{e^+e^-}-M_Z|$, the OSP1-W calculation provides predictions which are impressively close to the full ones (less than 0.7\% differences).
The issues related to the $Z$ resonant part of the amplitude are absent, since all the contributions relevant for the description of the $e^+e^-$ pair are included both in the full and in the OSP1-W calculation.

When applying OSP1 to $Z$ resonant diagrams, the situation is slightly different.
When leaving $|M_{e^+e^-}-M_Z|$ unconstrained, the approximate result is 5\% smaller
than the full, and very similar to the result obtained applying double On Shell
projections (RES OSP) on double resonant diagrams (see second column, first line
of Tab.~\ref{table:sigma2W2Z_WZ}). The 5\% discrepancy can be traced back (exactly
as for double OSP) to the missing photon decay diagrams, which give non negligible
contributions to the cross section. When applying a reasonable cut on $|M_{e^+e^-}-M_Z|$,
the OSP1-Z result agrees with the full one within 1\%. The sharper the cut, the
better the agreement.

We stress that the OSP1-W(Z) results presented in Tab.~\ref{tableosp1} assume
that we select only single $W$($Z$) resonant and double resonant diagrams.
This would be enough to separate polarizations of a single vector
boson at a time. Nevertheless, with a view to separating polarizations for both vector
bosons, it is even possible to treat only $W\!Z$ double resonant
contributions with OSP1 to describe the full result with reasonable accuracy, still
avoiding cuts on $M_{\ell\nu_\ell}$.
This can be done performing OSP1-W, and imposing a cut on $|M_{\ell^+\ell^-}-M_Z|$.
The results obtained with this approximation in the presence of a 30 and 5 GeV cut
on $|M_{\ell^+\ell^-}-M_Z|$ are almost identical to those shown in the rightmost
column of Tab.~\ref{tableosp1}.

In conclusion, the OSP1 procedure provides an alternative approach to separate and
treat $W(Z)$ resonant diagrams in $W\!Z$ scattering, which reproduces
correctly the full results in the presence of a reasonable cut on $|M_{\ell^+\ell^-}-M_Z|$,
and avoiding cuts on $|M_{\ell\nu_\ell}-M_W|$.

\section{$Z\!Z$ scattering}
\label{sec:ZZ}
The cross section for $Z\!Z$ production in VBS has been measured by the CMS Collaboration in the fully leptonic channel
\cite{Sirunyan:2017fvv}. In the fiducial region, at 13 TeV, it is less than 1 $fb$.
Therefore,  a detailed investigation of this process, even with the full luminosity of LHC Run 2, will be impossible.
The high luminosity run of the LHC, with an integrated luminosity of about 3 ${ab}^{-1}$, is more promising
and will hopefully allow for a separation of the longitudinal cross section
in this channel \cite{CMS:2018mbt, Azzi:2019yne}. The four charged leptons in the final state enable a precise
reconstruction of the $Z$ decays and of their angular distributions.
Furthermore, reducible backgrounds are small.

The $\gamma\gamma$ and $\gamma Z$ contributions, already mentioned in previous sections, can be controlled
requiring the $\ell^+\ell^-$ invariant mass to be close to the $Z$ pole mass: an experimentally viable cut
is $|M_{\ell^+\ell^-}-M_Z|<15\,\GeV$.

\subsection{Setup of the simulations}
\label{subsec:setupzz}
We consider the process $pp\rightarrow jje^+e^-\mu^+\mu^-$ at the LHC@13TeV.
All simulations have been performed at parton level with 
\texttt{PHANTOM} \cite{Ballestrero:2007xq,Ballestrero:1994jn}, employing
\texttt{NNPDF30\_lo\_as\_0130} PDFs, with factorization scale $\mu=M_{4\ell}/\sqrt{2}$,
in coherence with \rf{Ballestrero:2017bxn}, and as suggested in Sect.~I.8.3 of \rf{deFlorian:2016spz}.

We consider
tree level electroweak contributions only ($\mc O (\alpha^6)$) and neglect partonic processes involving
$b$ quarks, which account for less than 0.5\% of the total cross section.

The following set of kinematic cuts is applied for all results presented in this section:
\begin{itemize}
\item[-] maximum jet rapidity, $|\eta_j|<5$;
\item[-] minimum jet transverse momentum, $p_t^j>20$ GeV;
\item[-] minimum invariant mass of the system of the two tagging jets, $M_{jj}>500$ GeV;
\item[-] minimum rapidity separation between the two tagging jets, $|\Delta\eta_{jj}|> 2.5$;
\item[-] maximum difference between the invariant mass of each charged lepton pair (same flavor, opposite sign) and
the $Z$ pole mass, $|M_{\ell^+\ell^-}-M_Z|<15$ GeV;
\item[-] minimum invariant mass of the four charged lepton system, $M_{4\ell}>200$ GeV.
\end{itemize}
In Sect.~\ref{subsec:yeslepcutzz}, in addition to those detailed above, transverse momentum and rapidity cuts are
imposed on charged leptons kinematics: $p_t^\ell>20$ GeV, $|\eta_\ell|<2.5$.

Requiring the invariant mass of the four lepton system to be larger than 200 GeV, shields our results from 
the Higgs peak and selects the large diboson invariant mass region which is the most interesting one for 
studying the EWSB mechanism.  We have preferred a mild cut which gives us more flexibility in the search for 
observable signatures and makes it easier to compare with the $WW$ channel results in 
\rf{Ballestrero:2017bxn}, 
where the On Shell projection procedure forces the invariant mass of the four leptons to be larger
than $2\,M_W$.

\subsection{Single polarized results and their validation in the absence of lepton cuts}
\label{subsec:nolepcutzz}
In this section we provide Standard Model predictions for polarized $Z\!Z$ scattering, in the
absence of $p_t$ and $\eta$ cuts on charged leptons, and validate the results against the
polarization information which can be extracted from the full distributions. In the following we only separate the
polarized components of the $Z$ which decays to $e^+e^-$. The $Z$ which decays to $\mu^+\mu^-$ is always unpolarized.

Following the conclusions of Sect.~\ref{sec:separating}, in order to define polarized signals, we select
only double resonant $Z\!Z$ diagrams and impose  $|M_{\ell^+\ell^-}-M_Z|<15$ GeV for each lepton pair, without performing any on shell projection of the intermediate $Z$ bosons, since this would be numerically 
irrelevant.

The total cross section obtained from the full matrix element is 122.41(9) $\ab $.
In Table~\ref{tab:zzxsec} we show  the total cross sections corresponding to the underlying scattering reactions.

\begin{table}[th]
\begin{center}
\begin{tabular}{|c|c|}
\hline
\multicolumn{2}{|c|}{\cellcolor{blue!9} Cross sections [$\ab $]}\\
\hline
\hline
$Z\!Z\rightarrow Z\!Z$   & 0.979(2) (0.8\%) \\
\hline
$W\!W\rightarrow Z\!Z$   &  24.73(3) (20.2\%)\\
\hline
mixed   &  96.70(5) (79\%)\\
\hline
total   &  122.41(9) (100\%)\\
\hline
\end{tabular}
\end{center}
\caption{
Cross sections ($\ab $) for the VBS $Z\!Z$ production in the absence of lepton cuts. The mixed
contribution refers to partonic subprocesses which include
both $Z\!Z\rightarrow Z\!Z$ and $W\!W\rightarrow Z\!Z$ scattering subdiagrams. In parentheses we show the
relative fractions.
}
\label{tab:zzxsec}
\end{table}

We observe that the partonic processes embedding only $Z\!Z\rightarrow Z\!Z$
account for less than 1\% of the total. Furthermore, we have shown in Sect.~\ref{sec:separating}
that for mixed $e^+e^-\mu^+\mu^-$ processes the contribution of 4Z amplitudes
account for less than 1\% and that the interference  between 4Z and 2W2Z amplitudes is negligible.
Therefore, when including all partonic processes, any effect of $\gamma/Z$ mixing is very small.

First, we investigate how well the different approximations compare with the full result in the unpolarized
case. The unpolarized cross section obtained selecting only resonant diagrams  is $121.48(8)\,\ab$, which
reproduces the full cross section within 1\%.
If, in addition, we perform On Shell projections on resonant contributions,
the approximate cross section is $121.44(8)\,\ab $. Since
On Shell projections do not improve the approximate calculation, we do not apply them for $Z\!Z$ in
the following.

We now consider polarized signals.
As lepton cuts are not applied, the incoherent sum of  polarized cross sections
($121.52(5)\,\ab $) reproduces perfectly their coherent sum
presented above ($121.48(8)\,\ab $).
The results for a polarized $Z$ are shown in Table~\ref{tab:polzzxsec}.
\begin{table}[th]
\begin{center}
\begin{tabular}{|c|c|}
\hline
\multicolumn{2}{|c|}{\cellcolor{blue!9} Polarized cross sections [$\ab $]}\\
\hline
\hline
longitudinal  & 32.60(2) \\
\hline
left handed   & 56.55(4) \\
\hline
right handed  & 32.37(2) \\
\hline
sum  & 121.52(5)\\
\hline
\end{tabular}
\end{center}
\caption{Polarized cross sections ($\ab $) for the VBS $Z\!Z$ production in the absence of lepton cuts.}
\label{tab:polzzxsec}
\end{table}

In order to validate the separation of polarized components, we compare the polarized cross sections obtained
with the Monte Carlo with the results which, in the absence of lepton cuts, can be extracted from the full unpolarized
$\cos\theta_{e^-}$ distributions. $\theta_{e^-}$ is the angle, in the correspondent $Z$ rest frame, between the electron
direction and the $Z$ direction in the lab frame.
As a consequence of Eq.~\ref{eq:diffeqV}, each polarized decay cross section of a
$Z$ can be simply determined as it is a superposition of the first three Legendre polynomials.

\begin{figure}[!htb]
\centering
\subfigure[$M_{Z\!Z}>200 \,\GeV$, 200 GeV $< p_t^{Z_{e}}<$ 400 GeV\label{fig:mzz750}]
{\includegraphics[scale=0.37]{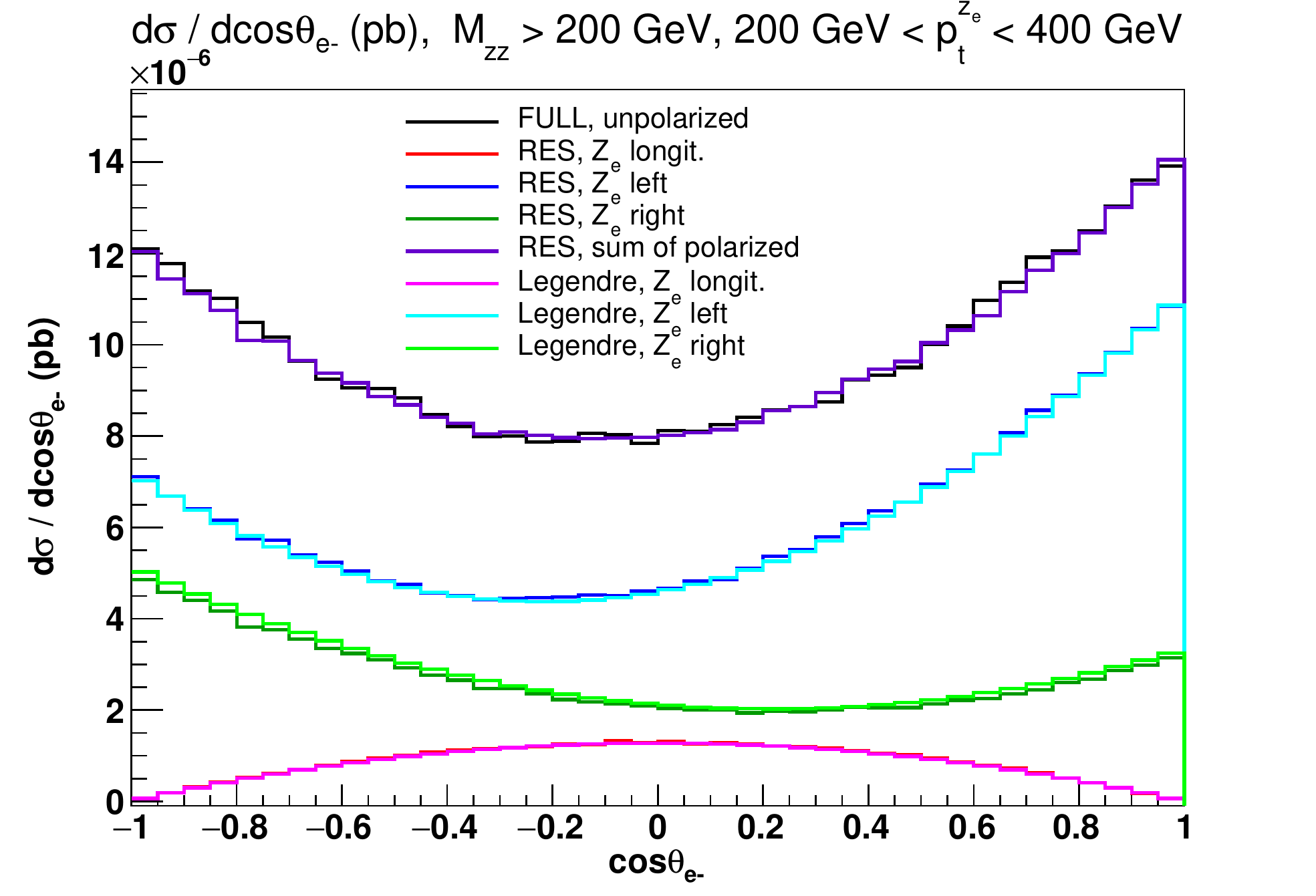}}
\subfigure[$M_{Z\!Z}> 750$ GeV\label{fig:200ptze400}
]{\includegraphics[scale=0.37]{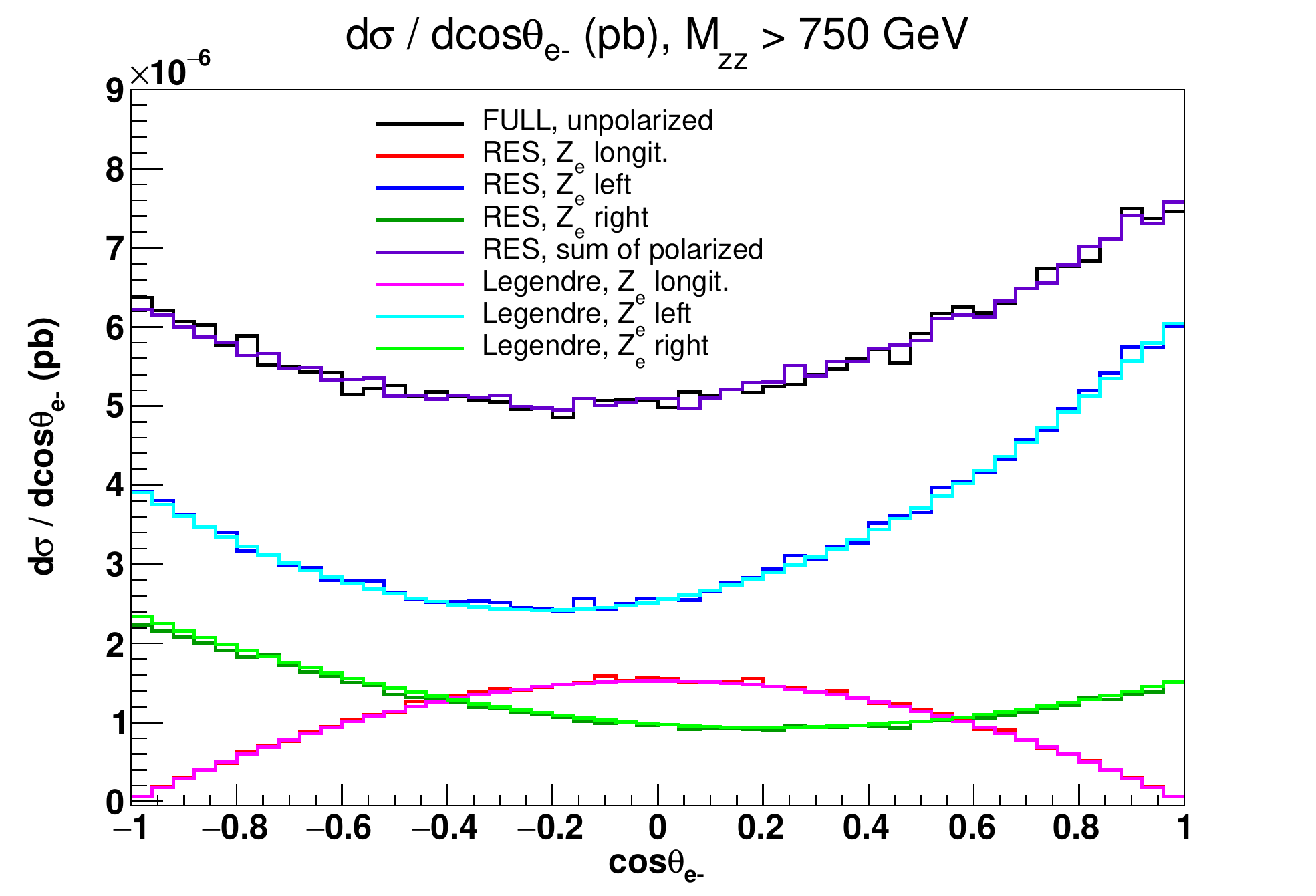}}
\caption{$Z\!Z$ scattering: $\cos\theta_{e^-}$ distributions in two different kinematic regions, in the absence of leptonic cuts, for a
polarized $Z$ decaying into $e^+e^-$. The pink, azure and light green curves represent respectively the
longitudinal, left and right contributions obtained by expanding the full angular distribution (black
curve) on the first three Legendre polynomials. The red, blue and dark green histograms represent respectively the
longitudinal, left and right contributions computed with polarized amplitudes. The
$Z$ decaying to $\mu^+\mu^-$ is unpolarized.}\label{fig:zzcostheta}
\end{figure}

In Fig.~\ref{fig:zzcostheta} we show the $\cos\theta_{e^-}$ distributions in two kinematic regions
which are most sensitive to new physics, namely at large boson $p_t$ and large mass of the diboson system.

In both figures, the full distribution is shown in black. The longitudinal, left and right distributions obtained with
the Monte Carlo polarized amplitudes are shown in red, blue and dark green, respectively. The violet curve is their sum.
The pink, cyan and light green curves represent the
polarized distributions obtained projecting the full distribution onto the first three Legendre polynomials.
The agreement between Monte Carlo polarized signals and the corresponding Legendre projections results is good,
both for the total cross sections and for the distribution shapes. The discrepancy
between the full distributions and the sum of polarized results, though $\lesssim 3\%$ bin per
bin in all the analyzed kinematic region, are slightly larger than those observed in $WW$ scattering. This may be
due to $\gamma$ contributions, as well as to non resonant contributions, which are missing in the resonant
approximate calculation. However, the agreement is satisfactory, therefore we can proceed to study polarized signals
in the presence of lepton cuts.

\subsection{Effects of lepton cuts on polarized distributions}
\label{subsec:yeslepcutzz}
The inclusion of $p_t$ and $\eta$ cuts on charged leptons defines
a fiducial region where it is possible to reconstruct the entire
final state.
The total cross section computed with full matrix elements is
61.02(4)$\,\ab $ . The result of the computation
including only double resonant diagrams is 60.59(4) $\ab $
which reproduces the full result with 1\% accuracy, meaning that,
also in this case, non resonant and $\gamma$ contributions are very
small.
Applying On Shell projections we get 60.46(4) $\ab $,
showing that they can be avoided, even in the presence of the full
set of lepton cuts, which spoils the cancellation of interference
terms among different polarization states.
This introduces an additional source of discrepancy between the
incoherent sum of polarized distributions and the full unpolarized
distribution. This is quantified in Table~\ref{tab:xseczzlepcut}.

\begin{table}[th]
\begin{center}
\begin{tabular}{|c|c|}
\hline
\multicolumn{2}{|c|}{\cellcolor{blue!9} Polarized cross sections [$\ab $]}\\
\hline
\hline
longitudinal   & 16.19(1) \\
\hline
left handed   & 26.76(2) \\
\hline
right handed  & 15.95(1) \\
\hline
sum  & 58.91(2)\\
\hline
\end{tabular}
\end{center}
\caption{Polarized cross sections ($\ab $) for the VBS $Z\!Z$ production in the presence of lepton cuts.}
\label{tab:xseczzlepcut}
\end{table}
The incoherent sum of polarized total cross sections underestimates the full cross section by 3.5\%:
interferences among polarization modes are small but non negligible.

\begin{figure}[!h]
\centering
\subfigure[{$\cos\theta_{e^-} $}\label{fig:distrib_cte_cut}]
{\includegraphics[scale=0.37]{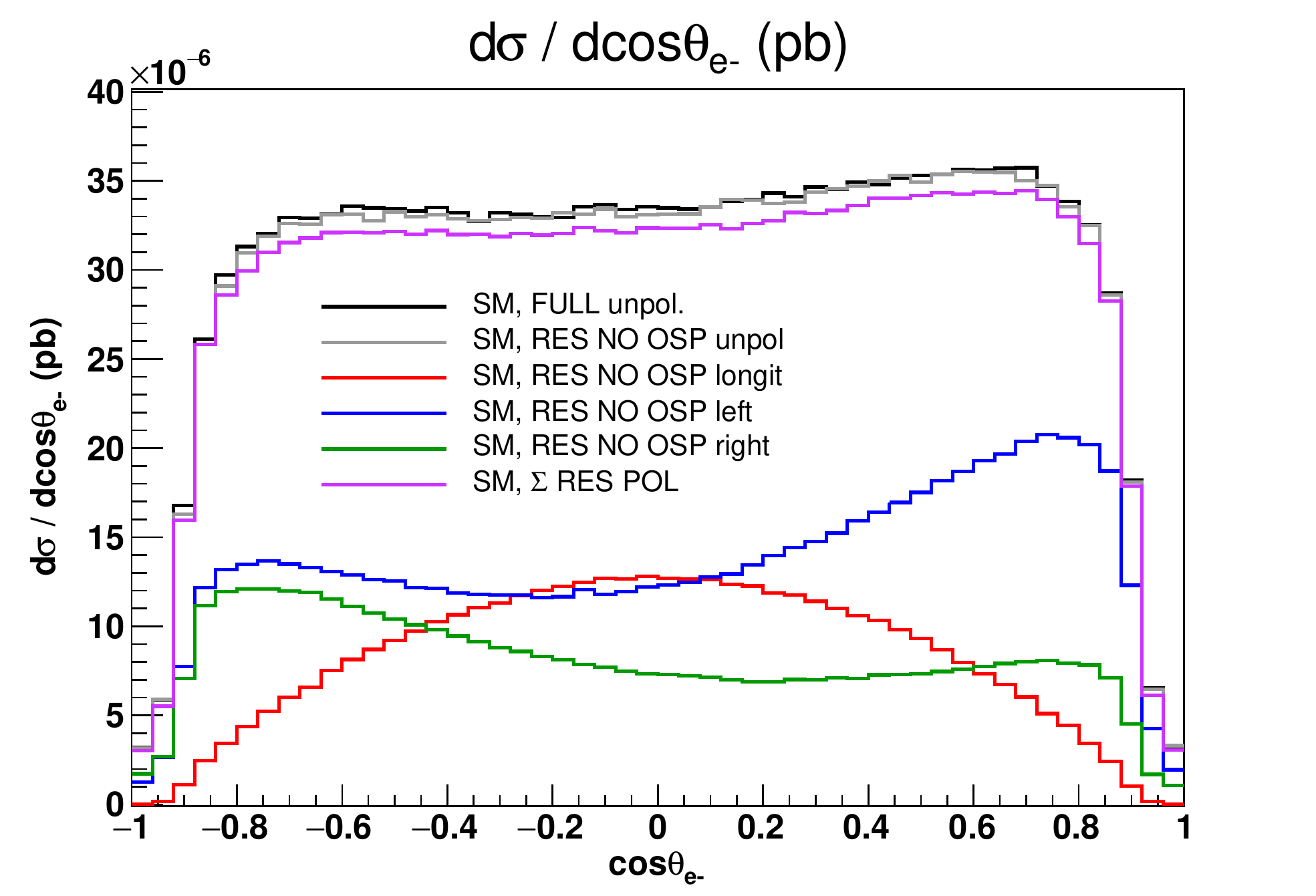}}
\subfigure[{$M_{Z\!Z}$}\label{fig:distrib_Mzz_cut}]
{\includegraphics[scale=0.37]{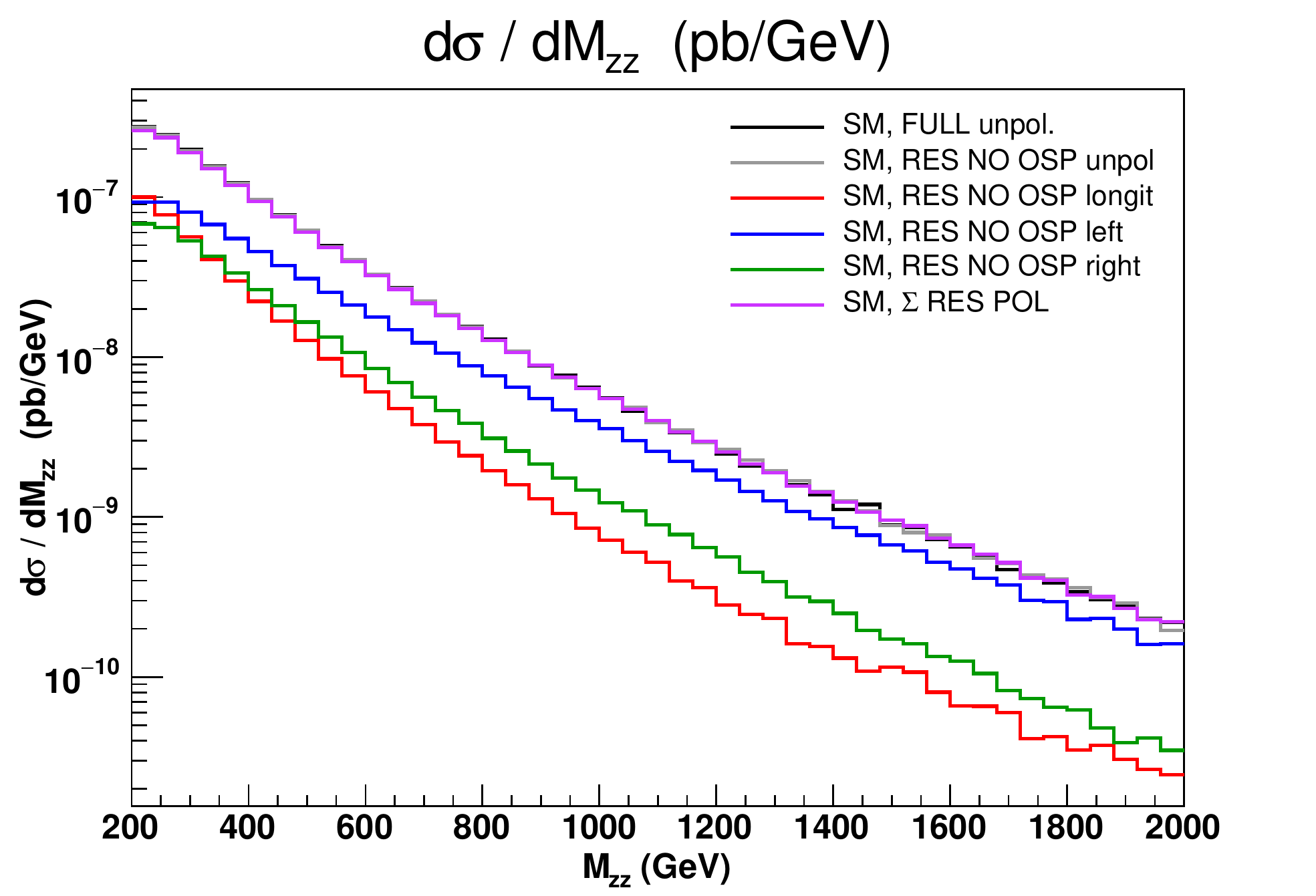}}\\
\subfigure[{$p_t^{Z_e}$}\label{fig:distrib_Ptze_cut}]
{\includegraphics[scale=0.37]{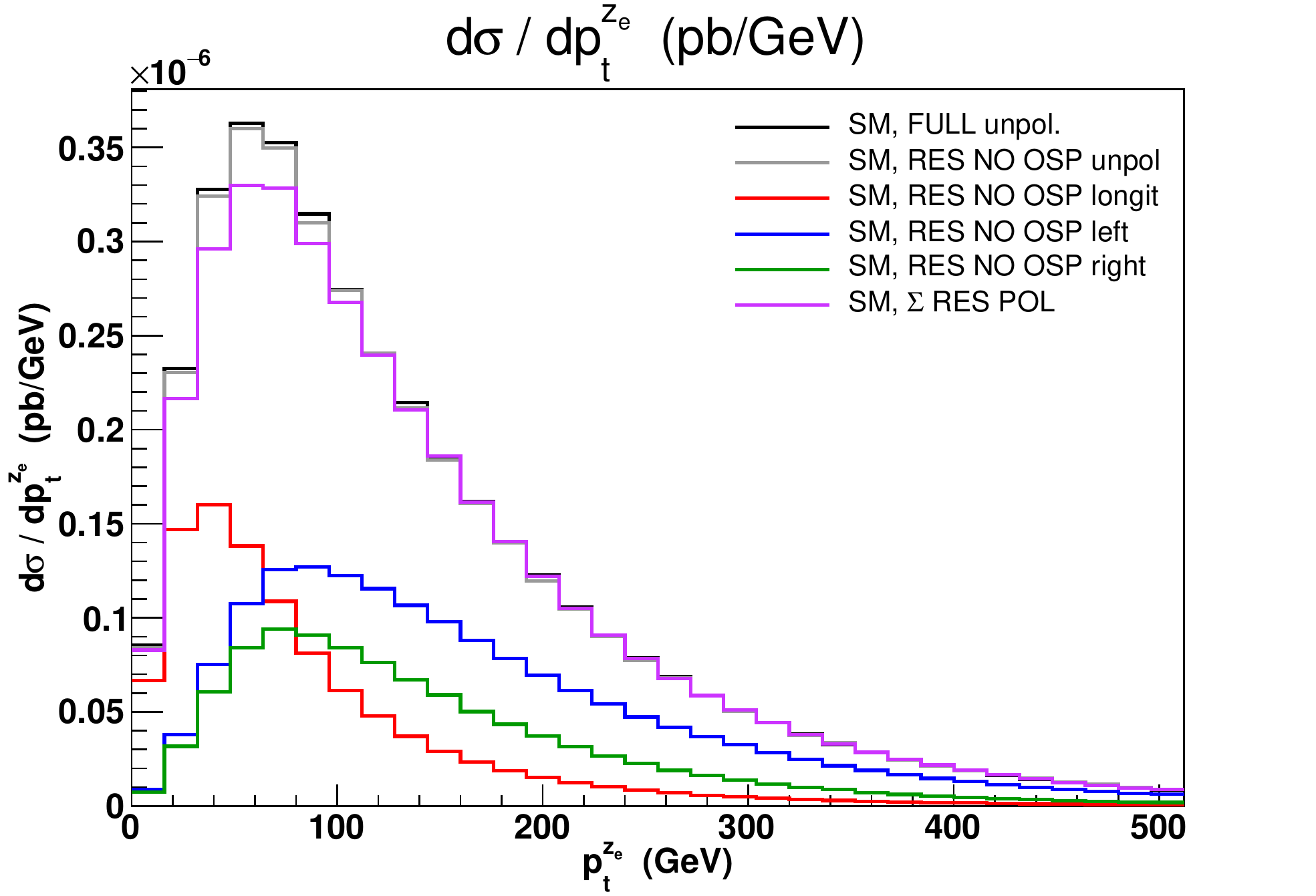}}
\subfigure[{$\eta_{Z_e}$}\label{fig:distrib_Etaze_cut}]
{\includegraphics[scale=0.37]{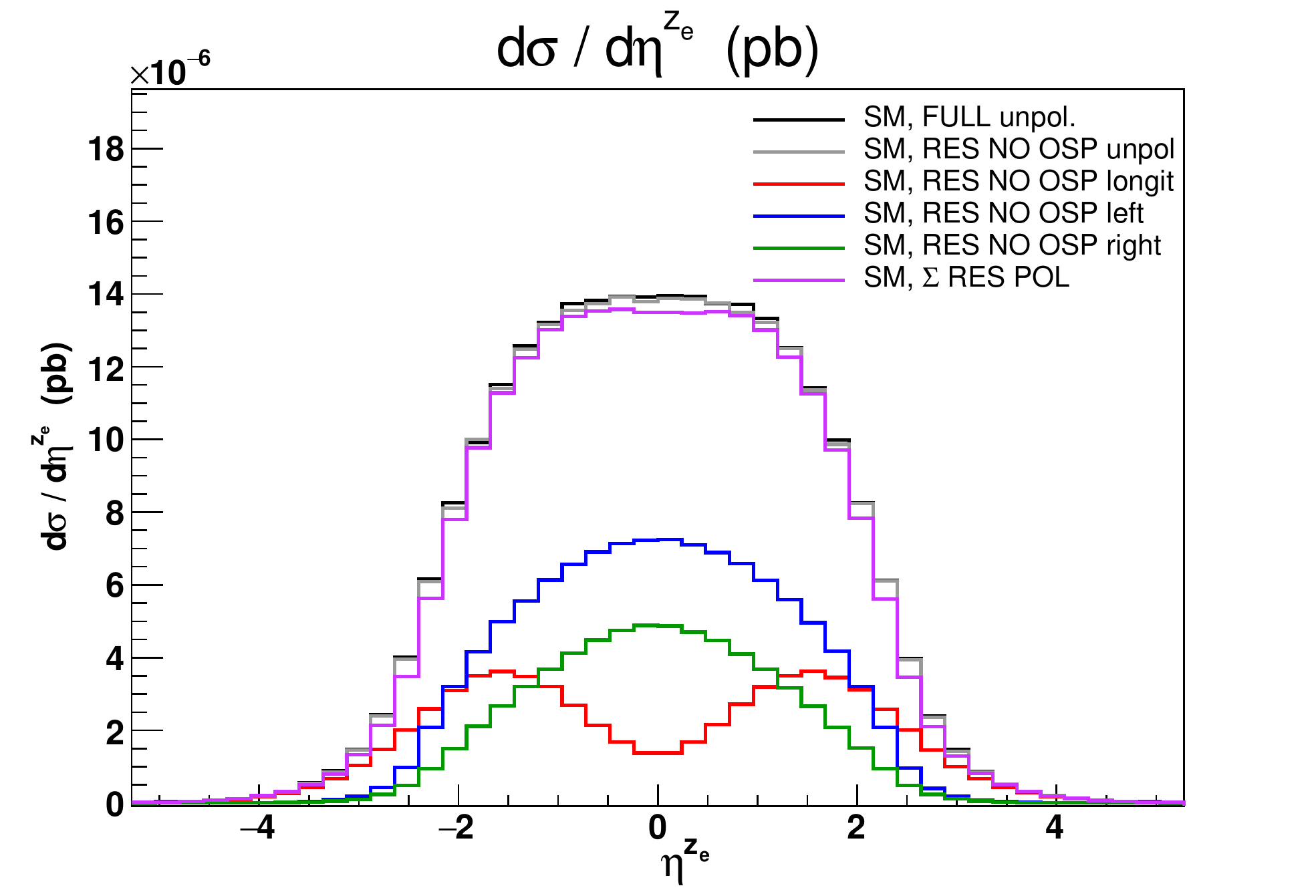}}
\caption{$Z\!Z$ scattering: polarized distributions in the presence of lepton cuts, obtained integrating the polarized amplitudes
squared. The $Z$ decaying to $e^+e^-$ is polarized, the one decaying to $\mu^+\mu^-$ is unpolarized. The sum of
polarized components (violet curve) is compared against the full unpolarized distribution (black curve). }
\label{fig:lepcutzz}
\end{figure}

These interferences are even more evident in differential cross sections.
In Fig.~\ref{fig:distrib_cte_cut} we show the distributions in $\cos\theta_{e^-}$, which are to be compared with those in
Fig.~\ref{fig:zzcostheta}. The effect of $p_t$ and $\eta$ cuts on charged leptons is to deplete the
forward and backward regions at $\theta=0,\,\pi$:
this mainly induces a strong modification of the transverse distribution shapes,
which would feature a maximum in those regions, if lepton cuts were not applied. The sum of the three polarized
distributions (violet curve) reproduces the full result fairly well. There is an essentially constant 4\% shift in each bin
which reflects the overall cross section difference.

In Fig.~\ref{fig:distrib_Mzz_cut} we present the differential cross sections of the four lepton system invariant mass: the
left component is the largest, with the exception of the very first bin, and this effect increases at large invariant masses,
where the longitudinal and right contributions
each accounts for about 10\% of the full cross section.
A similar behaviour can be observed in the distribution of the transverse momentum of the $Z$ boson which decays
into $e^+e^-$ (Fig.~\ref{fig:distrib_Ptze_cut}): the longitudinal component features a peak around $40\,\GeV$, then it
decreases much faster than the transverse ones.

The $Z$ pseudorapidity is shown in Fig.~\ref{fig:distrib_Etaze_cut}. The longitudinal component has the usual dip
in the central region, while its tails at $\vert \eta_{Z_e} \vert >$ 2 are larger than the transverse ones.\\
In general, the left and right distributions are characterized by a very similar shape for many kinematic distributions
(Figs.~\ref{fig:distrib_Mzz_cut}, \ref{fig:distrib_Ptze_cut} and \ref{fig:distrib_Etaze_cut}), with the obvious exception of the
$\cos\theta_{e^-}$ distribution.

The incoherent sum of polarizations at the amplitude level works reasonably well even in the presence of a complete set of
kinematic cuts,
provided a sufficiently narrow window for the $\ell^+\ell^-$ pair invariant mass around $M_Z$ is selected.
The longitudinal component accounts for  26.5\% of the total. The left  and right handed contributions account for 43.9\%
and 26.1\%, respectively. Interferences are not negligible, about 3.5\%, and should be taken into account in
experimental analyses.

For analogous predictions in the presence of BSM dynamics, and a discussion on the model independent
extraction of $Z$ polarization fractions from LHC data, we refer to Sect.~\ref{subsec:extracting_ZW_ZZ}.

\section{$W\!Z$ scattering}
\label{sec:WZ}
The measurements by the ATLAS and CMS Collaborations \cite{Aaboud:2018ddq,Sirunyan:2019ksz},
as well as the recent calculation of the NLO EW and QCD corrections \cite{Denner:2019tmn} and of parton shower effects \cite{Jager:2018cyo} highlight a growing interest in the $W\!Z$ scattering channel. In this section we investigate
the phenomenology of polarized $W^{+}\!Z$ scattering in the fully leptonic decay channel at the LHC.
We consider both the case in which the $W^+$ boson has definite polarization and the $Z$ is
unpolarized, and the case in which the $Z$ boson has definite polarization and the $W^+$ is unpolarized.

The $W\!Z$ channel is strongly sensitive to the EWSB mechanism, as it can be proved
computing the Feynman diagrams of the tree level amplitude for on shell scattering between longitudinal bosons.
Similarly to $W\!W\rightarrow Z\!Z$ and $W\!W\rightarrow W\!W$, pure gauge diagrams grow like $s^2$, while their sum
grows linearly with $s$ (more precisely with $t = s\, (1 -\cos\theta)$),
violating perturbative unitarity at high energies. The Higgs contribution regularizes the full amplitude, restoring unitarity.
The presence of a new resonance coupling to $W$ and $Z$ bosons or a modified Higgs sector
would interfere with this delicate cancellation of large contributions,
enhancing the longitudinal cross section at high energies.

The fully leptonic $W\!Z$ scattering is more appealing than the $W\!W$ channel because
the presence of only one neutrino in the final state allows to reconstruct, at least approximately, the center of mass frame of
the $W$ boson. In the $WW$ case the presence of two neutrinos makes the reconstruction impossible.
The cross section  for $W\!Z$ production in VBS is expected to be larger than the $Z\!Z$ one, enabling more
accurate analyses with the LHC Run II luminosity.

\subsection{Setup of the simulations}
\label{subsec:setupwz}
In the following, we consider the process $pp\rightarrow jje^+e^-\mu^+\nu_\mu $ at the LHC@13TeV.
We use the  same PDF set and factorization scale described in \ref{subsec:setupzz}.
We only consider tree level electroweak contributions and neglect partonic processes involving $b$ quarks,
which account for less than 1.5\% of the total cross section.

We have applied the following kinematic cuts:
\begin{itemize}
\item[-] maximum jet rapidity, $|\eta_j|<5$;
\item[-] minimum jet transverse momentum, $p_t^j>20$ GeV;
\item[-] minimum invariant mass of the two tagging jet system, $M_{jj}>500$ GeV;
\item[-] minimum rapidity separation between the two tagging jets, $|\Delta\eta_{jj}|> 2.5$;
\item[-] maximum difference between the invariant mass of the $e^+e^-$ pair and
the $Z$ pole mass, $|M_{e^+e^-}-M_Z|<15$ GeV;
\item[-] minimum invariant mass of the four lepton  system, $M_{W\!Z}>200$ GeV.
\end{itemize}
The results presented in Sect.~\ref{subsec:lepcutwz} include three additional cuts
\begin{itemize}
\item[-] maximum charged lepton rapidity, $|\eta_\ell|<2.5$;
\item[-] minimum charged lepton transverse momentum, $p_t^\ell>20$ GeV;
\item[-] minimum missing transverse momentum, $p_t^{\rm miss}>40\,\GeV$.
\end{itemize}
In Sect.~\ref{subsec:wznocut} the $M_{W\!Z}$ cut is imposed directly on the generated, not reconstructed, momenta.
In  Sect.~\ref{subsec:lepcutwz} the cut is applied after neutrino reconstruction.
For more details on neutrino reconstruction the reader is referred to Appendix~\ref{subsec:vreco}.

\subsection{Single polarized results and their validation in the absence of lepton cuts}
\label{subsec:wznocut}
In order to verify that polarizations can be separated at the amplitude level  while reproducing
properly the full result, we consider the ideal kinematic setup in which no cut on charged
leptons and neutrinos is applied, apart from $|M_{e^+e^-}-M_Z|<15$ GeV and $M_{W\!Z}>200\,\GeV$.

Following the discussion in Sect.~\ref{sec:separating}, in order to isolate the
polarizations of the $W$($Z$) boson in $W^+\!Z$ scattering, we select only
the $W$($Z$) resonant diagrams (single and double resonant) out of the
full set of contributions. Then we apply the OSP1-W(Z) projection on the
$W$($Z$) boson, to avoid any cut on the $\mu^+\nu_\mu $ system invariant
mass. We have shown that for $Z$ resonant diagrams, OSP1 has no visible
effect, but, nonetheless, we apply it for consistency. In all the following we will refer
to OSP1-W(Z) projected $W$($Z$) resonant calculation simply as resonant
calculation.

The full matrix element includes both resonant and non resonant diagrams,
therefore in principle it would not be possible to cast the full $\cos\theta_{\ell}$
distribution in the form of Eq.~\ref{eq:diffeqV}.
On the contrary, the unpolarized resonant amplitude features a vector boson
(either the $W$ or the $Z$) which is radiated and then decays leptonically,
making it possible to apply Eq.~\ref{eq:diffeqV}.
We have checked that the unpolarized resonant distributions describe accurately
the full distributions. This enables to treat equivalently the full or the
resonant $\cos\theta_{\ell}$ distributions by means of Eq.~\ref{eq:diffeqV}.

\begin{figure}[!htb]
\centering
\subfigure[\label{fig:legcw200}]{\includegraphics[scale=0.37]{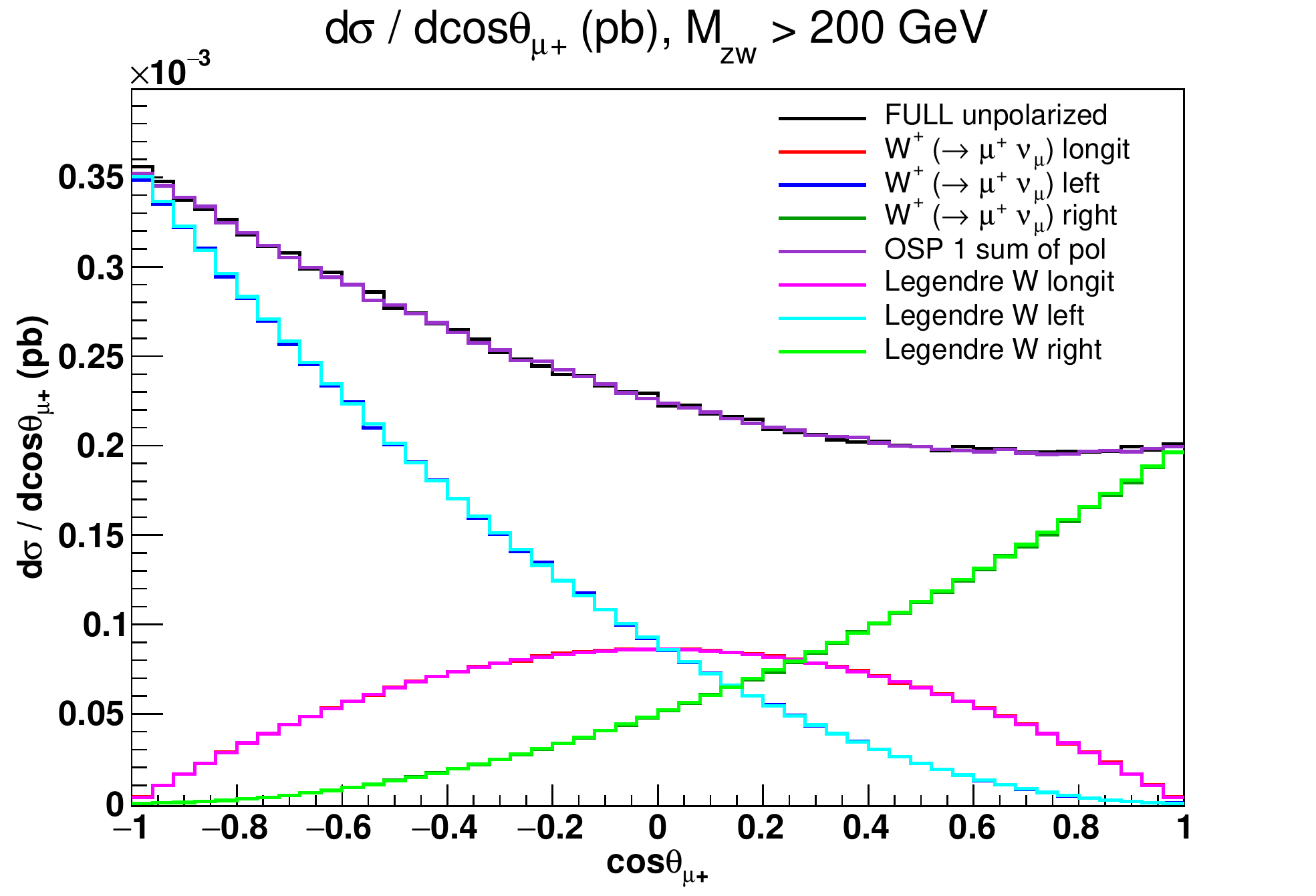}}
\subfigure[\label{fig:wpolfr}]{\includegraphics[scale=0.35]{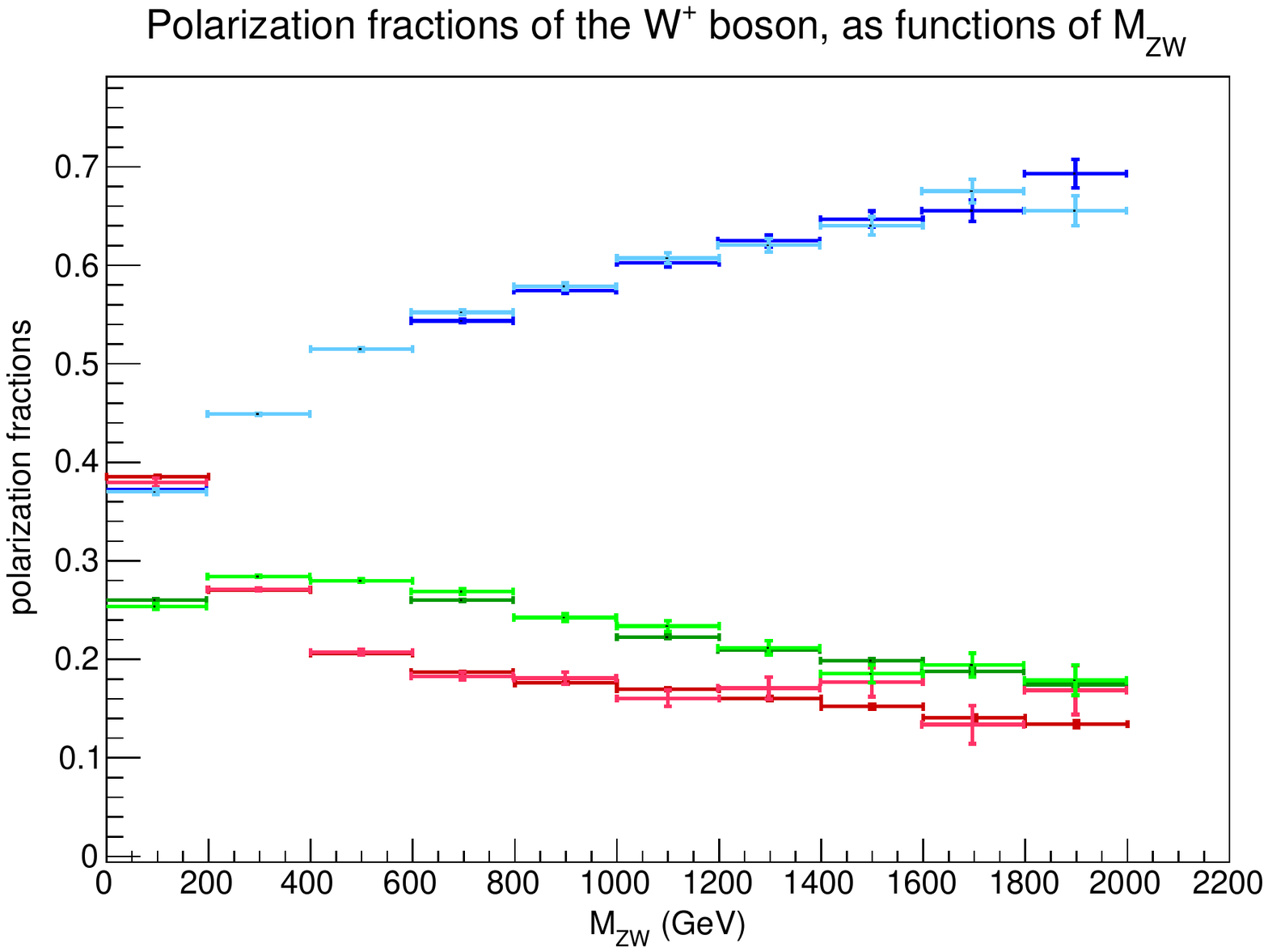}}
\caption{$W^+\!Z$ scattering: $\cos\theta_{\mu^+}$ distributions for a polarized $W^+$, in the region
$M_{W\!Z}>200\,\GeV$ (left), and
polarization fractions as functions of $M_{W\!Z}$ (right). Comparison between Monte Carlo distributions and
results extracted from the full $\cos\theta_{\mu^+}$ distribution by projecting into the first three
Legendre polynomials.
No lepton cuts, no neutrino reconstruction.}\label{fig:legw}
\end{figure}

The total cross section computed with full matrix elements is $486.4(2)\,\ab$.
The unpolarized OSP1-W resonant result is only 0.2\% smaller.
Similarly, the OSP1-Z resonant computation underestimates by
0.7\% the full result.
Differential distributions are also in
good agreement. Discrepancies are smaller than 2\% bin by bin.

We first separate the polarizations of the $W^+$ boson. In \Fig{fig:legcw200} we consider the
$\cos\theta_{\mu^+}$ distributions in the full $M_{W\!Z}>200\,\GeV$ range
in the absence of lepton cuts and without neutrino reconstruction.

\begin{figure}[!htb]
\centering
\subfigure[\label{fig:legcz200}]{\includegraphics[scale=0.37]{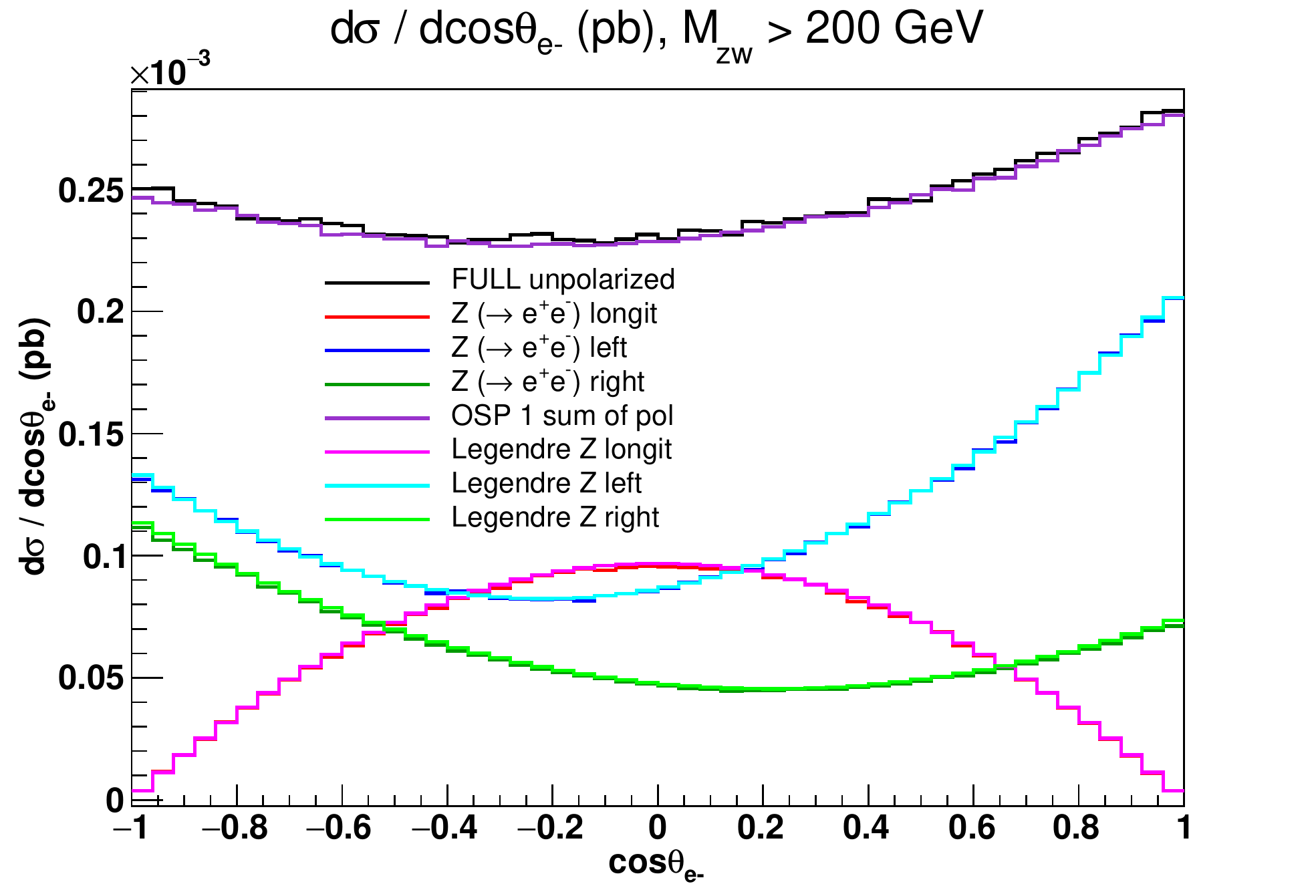}}
\subfigure[\label{fig:zpolfr}]{\includegraphics[scale=0.35]{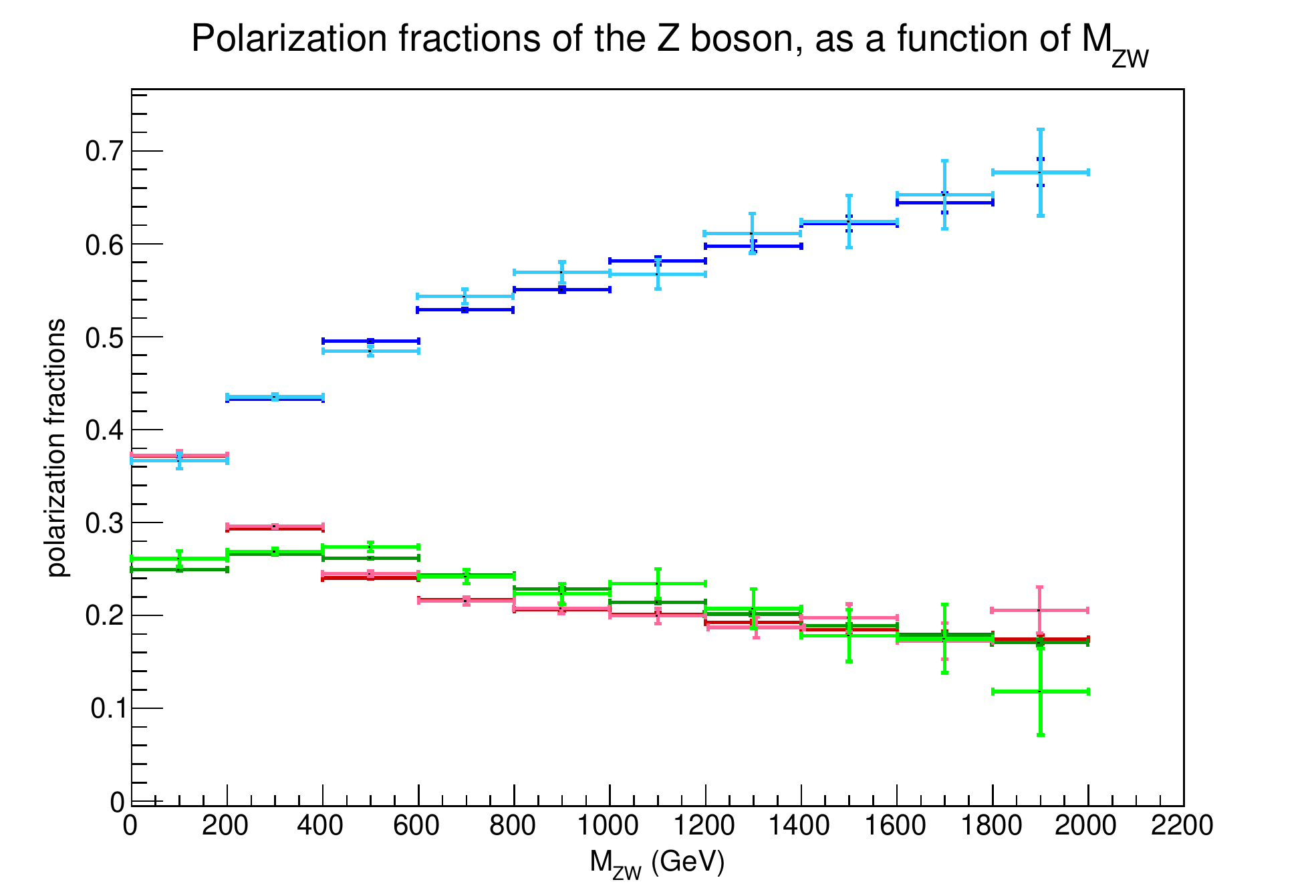}}
\caption{$W^+\!Z$ scattering: $\cos\theta_{e^-}$ distributions for a polarized $Z$, in the region $M_{W\!Z}>200\,\GeV$ (left), and
polarization fractions as functions of $M_{W\!Z}$ (right). Comparison between Monte Carlo distributions and
results extracted from the full $\cos\theta_{e^-}$ distribution by projecting into the first three Legendre polynomials.
No lepton cuts, no neutrino reconstruction. }\label{fig:legz}
\end{figure}

The distributions obtained with polarized amplitudes (red, blue and dark green, for longitudinal,
left and right polarization, respectively) are compared with the components extracted from
the full distribution (magenta, azure and light green) by projecting onto the first three Legendre polynomials.
The agreement is very good: both the normalization and the quadratic dependence on $\cos\theta_{\mu^+}$ is perfectly
reproduced for each polarization state.
We have performed the same study in several $W\!Z$ invariant mass regions, and we have compared the polarization
fractions extracted from the full result with the ratios of polarized Monte Carlo cross sections to the
full cross section. The results are shown in \Fig{fig:wpolfr}. Error bars show the statistical
uncertainties on the polarization fractions. The agreement is good in all invariant mass regions.

Very similar conclusions can be drawn when separating the polarization of the $Z$ boson.
We show in \Fig{fig:legz} the $\cos\theta_{e^-}$ distributions for a polarized $Z$ boson in the
$M_{W\!Z}>200\,\GeV$ region, as well as the polarization fractions as functions of the $W^+\!Z$
invariant mass.
Also for the $Z$ decay, the polarized distributions extracted from the full
result and those produced directly with the Monte Carlo are in very good agreement in all
kinematic regions, both for the total rates, \Fig{fig:zpolfr}, and for the distribution shapes,
\Fig{fig:legcz200}.

\subsection{Effects of lepton cuts and neutrino reconstruction on polarized distributions}
\label{subsec:lepcutwz}
In this section we present polarized differential distributions in the presence of lepton
cuts and neutrino reconstruction for a number of relevant kinematic variables.
{The specific neutrino reconstruction scheme that is applied in the following
(\texttt{CoM} + \texttt{transvMlv}) is described in Appendix~\ref{subsec:vreco}.}
We provide results
for a polarized $W^+$ produced in VBS together with an unpolarized $Z$ boson, as well as
for a polarized $Z$ produced in association with an unpolarized $W^+$.

We start from the total cross section. In order to evaluate separately the effect of dropping
the non resonant diagrams and the effect of neglected interferences among different polarization modes,
we have computed the cross section with the full matrix element and with OSP1-W(Z) projected resonant
diagrams (see Sect.~\ref{sec:separating}).
The difference between these two results provides an estimate of non resonant effects. The difference between the resonant unpolarized cross section and the
sum of the single polarized ones (either for a polarized $W^+$ or for a polarized $Z$) provides an
estimate of the interference among polarizations, which is non zero because of the leptonic cuts. Numerical
results are shown in Tab.~\ref{tab:WZSM}.

\begin{table}[hbt]
\begin{center}
\begin{tabular}{|c|c|c|}
\hline
\multicolumn{3}{|c|}{\cellcolor{blue!9} Total cross sections [$\ab $]}\\
\hline
& {polarized $W^+$} & { polarized $Z\,\,$}\\
\hline
\hline
longitudinal (res. OSP1) &   33.21(3) & 42.56(3)\\
\hline
left handed  (res. OSP1) &  96.31(8)  & 76.87(6)\\
\hline
right handed (res. OSP1) &  30.93(2)  & 40.54(3)\\
\hline
sum of polarized &  160.45(9) & 159.97(8)\\
\hline
unpolarized (res. OSP1)  &  164.2(2 ) & 164.0(2) \\
\hline
non res. effects     &  0.9(2)   & 1.1(2)\\
\hline
pol. interferences     &  3.8(2)   & 4.0(2)\\
\hline
full         &  165.1(1)  &165.1(1)\\
\hline
\end{tabular}
\end{center}
\caption{Polarized and unpolarized total cross sections ($\ab $) for $W^+\!Z$ scattering in the fiducial region
(see Sect.~\ref{subsec:setupwz}).} \label{tab:WZSM}
\end{table}

The resonant unpolarized calculation has been performed selecting single $W$($Z$) resonant diagrams, and
then applying the corresponding
single On Shell projection. In both cases non resonant effects are smaller than 1\% , implying that the
resonant approximation works rather well.
Interference among polarization states amounts to
2.5\% . We are going to show in Sect.~\ref{sub:trans}  that the largest interference is between the left
polarization and the right one.
The combination of the two effects give a 3\% contribution to the full result, which is small but non negligible,
and should be taken into account for a proper determination of polarized signals.

Concerning polarized total cross sections, the $W^+$ is mainly left handed (58.3\%),
while the longitudinal and right handed contributions are of the same order of magnitude (20.1\% and
18.7\%, respectively). For the $Z$ boson, the left polarization is again the largest (46.6\%) while the
longitudinal and right components account respectively for 25.8\% and 24.6\%.

\begin{figure}[!h]
\centering
\subfigure[$M_{W\!Z}$\label{fig:lcmwz}]
{\includegraphics[scale=0.35]{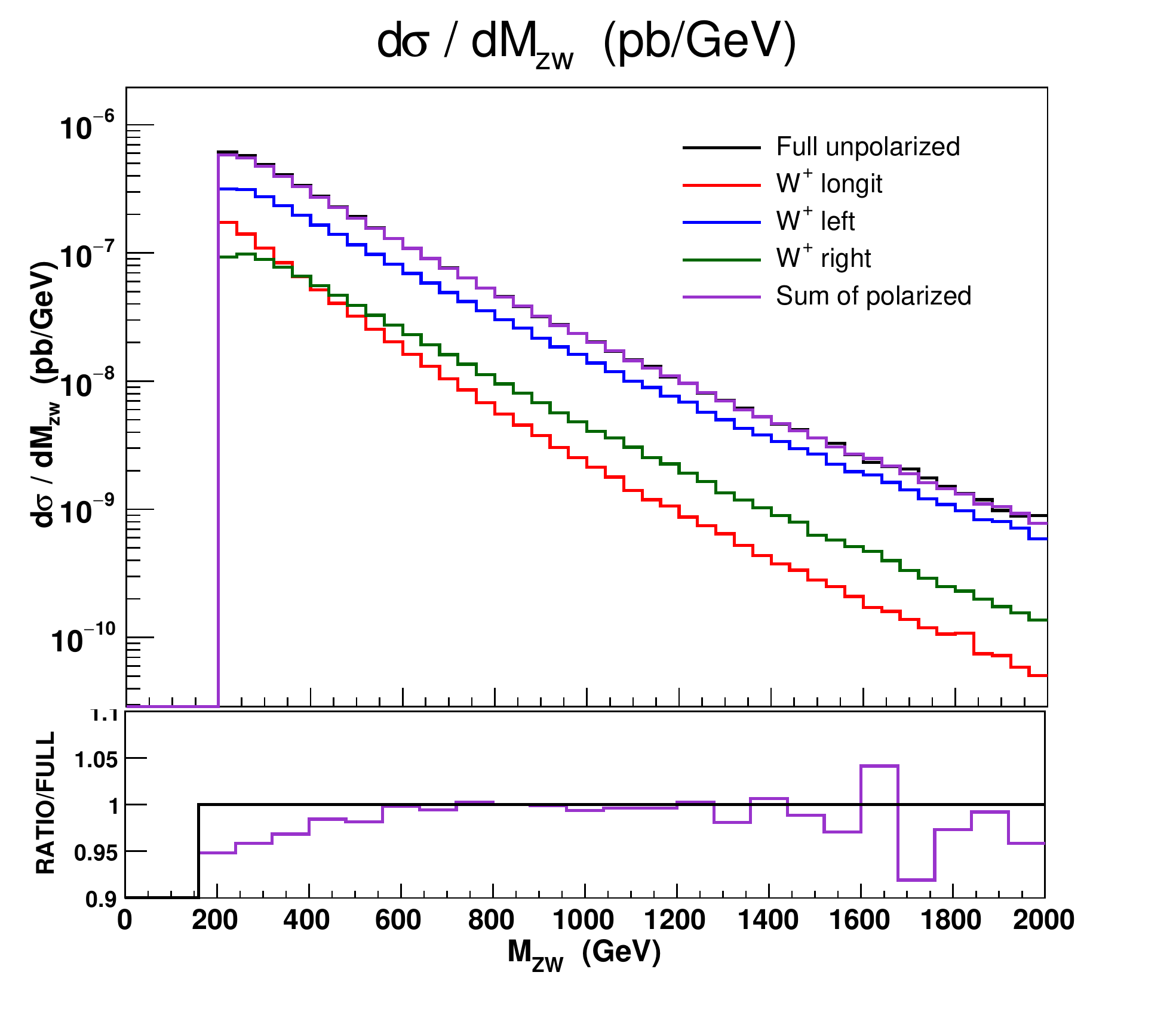}}
\subfigure[$\cos\theta_{\mu^+}$\label{fig:lccth}]
{\includegraphics[scale=0.35]{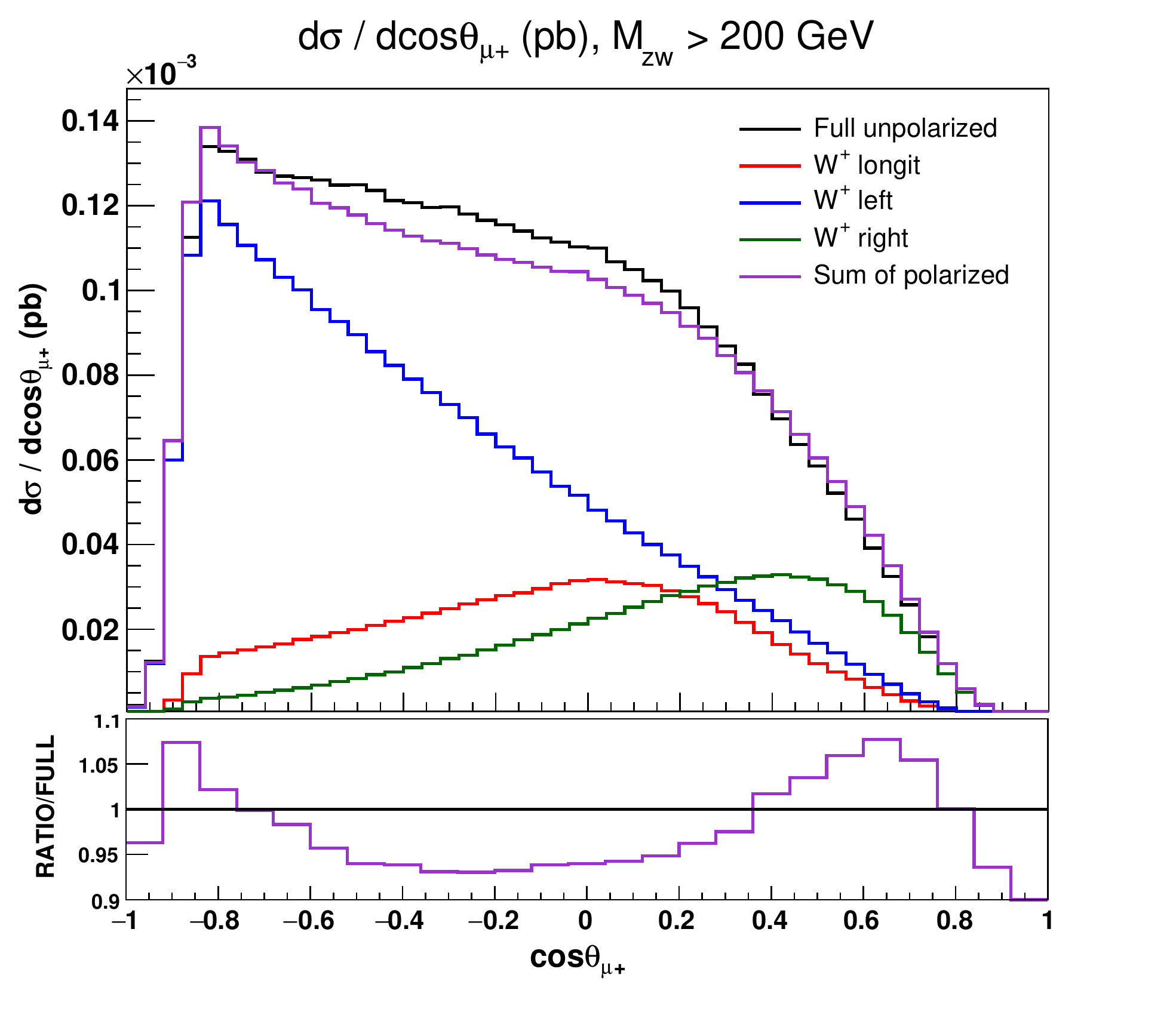}}\\
\subfigure[$\eta_{W}$\label{fig:lcetaw}]
{\includegraphics[scale=0.35]{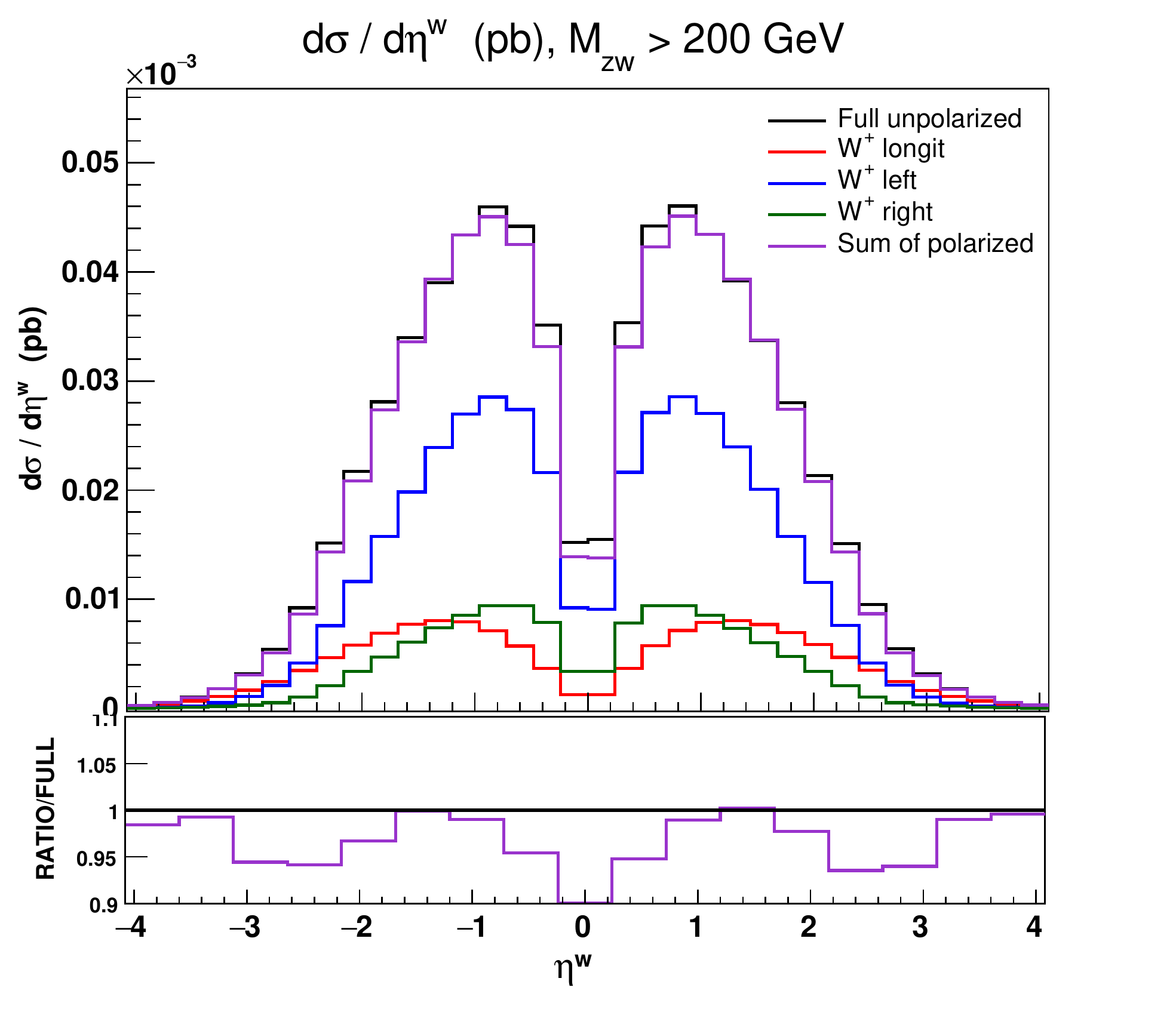}}
\subfigure[$p_t^{\mu^+}$\label{fig:lcptmu}]
{\includegraphics[scale=0.35]{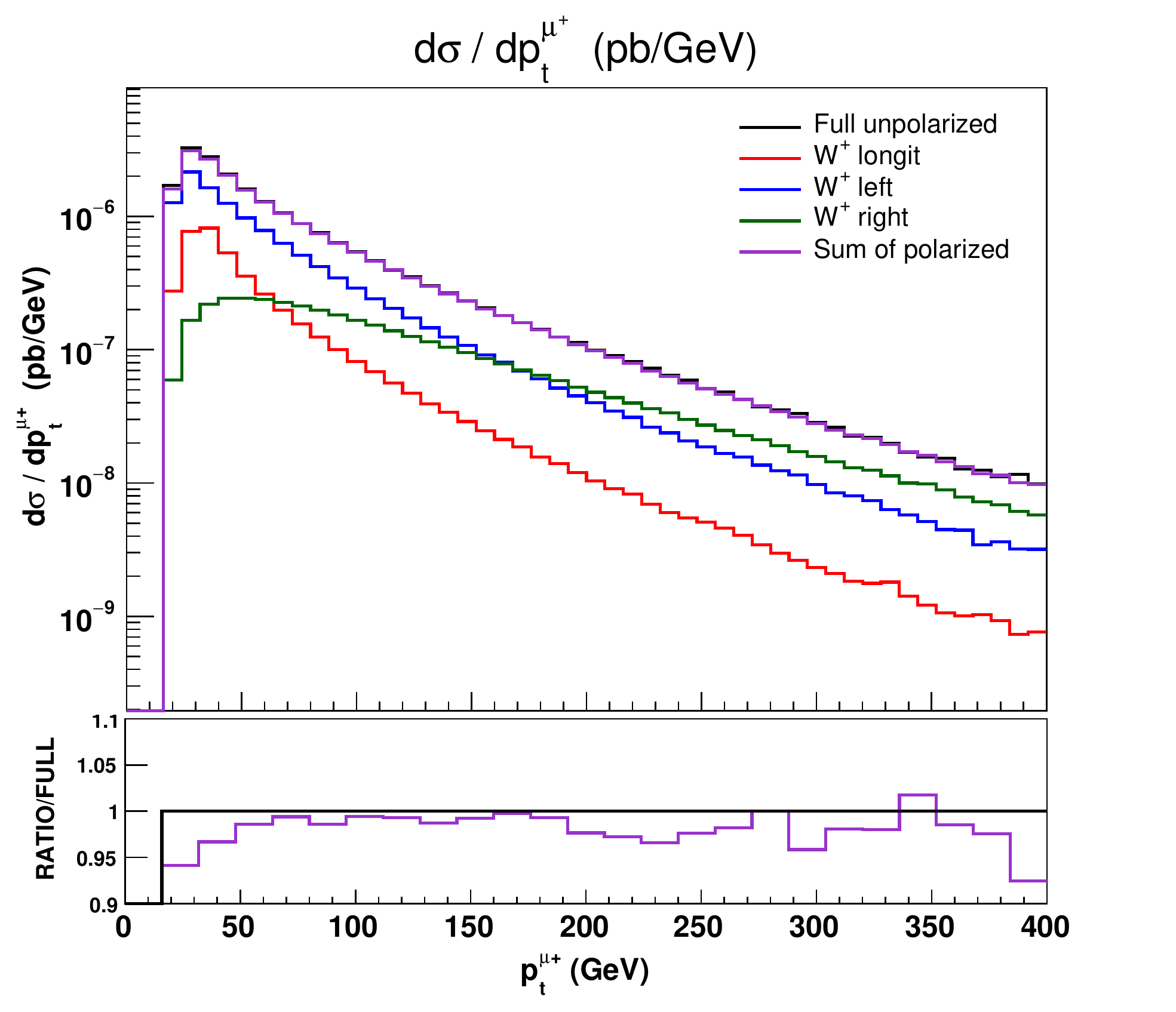}}
\caption{$W^+\!Z$ scattering: differential cross sections for a polarized $W^+$ boson, in the presence of lepton cuts
and neutrino reconstruction.
We show the full result (black), the single polarized distributions (red: longitudinal, blue: left handed, green:
right handed) and the incoherent sum of the polarized results (violet).
The pull plot shows
the ratio of the sum of polarized distributions to the full one.}\label{fig:wzlepcut_w}
\end{figure}

\begin{figure}[!htb]
\centering
\subfigure[$M_{W\!Z}$\label{fig:lcz_m}]
{\includegraphics[scale=0.35]{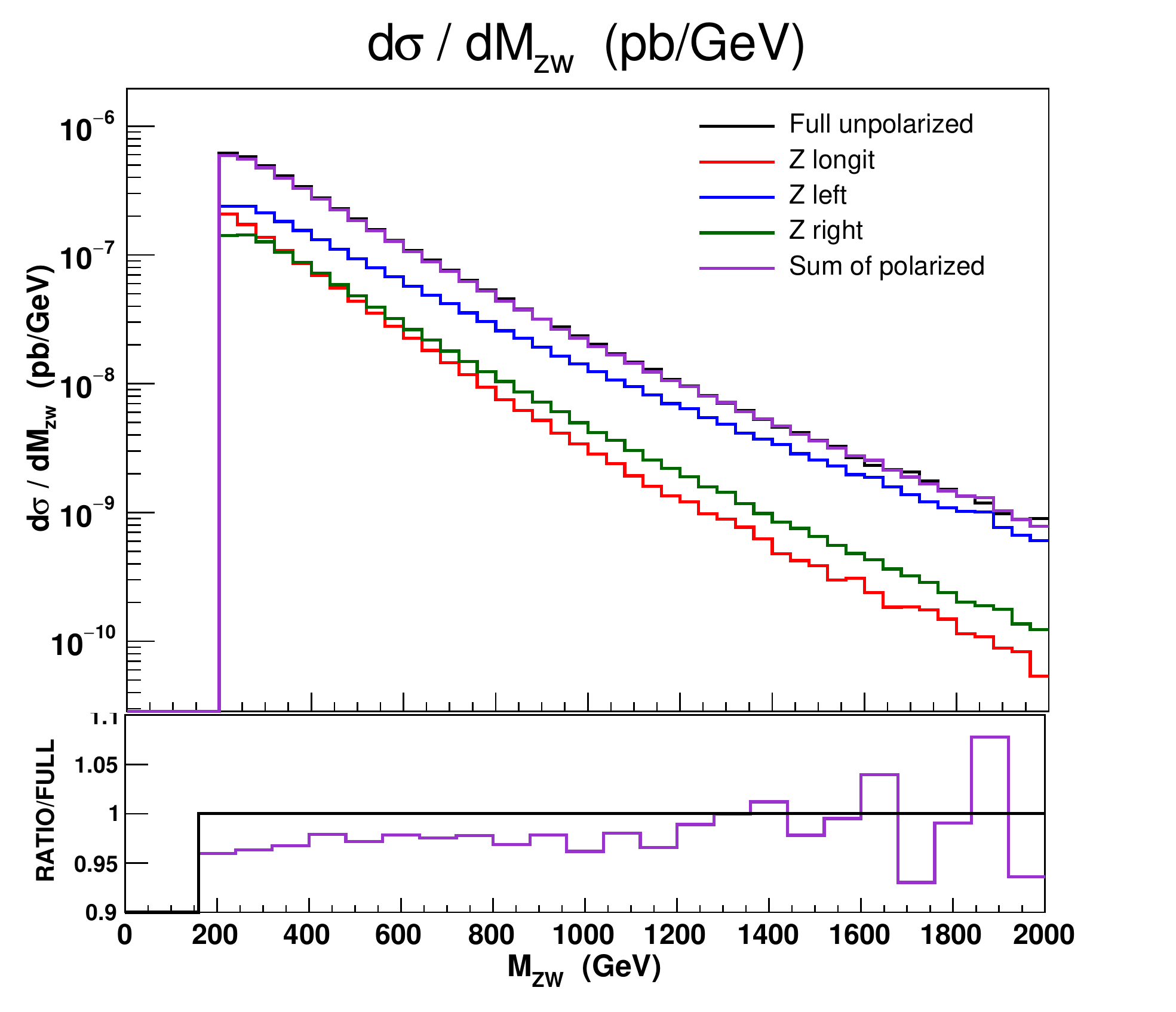}}
\subfigure[$\cos\theta_{e^-}$\label{fig:lcz_cth}]
{\includegraphics[scale=0.35]{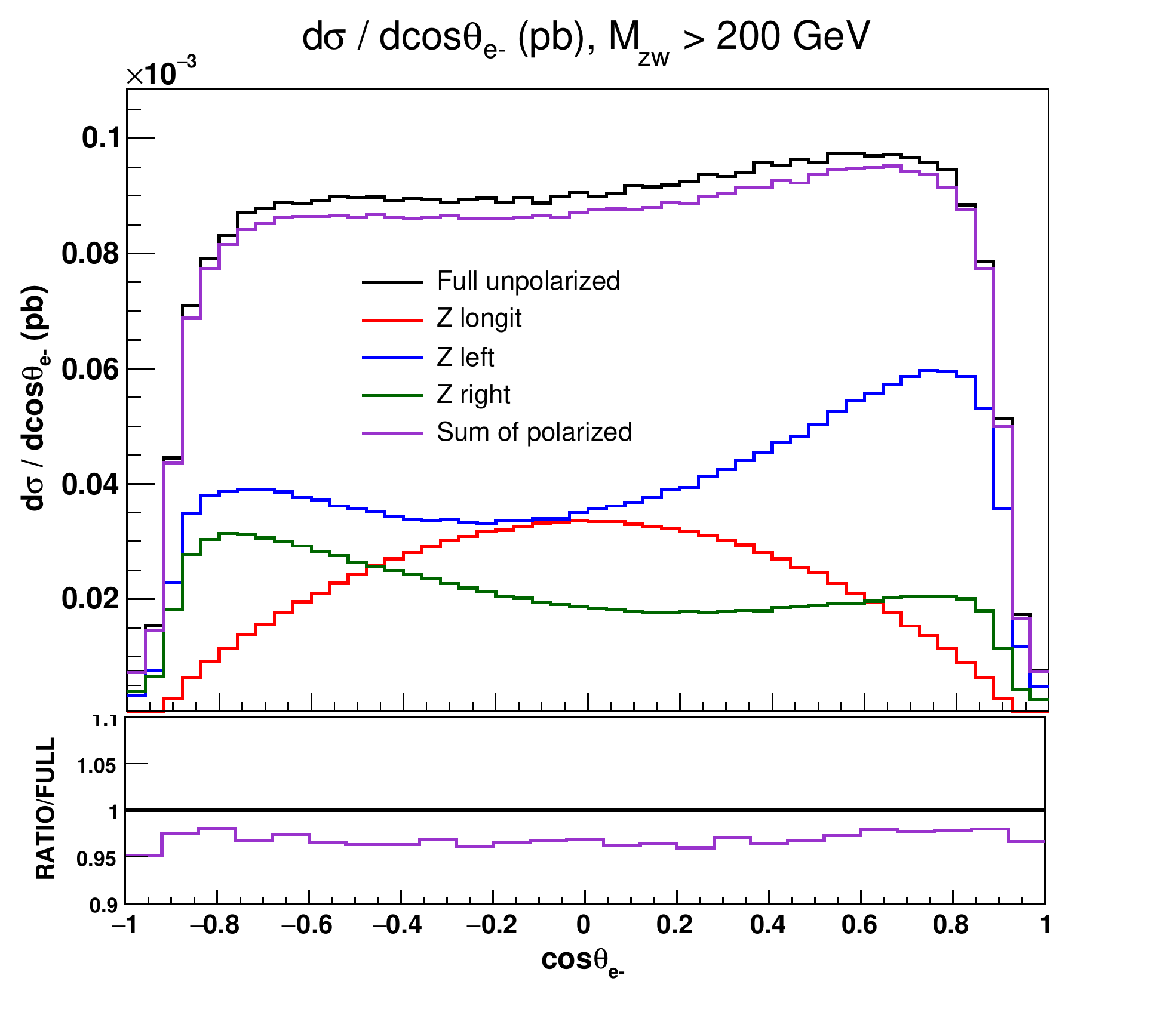}}\\
\subfigure[$\eta_Z$\label{fig:lcz_eta}]
{\includegraphics[scale=0.35]{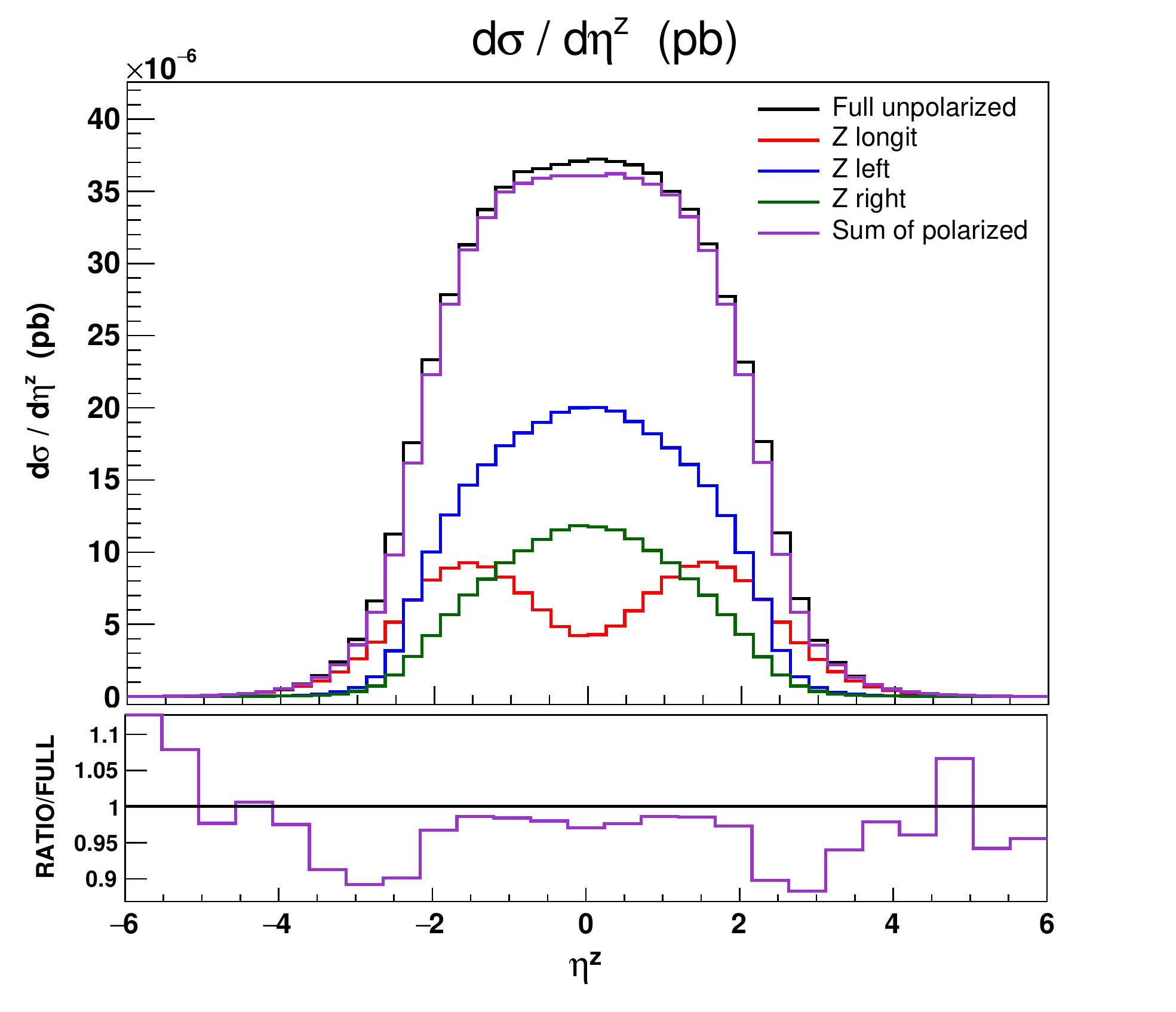}}
\subfigure[$p_t^{e^-}$\label{fig:lcz_pt}]
{\includegraphics[scale=0.35]{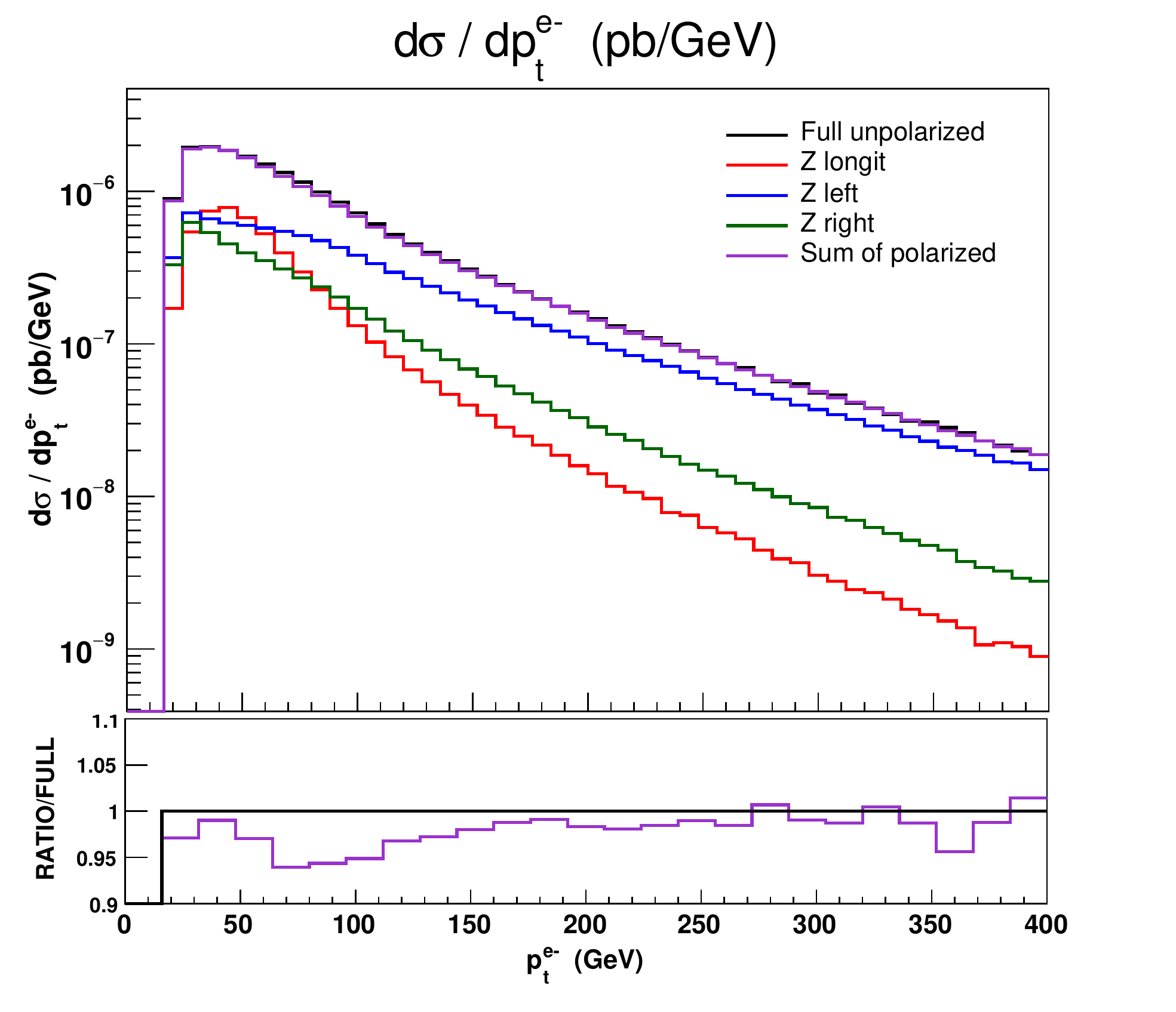}}
\caption{$W^+\!Z$ scattering: differential cross sections for a polarized $Z$ boson, in the presence of lepton cuts
and neutrino reconstruction.
We show the full result (black), the single polarized distributions (red: longitudinal, blue: left handed, green:
right handed) and the incoherent sum of the polarized results (violet).
The pull plot shows the ratio of the sum of polarized distributions to the full one.}\label{fig:wzlepcut_z}
\end{figure}

In \fig{fig:wzlepcut_w} and \fig{fig:wzlepcut_z} we present differential distributions for a variety of
kinematic variables, which provide a more detailed description of the polarized signals.
For each variable, we show single polarized distributions,
their incoherent sum, and the distribution of the full result.
In both figures we use the same color code: the full result is in black;
the longitudinal, left and right single polarized distributions are in red blue and green respectively; the
incoherent sum of the polarized results is in violet.
Pull plots show the bin by bin ratio of the  incoherent sum of polarized distributions
to the full one.

In \Fig{fig:wzlepcut_w} we show distributions for a polarized $W^+$ boson.
\Fig{fig:lcmwz} presents the distribution of the invariant mass of the four leptons.
The interference and non resonant effects account for less than 5\% of the full result (bin by bin) in the
whole $W^+\!Z$
invariant mass spectrum, up to fluctuations due to low statistics in the large mass region
(\Fig{fig:lcmwz}). The longitudinal fraction decreases rapidly with increasing energy. The left handed
component is the largest one over the whole range. For $M_{W\!Z} >1600\,\GeV$ it is about ten times
larger than the longitudinal one.

The angular distributions in $\cos\theta_{\mu^+}$ are strongly affected by the neutrino reconstruction
and the lepton cuts, as can be seen comparing \Fig{fig:lccth} with \Fig{fig:legcw200}.
The difference is mainly due to the $p_t$ cuts on the muon and the corresponding neutrino,
which deplete the peaks
at $\theta_{\mu^+}=0,\,\pi$ of the transverse modes and make the longitudinal shape asymmetric.
The sum of polarized
distributions underestimates the full result by at most 5\% bin by bin, apart from the
regions
$\cos\theta_{\mu^+}\approx -0.8$  and $\cos\theta_{\mu^+}\approx +0.7$, where the interferences become
large and negative, inducing a discrepancy of about $ 8-10\%$.

In \Fig{fig:lcetaw} we show distributions of the reconstructed $W^+$ pseudorapidity. Neutrino
reconstruction leads to a marked depletion of the central region.
Without neutrino reconstruction, the unpolarized and transverse distributions
would have a maximum in $\eta_{W}=0$. The longitudinal component is less affected since,
even in the absence of neutrino reconstruction, it shows
a dip in the central region. Interferences and non resonant effects account for less than
6\% of the full
result over all the pseudorapidity range, apart from the central bin where they reach 10\%.

The transverse momentum of the muon (\Fig{fig:lcptmu}) is minimally affected by neutrino reconstruction.
The longitudinal component is of
the same order of magnitude than the left handed one for $p_t$ values slightly above the cut threshold, while
for large values it decreases faster than the two transverse distributions.
For $p_t^{\mu^+}>160\,\GeV$ the right handed component
becomes larger than the left handed one.
Interferences are small (less than 6\% bin by bin) over the full range.

In \Fig{fig:wzlepcut_z}, we present distributions for a polarized $Z$ boson.
The $M_{W\!Z}$ distributions shown in \Fig{fig:lcz_m} are very similar to
those of Fig.~\ref{fig:lcmwz}, apart from a less pronounced difference between the
longitudinal and the right handed contribution at large mass.

The variables related to the $Z$ boson kinematics are not directly affected by neutrino reconstruction,
apart from a small shift in the total cross section due to the minimum $M_{W\!Z}$ cut.

In \Fig{fig:lcz_cth} we show how $\cos\theta_{e^-}$ distributions are affected by lepton cuts.
This variable depends on the kinematics of two same flavour opposite sign charged leptons, whose
$p_t$ is cut at 20 GeV: these symmetric cuts result in a less pronounced effect in the $\theta_{e^-}=0,\pi$
regions depletion, if compared with the $\cos\theta_{\mu^+}$ distributions of \Fig{fig:lccth},
where the effects are more prominent due to an asymmetry in the $p_t$ cuts on the two objects that
reconstruct the $W^+$ ($p_t^{\mu^+}>20\,\GeV, \,p_t^{\rm miss}=p_t^{\nu_{\mu}}>40\,\GeV$).
The incoherent sum of polarized $\cos\theta_{e^-}$ distributions reproduces quite well the full one:
interferences and non resonant effects are roughly constant over the kinematic range, accounting for
4\% of the full.

The $\eta_Z$ distributions in \Fig{fig:lcz_eta} show larger interferences among polarization
modes
in the forward regions $2<|\eta_Z|<3$ where they account for 10\% of the total. The longitudinal
component features the typical depletion at $\eta_Z=0$, where the transverse modes show a peak. The
longitudinal fraction
becomes larger than the transverse ones in the forward regions, for $\eta_{Z}>2.5$.

In \Fig{fig:lcz_pt} we show the transverse momentum distributions of the electron.
Interferences and non resonant effects are small ($\lesssim 5\%$ bin by bin). In contrast with the $W^+$ case,
the $Z$
is mainly left handed at large $p_t$. In the soft $p_t$ region the three polarizations are of the same order of
magnitude. We have checked that if we allow for
small $W\!Z$ invariant masses ($M_{W\!Z}<200\,\GeV$), the peak of the longitudinal component becomes
more pronounced.
In general, the longitudinal contribution decreases more rapidly than the transverse ones
in the high energy, high $p_t$ region.

In conclusion, the separation of polarized signals at the amplitude level gives reliable predictions
both for a polarized $W^+$ and for a polarized $Z$, even in the presence of a realistic set of kinematic
cuts and when neutrino reconstruction is applied. Interference and reconstruction effects are non negligible,
but small (few percent) and well under control in our framework.

\section{Polarized amplitudes and reweighting approach}
\label{subsec:reweight}
Reweighting is an approximate procedure which has been
widely used by experimental collaborations to obtain polarized
samples, starting from unpolarized Monte Carlo events.
It has been employed for the extraction of polarization fractions of
$W$ bosons \cite{Bittrich:2015aia,ATLAS:2012au,Chatrchyan:2011ig},
$Z$ bosons \cite{Khachatryan:2015paa} and top quarks \cite{Aad:2013ksa}.
In this section we evaluate how well the reweighting method can
separate polarized samples and describe polarized distributions in the case
of $W^+\!Z$ scattering, by comparing its results with those presented in
Sect.~\ref{sec:WZ}, which have been obtained using polarized
amplitudes computed by the Monte Carlo.

Let's consider a generic process which involves a $W^+$ boson decaying into leptons
(similar considerations apply to the $Z$). The reweighting
procedure is based on the partition of the $W^+$ boson phase space in two dimensional
$\{p_t,\eta\}$ regions, as narrow as possible. In the absence of lepton cuts and
neutrino reconstruction, polarization fractions $f_0^{(i)},\,f_L^{(i)}$ and $f_R^{(i)}$
are computed in each $\{p_t^W,\eta_W\}$ region labelled by index $i$, expanding
the full, unpolarized $\cos\theta_{\mu^+}$ distribution in Legendre
polynomials (or, equivalently, fitting it with the distribution of Eq.~\ref{eq:diffeqV}).
The unpolarized sample is then divided as follows.
If an event belongs to region $i$ and has $\cos\theta_{\mu^+}=x$, three weights are
computed,
\beq
w_{0,L,R}=\frac{
\frac{1}{\sigma}\frac{d\sigma}{dx}\Big|_{0,L,R}
}{ \frac{3}{4}(1-x^2)f_{0}^{(i)}+ \frac{3}{8}(1-x)^2f_{L}^{(i)} + + \frac{3}{8}(1+x)^2f_{R}^{(i)}}
\label{eq:weightfactor}
\eeq
where,
\beq
\frac{1}{\sigma}\frac{d\sigma}{dx}
\Big|_{0}=\frac{3}{4}(1-x^2)f_{0}^{(i)},
\qquad \frac{1}{\sigma}\frac{d\sigma}{dx}
\Big|_{L/R}=\frac{3}{8}(1\mp x)^2f_{L/R}^{(i)}\,. \nnb
\eeq
Finally, the event is assigned to the longitudinal, left or right polarized sample with probability
$w_0,\,w_L,\,w_R$, respectively. The three samples are then analyzed separately, applying
lepton cuts and performing neutrino reconstruction.

\begin{figure}[!htb]
\centering
\subfigure[Differential cross sections \label{fig:rew1}]{\includegraphics[scale=0.37]{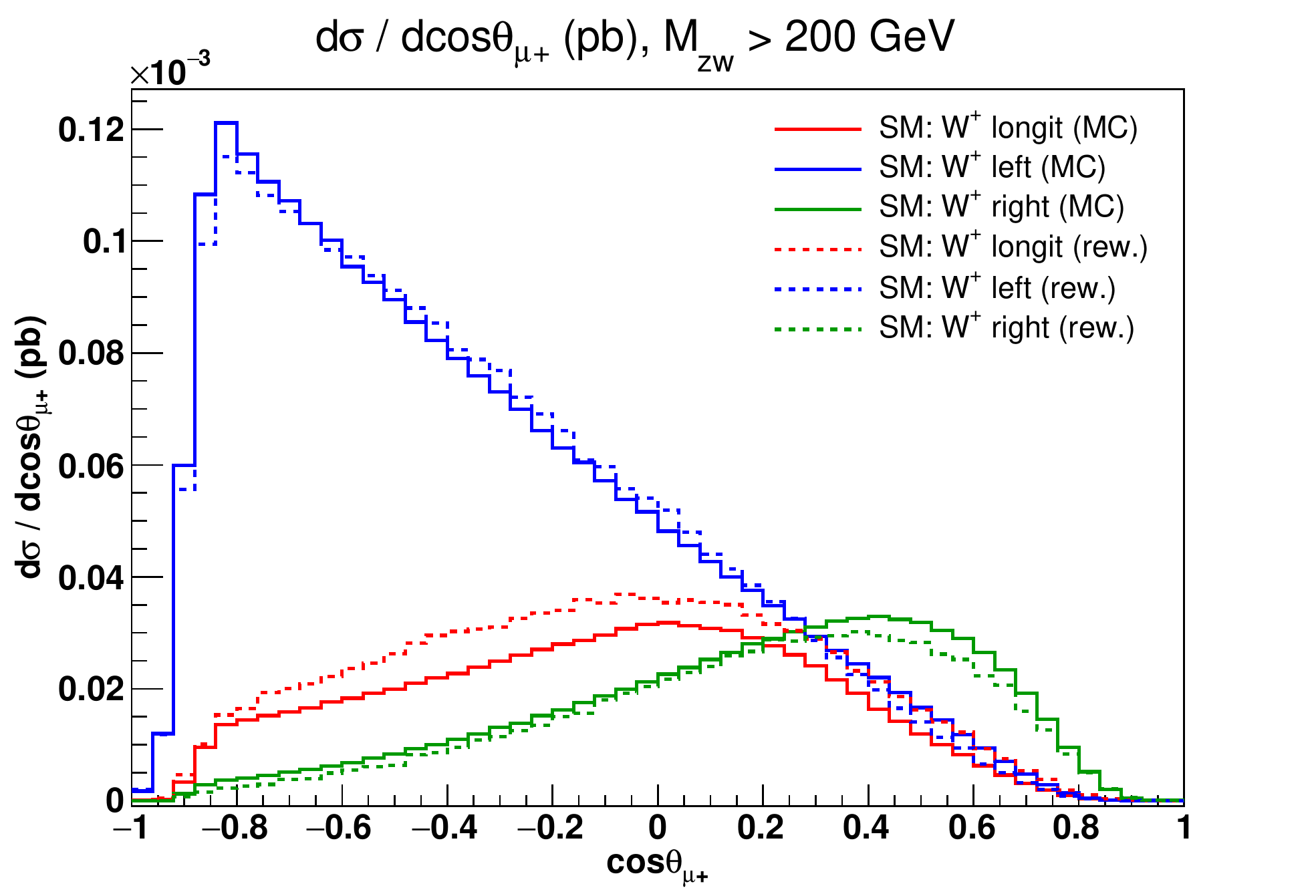}}
\subfigure[Normalized shapes \label{rew2}
]{\includegraphics[scale=0.37]{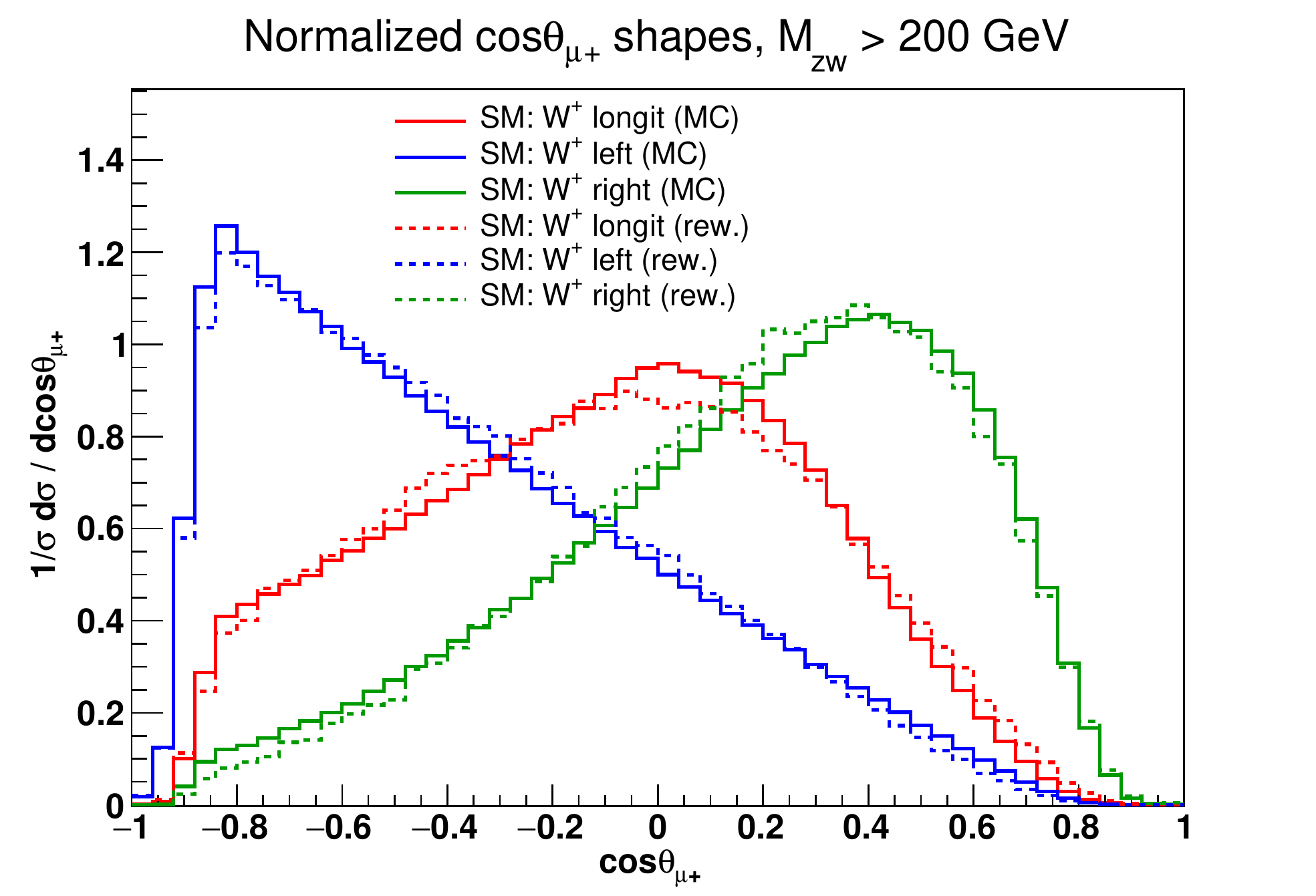}}
\caption{$W^+\!Z$ scattering: polarized $\cos\theta_{\mu^+}$ distributions in the region $M_{W\!Z}>200\,\GeV$. Results of the reweighting procedure compared with results of the MC calculation with polarized amplitudes. The full set of cuts and neutrino reconstruction are understood.} \label{fig:MCvsREW1}
\end{figure}

\begin{figure}[!htb]
\centering
\subfigure[Distributions \label{fig:rew3}]{\includegraphics[scale=0.37]{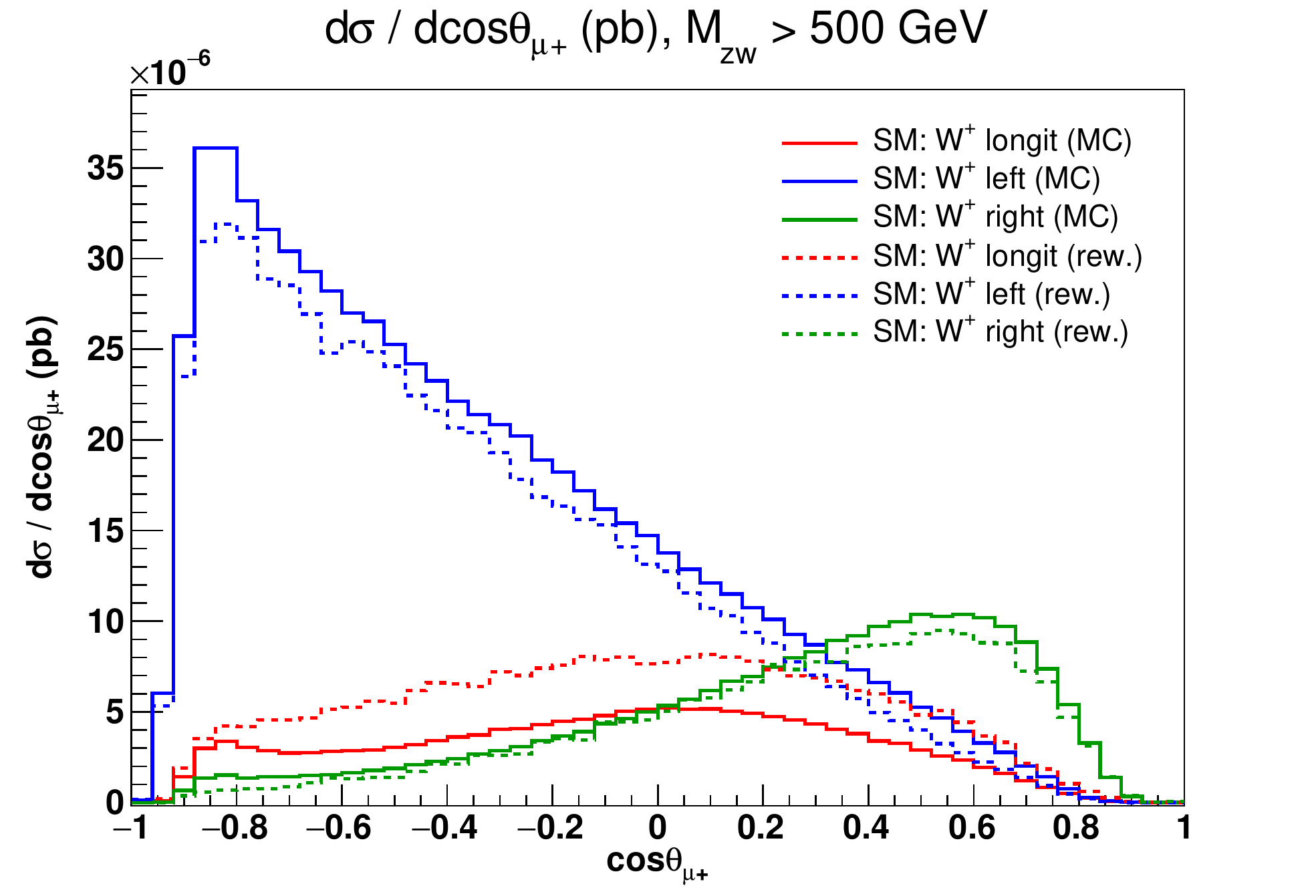}}
\subfigure[Normalized shapes \label{rew4}]
{\includegraphics[scale=0.37]{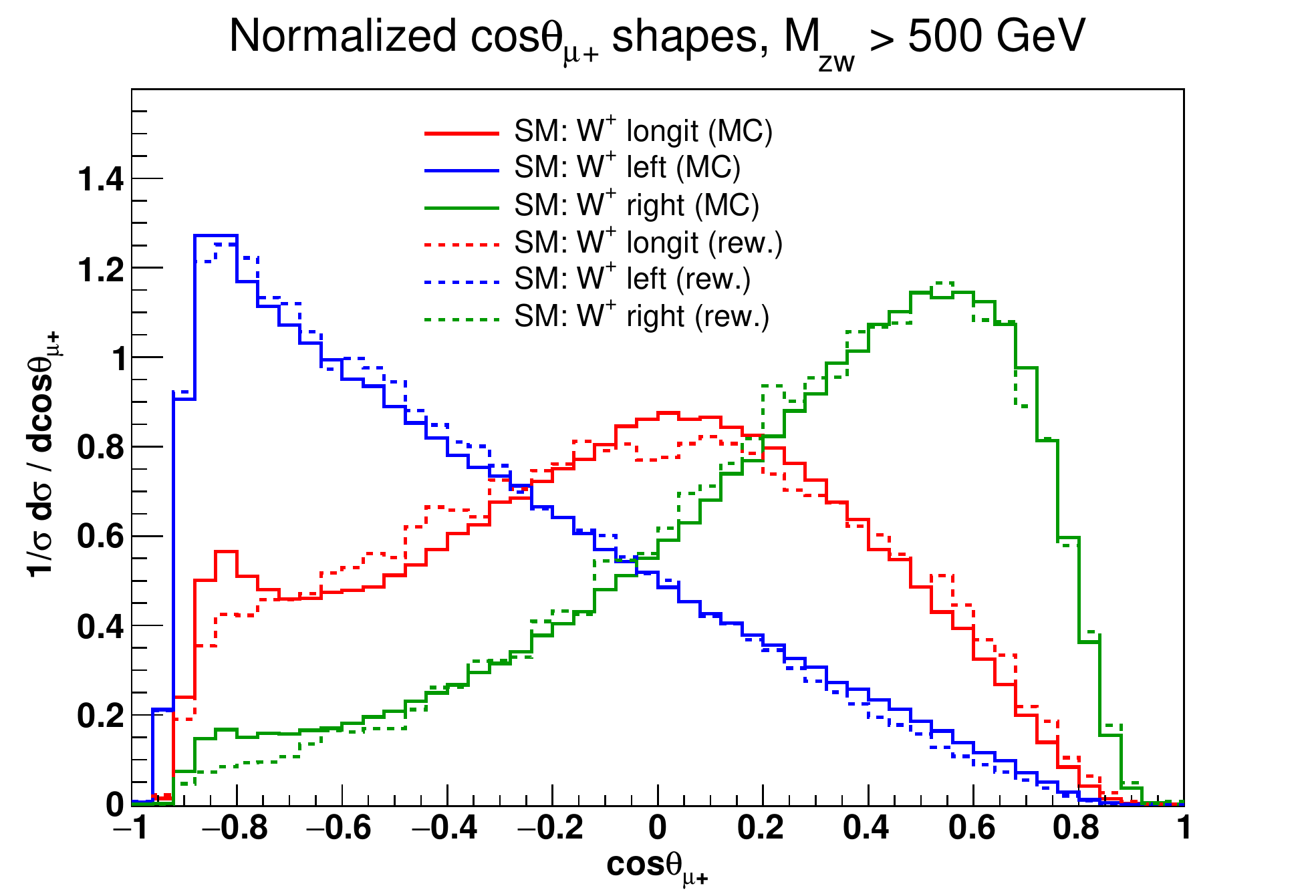}}
\caption{$W^+\!Z$ scattering: polarized $\cos\theta_{\mu^+}$ distributions in the region $M_{W\!Z}>500\,\GeV$. Results of the reweighting procedure compared with results of the MC calculation with polarized amplitudes. The full set of cuts and neutrino reconstruction are understood.} \label{fig:MCvsREW2}
\end{figure}

We have applied the reweighting method to $pp\rightarrow jj  e^+e^- \mu^+\nu_\mu$. In the
absence of lepton cuts and neutrino reconstruction (see Sect.~\ref{subsec:wznocut}), we have
computed polarization fractions for the full process with the following partitioning of the
$\{p_t^W,\eta_W\}$ phase space, as done in Ref.~\cite{Bittrich:2015aia}:
\begin{itemize}
\item[-] $p^{W}_t<30\GeV$, $30\GeV\,<p^W_t<60\GeV$, $60\GeV\,<p^W_t<90\GeV$, $p^W_t>90\GeV$;
\item[-] $|\eta_{W}|<1$, $1<|\eta_{W}|<2$, $2<|\eta_{W}|<3$, $|\eta_{W}|>3$.
\end{itemize}
In each region, we have separated the full unpolarized sample into three polarized samples, using
the algorithm described above. Then we have applied
the full set of leptonic cuts and performed neutrino reconstruction, obtaining
approximate polarized distributions which can be compared with those presented
in Sect.~\ref{subsec:lepcutwz}.
We have compared total cross sections (Tabs.~\ref{tab:MCvsREW1}, \ref{tab:MCvsREW2})
and reconstructed $\cos\theta_{\mu^+}$ differential distributions (Figs.
~\ref{fig:MCvsREW1}, \ref{fig:MCvsREW2}).
\begin{table}[h!]
\begin{center}
\begin{tabular}{|c|c|c|}
\hline
\multicolumn{3}{|c|}{\cellcolor{blue!9} Polarized cross sections [$\ab $], $M_{W\!Z}> 200$ GeV}\\
\hline
polarization & MC polarized & Reweighting \\
\hline
\hline
longit. & 33.21(3)  &     41.02(3) \\
\hline
left &  96.31(8)   &     95.97(2)  \\
\hline
right & 30.93(2)  &     27.87(3)  \\
\hline
\end{tabular}
\caption{Polarized total cross sections ($\ab $) for $W^+\!Z$ scattering in the region $M_{W\!Z}>200\,\GeV$: results of the reweighting procedure compared with results of the MC calculation with polarized amplitudes. The full set of cuts and neutrino reconstruction are understood.} \label{tab:MCvsREW1}
\end{center}
\end{table}

In the whole fiducial region ($M_{W\!Z}>200\,\GeV$), the left polarized $\cos\theta_{\mu^+}$ distribution
obtained with the reweighting procedure describes fairly well the analogous
distribution obtained with polarized amplitudes, both in total cross section
and in shape (${\sigma}^{-1}\,{d\sigma(X)}/{dX}$).
On the contrary, the longitudinal total cross section is
overestimated by 23\% and the right polarized cross section
is underestimated by 10\%, as shown in Tab.~\ref{tab:MCvsREW1}. Even larger
discrepancies show up when analyzing the $\cos\theta_{\mu^+}$ differential
cross section and shape (Fig.~\ref{fig:MCvsREW1}).

It is important to observe that the sum of the three cross sections
obtained with polarized amplitudes (central column of Tab.~\ref{tab:MCvsREW1}),
is not equal to the full unpolarized cross section, since the interferences
among polarizations account for 5\% of the full result.

Interferences are completely neglected in the reweighting method
(rightmost column of Tab.~\ref{tab:MCvsREW1}). As a consequence,
the sum of the three polarized cross sections is, by construction,
equal to the full unpolarized one.

The inaccuracy of the reweighting procedure becomes even more evident at high
energies, as Fig.~\ref{fig:MCvsREW2} and Tab.~\ref{tab:MCvsREW2} show. For
$M_{W\!Z}>500\,\GeV$, the  reweighting predictions are absolutely unreliable. In
particular, the longitudinal cross section is overestimated by 70\%, and
the corresponding $\cos\theta_{\mu^+}$ shape is rather different from the Monte Carlo polarized
prediction. At large diboson masses, the polarization interferences are smaller than at lower masses, however
neglecting them contributes to the low precision of the reweighting method.
\begin{table}[h!]
\begin{center}
\begin{tabular}{|c|c|c|}
\hline
\multicolumn{3}{|c|}{\cellcolor{blue!9} Polarized cross sections [$\ab $], $M_{W\!Z}> 500$ GeV}\\
\hline
polarization & MC polarized & Reweighting \\
\hline
\hline
longit. &  5.96(2)  &     9.94(4) \\
\hline
left    &  28.38(3)  &     25.49(3)  \\
\hline
right   &  9.06(3)  &     8.13(3)  \\
\hline
\end{tabular}
\caption{Polarized total cross sections ($\ab $) for $W^+\!Z$ scattering in the region $M_{W\!Z}>500\,\GeV$: results of the reweighting procedure compared with results of the MC calculation with polarized amplitudes. The full set of cuts and neutrino reconstruction are understood.} \label{tab:MCvsREW2}
\end{center}
\end{table}

The main bottleneck of the reweighting procedure is represented by
the phase space dependence of the polarization fractions.
In the absence of lepton cuts, each polarization gives the same lepton
angular distribution in the $W$ rest frame in any phase space point.
However, the relative weight of the three polarizations varies from point to point.
When assigning a polarization state to a single event, the reweighting
procedure assigns to each event belonging to a $\{p_t^{W},\,\eta_W\}$ cell
the average weight over the whole region. As a consequence, the reweighting method
is not capable of reproducing the correct dependence on kinematic variables different
from $\cos\theta_{\ell^+}$.

\begin{figure}[!htb]
\centering
\subfigure[Monte Carlo \label{fig:rew1dejj}]{\includegraphics[scale=0.37]{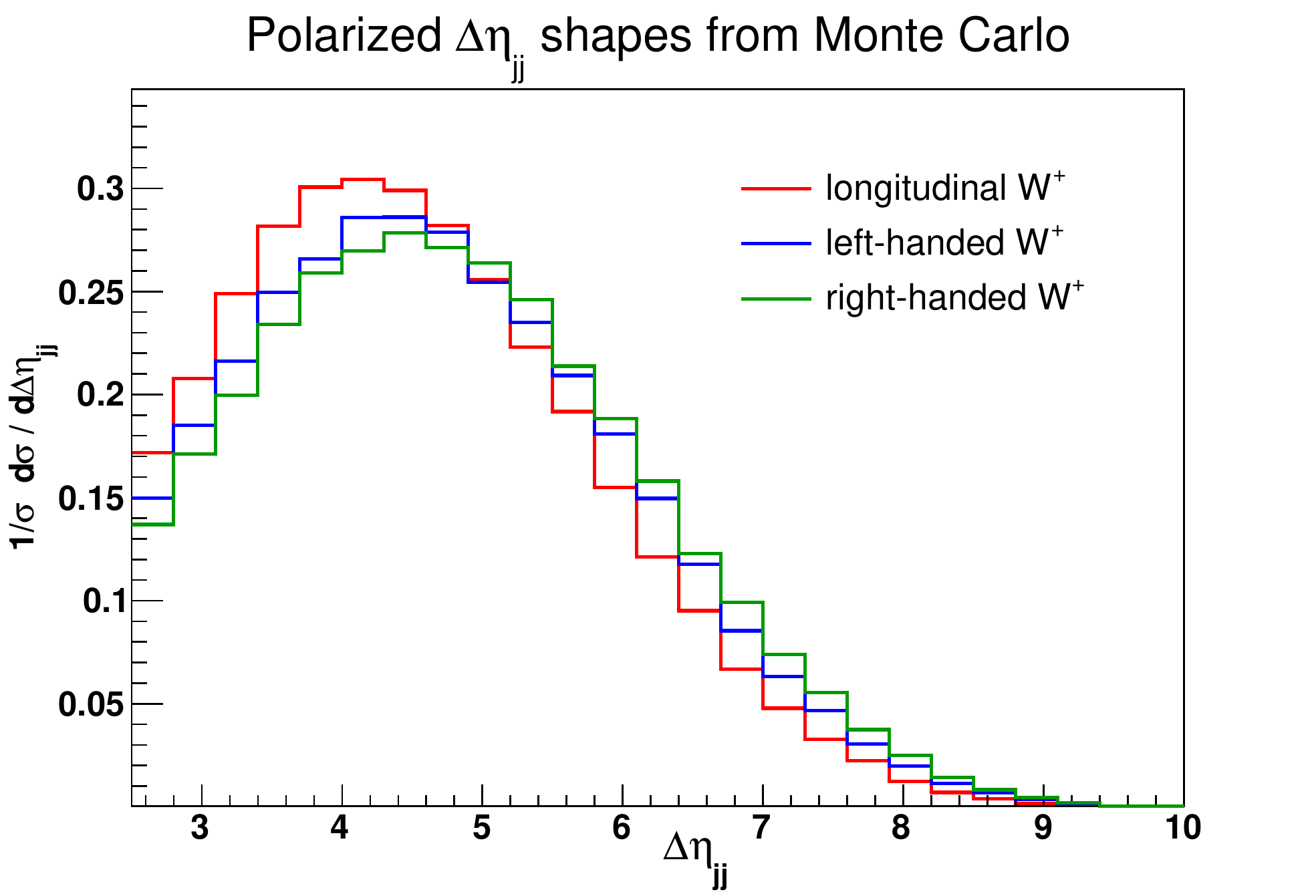}}
\subfigure[Reweighting \label{fig:rew2dejj}]{\includegraphics[scale=0.37]{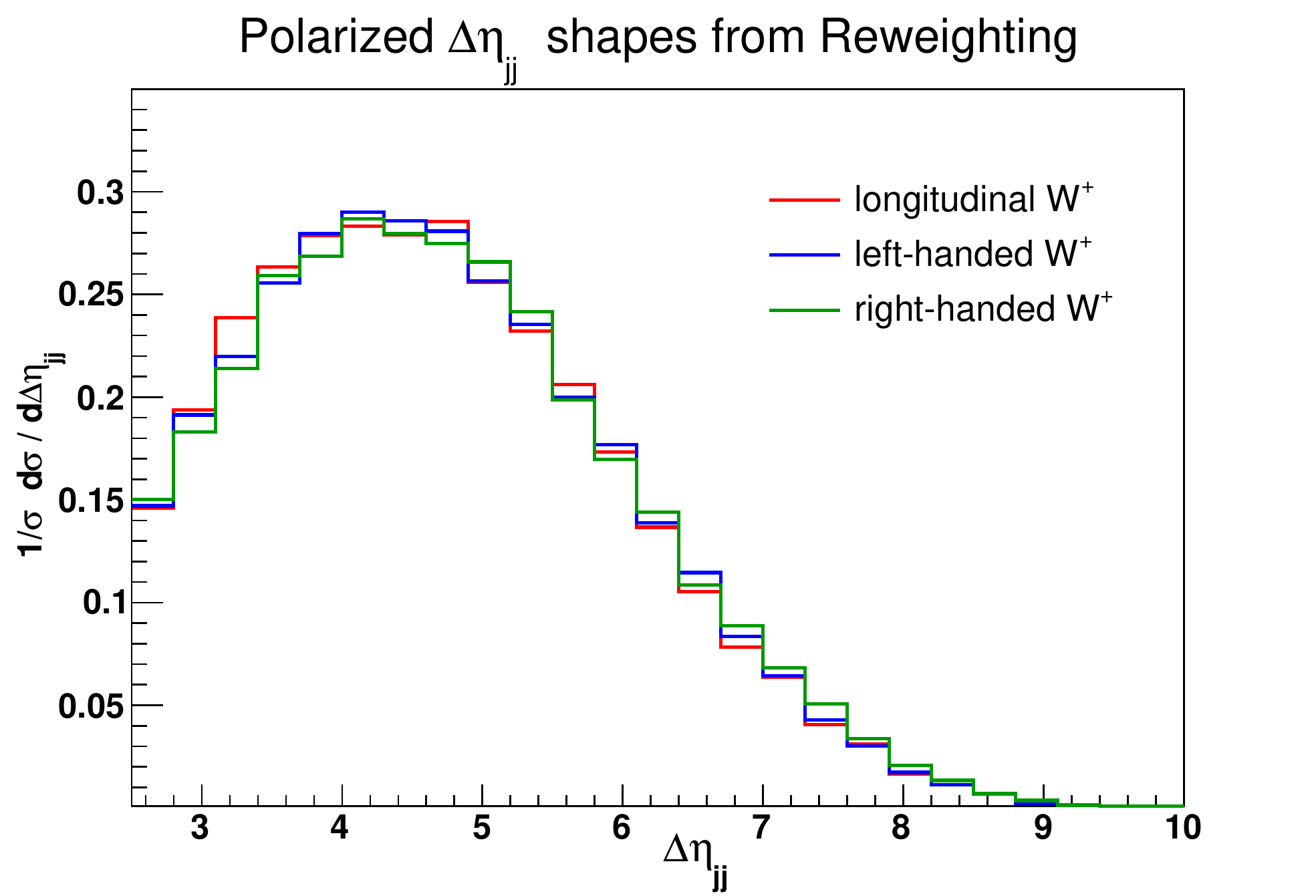}}
\caption{$W^+\!Z$ scattering: $\Delta\eta_{jj}$ normalized distributions for a polarized $W^+$,
obtained with polarized amplitudes (left side) and with
the reweighting procedure (right side), in the region $60\,\GeV\! <\! p_t^W \!<\!90\,\GeV,1\!<\!|\eta_W|\!<\!2$, in the absence of lepton cuts and
without neutrino reconstruction.} \label{fig:deltaetajj_sec6}
\end{figure}

To show that this is the case even in the absence of lepton cuts and neutrino reconstruction,
we have compared, in the region $60\,\GeV\! <\! p_t^W \!<\!90\,\GeV,1\!<\!|\eta_W|\!<\!2$,
the longitudinal, left, and right distributions obtained from reweighting
with those computed directly with polarized amplitudes, for a number of variables
which do not depend on the decay products of the polarized $W^+$.
In Fig.~\ref{fig:deltaetajj_sec6}, we show the normalized distributions of
the rapidity difference between the two tagging jets.
The polarized shapes on the left, obtained with polarized amplitudes,
are clearly different from each other: the longitudinal one is peaked at a
smaller value of $\Delta\eta_{jj}$ than the two transverse components, and
decreases faster in the distribution tail.
The analogous polarized shapes from reweighting,  on
the right of Fig.~\ref{fig:deltaetajj_sec6}, are similar to each other,
confirming that, even when considering a small  $\{p_t^W,\,\eta_W\}$
region, reweighting corresponds to averaging on the dependence on other variables, washing out
the differences, even in the absence of leptonic cuts.

Also the $\eta_W$ and $p_t^W$ distributions cannot be described perfectly.

This becomes even more problematic when lepton cuts are imposed on
the polarized samples, since selection cuts have different effects on different
polarizations.
The conceptual issue is that the polarized samples are obtained without lepton cuts, and
then are analyzed in the presence of cuts. The computation of polarization fractions
and the application of lepton cuts are non commuting procedures.

Notice that the correct description of all kinematic variables is
mandatory for a Multi Variate Analysis.

We have shown that the reweighting method to separate an unpolarized
event sample into three polarized samples provides only approximate predictions,
which can be quite far from being accurate, particularly at high energies.
Therefore it would be better, both for phenomenological and for experimental analyses,
to produce polarized event samples employing directly polarized amplitudes.

\section{Extracting polarization fractions}
\label{subsec:extracting_ZW_ZZ}
In this section we investigate the possibility of extracting polarization fractions from VBS
events without prior knowledge of the underlying dynamics.
We present results both for $W^+\!Z$ and for $Z\!Z$ scattering.
As instances of theories beyond the Standard Model (BSM), we consider a Standard Model
with no Higgs boson, \emph{i.e.} $M_h\rightarrow \infty$, and a Singlet extension of the Standard Model.

After the discovery of a 125 GeV mass scalar particle compatible with a Higgs boson
\cite{Chatrchyan:2012xdj,Aad:2012tfa}, the Higgsless model is not viable anymore.
However, it can be considered as an extreme case of strongly coupled models: there is a
large class of models whose phenomenology lies in between the SM and the Higgsless model.
Deviations from the SM are expected in the large $VV$ invariant mass region, where the
vector boson longitudinal mode becomes dominant with respect to the transverse ones.
The other BSM model we consider is a $\mathcal{Z}_2$ symmetric Singlet extension of the
SM \cite{Silveira:1985rk,Schabinger:2005ei,O'Connell:2006wi,BahatTreidel:2006kx,Barger:2007im,
Bhattacharyya:2007pb,Gonderinger:2009jp,Dawson:2009yx,Bock:2010nz,Fox:2011qc,Englert:2011yb,
Englert:2011us,Batell:2011pz,Englert:2011aa,Gupta:2011gd},
which features a heavy scalar particle in addition to the (light) 125 GeV Higgs boson.
The new heavy Higgs, is characterized by $M_H=600\,\GeV,\,\Gamma_H = 6.45\,\GeV$.
Its interactions are related by simple multiplicative factors to those of the light Higgs.
Both sets of couplings are determined by the mixing angle $\alpha$, while the ratio
between the two vacuum expectation values is parametrized by an angle $\beta$.
In the following, we assume $\sin\alpha=0.2$ and
$\tan\beta=0.3$ \cite{Pruna:2013bma,Lopez-Val:2014jva,Robens:2015gla}.
In addition to the SM like couplings, the heavy Higgs couples to a light Higgs bosons pair.
Deviations from the SM are expected in the $VV$
invariant mass spectrum, around the heavy Higgs pole mass, if the scalar particle propagates
in $s$ channel. As for the light Higgs, the heavy Higgs couples mainly to the longitudinal
modes of $V$ bosons, therefore the deviations will concern especially the longitudinal
polarization mode.

Polarization fractions are determined with two different methods.
The first one relies on the expectation that the shapes of the
decay angular distributions are not too sensitive to the underlying
dynamics. If this is the case, one can  fit the unpolarized
distribution of a BSM model with a superposition of SM templates, as done in Ref.~\cite{Ballestrero:2017bxn}.
The second exploits the similarity, in shape and normalization, of the transverse distributions across different
models, which allows to extract the longitudinal component by subtracting the SM transverse contribution.
Both methods give acceptable results within a few percent. The difference between the two determinations
provides a rough estimate of the uncertainty in the extraction procedure.

All the results presented in this section have been obtained applying the complete set of cuts.
For $W^+\!Z$, neutrino reconstruction is always understood.

\subsection{Transverse polarizations}\label{sub:trans}
\begin{figure}[!hbt]
\centering
\subfigure[$\cos\theta_{\mu^+}$, for a polarized $W$\label{fig:trans_vs_RL_W}]{\includegraphics[scale=0.37]{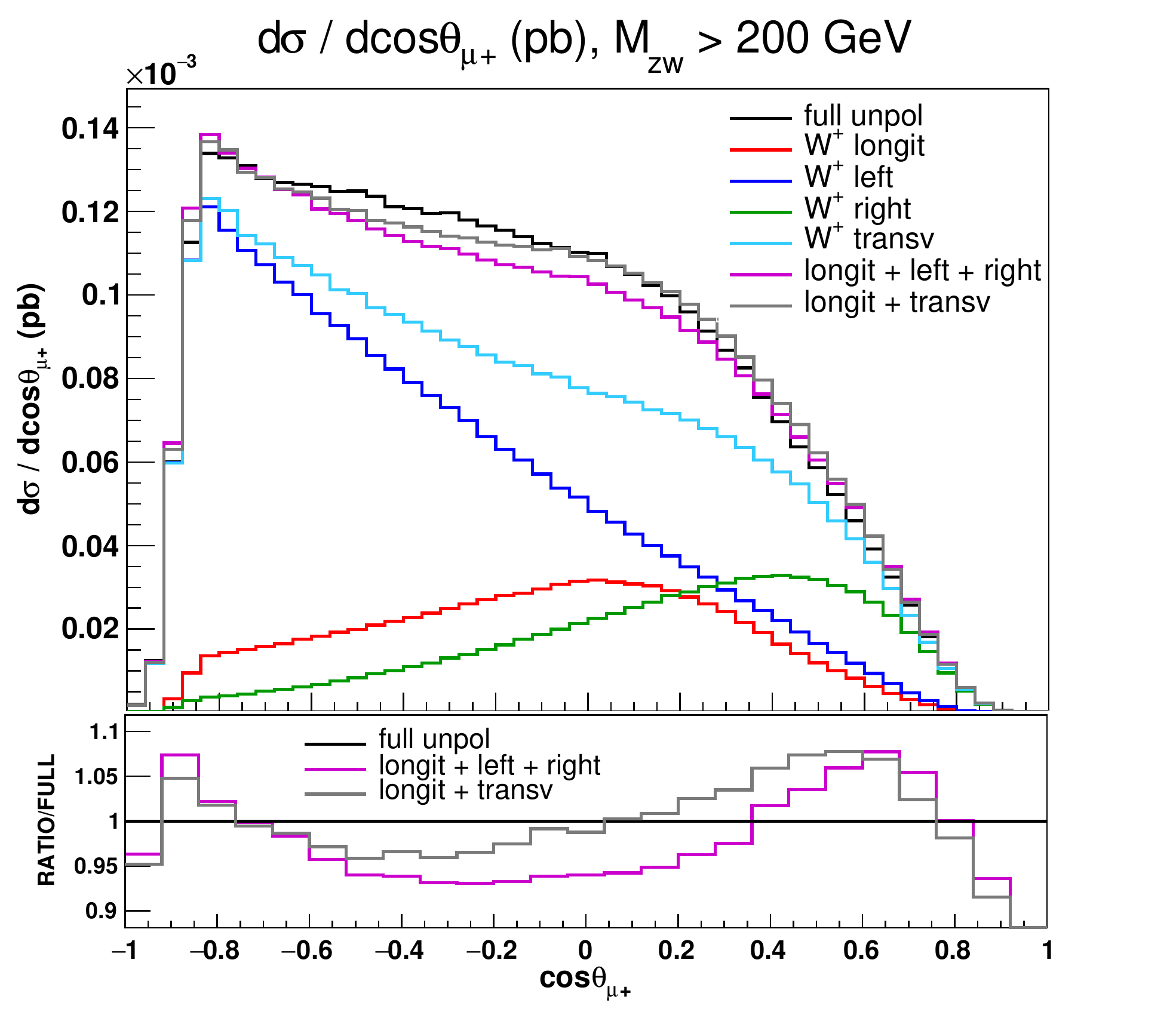}}
\subfigure[$\cos\theta_{e^-}$, for a polarized $Z$\label{fig:trans_vs_RL_Z}]{\includegraphics[scale=0.37]{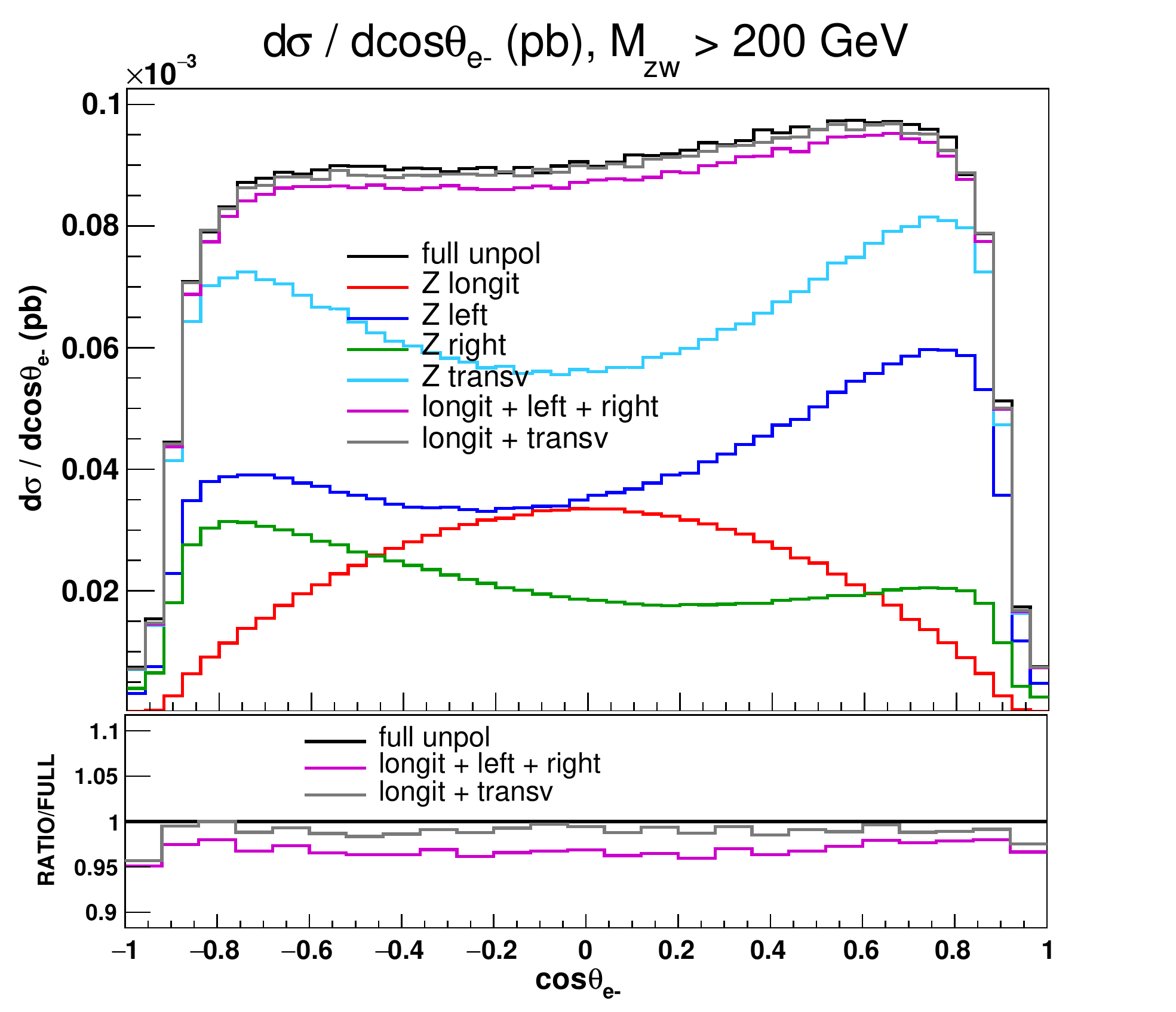}}
\caption{$W^+\!Z$ scattering: comparison of transverse to left+right Standard Model distributions in $\cos\theta_{\mu^+}$ and
$\cos\theta_{e^-}$ distributions, in the fiducial region ($M_{W\!Z}>200\,\GeV$). The full set of kinematic cuts
(see Sect.~\ref{subsec:setupwz}) is understood.}\label{fig:trans_vs_RL}
\end{figure}

In \sects{sec:ZZ}{sec:WZ} we have shown the Standard Model predictions for polarized $W/Z$ bosons
in $Z\!Z$ and $W^+\!Z$ scattering. In particular, we have considered left and right contributions
separately. If we consider the coherent sum of left and right polarizations (which we refer to as
transverse), we include the left-right interference term. Therefore, separating only the longitudinal from the
transverse mode is expected to minimize the total interferences among different polarizations.
That this is indeed the case can be seen in Fig.~\ref{fig:trans_vs_RL}, where
we show the $\cos\theta_{\mu^+}$
distributions for a polarized $W^+$ and the $\cos\theta_{e^-}$ distributions for a polarized $Z$
in $W^+\!Z$ scattering.
If we compare the incoherent sum of left, right and longitudinal contributions
(magenta curve) to the sum of transverse and longitudinal (gray curve), we find that the interferences
among polarizations, defined as the difference between the full and the sum of polarized distributions,
are smaller in the second case. This holds for a number of other kinematic variables.
Very similar results have been obtained for a polarized $Z$ boson in $Z\!Z$ scattering.
Even in the presence of BSM dynamics (either Higgsless or Singlet Extension), we reach the same
conclusion.
We have verified that fitting the BSM unpolarized event samples with a combination of
transverse and longitudinal SM distributions gives better results than using a combination of
three separate single polarized ones.
Therefore, in the following we consider transverse distributions instead of left and right ones separately.


\subsection{$Z\!Z$ channel}\label{sub:zzextr}
We investigate how a different dynamics affects the longitudinal and transverse
polarizations of a $Z$ boson ($\rightarrow e^+e^-$) produced in VBS together with
an unpolarized $Z$ ($\rightarrow \mu^+\mu^-$). First, we compare the SM and the Higgsless
model. Second, we compare the SM and its Singlet
extension in the $Z\!Z$ invariant mass window around the heavy Higgs
boson pole mass.

The SM and Higgsless total cross sections for the polarized processes are shown in
Tab.~\ref{tab:ZZnoh_xsec},
\begin{table}[hbt]
\begin{center}
\begin{tabular}{|c|c|c|c|c|}
\hline
\multicolumn{5}{|c|}{\cellcolor{blue!9} Total cross sections [$\ab $]}\\
\hline
& \multicolumn{2}{c|}{ $M_{Z\!Z}>200\,\GeV$} & \multicolumn{2}{c|}{$M_{Z\!Z}>500\,\GeV$}\\
\hline
Polarization & SM & NoH & SM & NoH \\
\hline
\hline
longitudinal &16.19(2) & 27.66(3)& 2.19(1)& 9.72(1)\\
\hline
transverse & 44.11(4)& 46.56(5) & 10.72(5)& 10.95(6)\\
\hline
longit.+transv.  & 60.31(5)   &74.22(6) & 12.91(5) & 20.67(6)\\
\hline
full unpolarized & 61.00(6) & 75.10(8) & 12.87(1) & 20.68(2)\\
\hline
\end{tabular}
\end{center}
\caption{SM and Higgsless total cross sections ($\ab $) for $Z\!Z$ scattering, in the whole fiducial region and for $M_{Z\!Z}>500\,\GeV$.}
\label{tab:ZZnoh_xsec}
\end{table}
in the complete fiducial region ($M_{Z\!Z}>200\,\GeV$) and in the large invariant mass regime ($M_{Z\!Z}>500\,\GeV$).

The bulk of the difference between the SM and the Higgsless model is due to
the longitudinal contribution, while the transverse one is only weakly sensitive to the underlying
dynamics, as it has been already observed in $W^+W^-$ scattering \cite{Ballestrero:2017bxn}.
The Higgsless transverse component is just 4\% larger than the SM one in the full fiducial volume,
2\% when considering only four lepton invariant masses larger than 500 GeV.
The longitudinal cross section in the Higgsless model, by contrast, is larger than the SM one by 70\% in the
$M_{Z\!Z}>200\,\GeV$ region and by more than a factor four for  $M_{Z\!Z}>500\,\GeV$.

\begin{figure}[!htb]
\centering
\subfigure[$M_{Z\!Z}$\label{fig:extr_compzz_noh1}]{\includegraphics[scale=0.37]{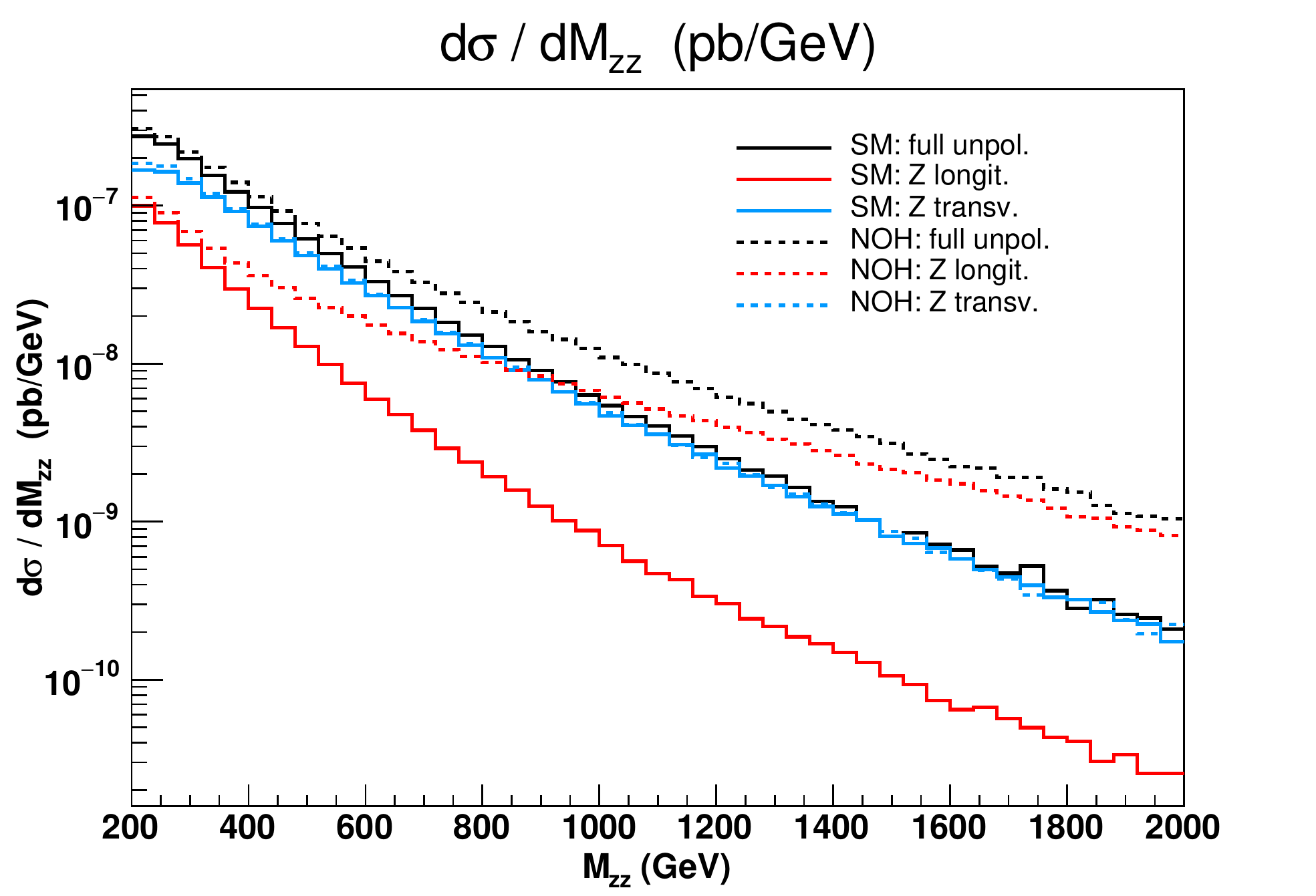}}
\subfigure[$\cos\theta_{e^-}$\label{fig:extr_compzz_noh2}]{\includegraphics[scale=0.37]{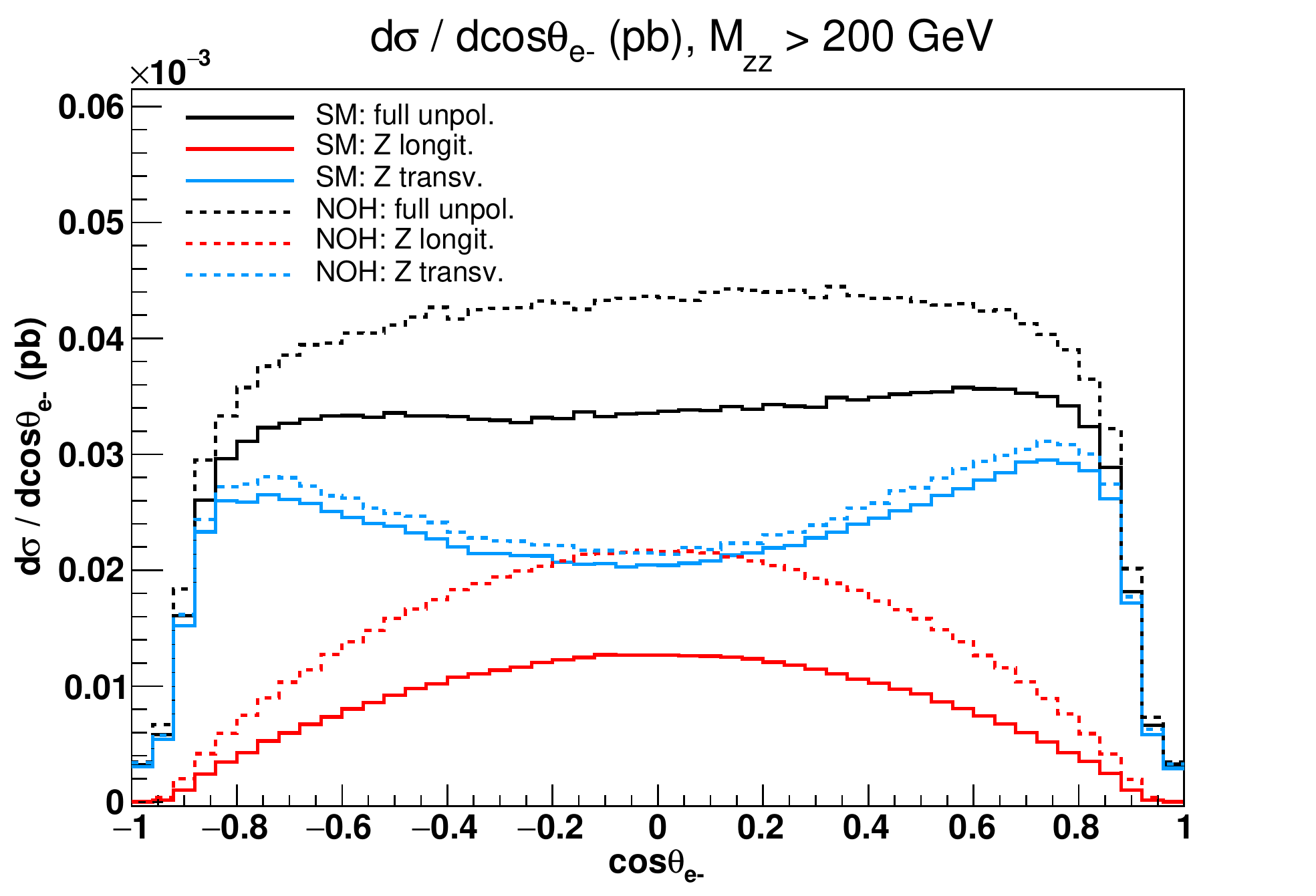}}
\caption{$Z\!Z$ scattering: comparison of Standard Model (solid) and Higgsless model (dashed) distributions in
$M_{Z\!Z}$ and $\cos\theta_{e^-}$, in the fiducial region ($M_{Z\!Z}>200\,\GeV$). The full set of kinematic
cuts (see Sect.~\ref{subsec:setupzz}) is understood.}\label{fig:extr_comp_nohzz}
\end{figure}

These effects are even more evident in the differential cross sections shown in Fig.~\ref{fig:extr_comp_nohzz}.
The SM longitudinal $M_{Z\!Z}$ distribution decreases much more rapidly than the corresponding
Higgsless one at large invariant boson boson invariant mass. A very similar, though milder,
effect is present also in $W\!Z$ scattering, as it can be observed comparing Figs.~\ref{fig:extr_compzz_noh1},
\ref{fig:extr_compmwz}, and \ref{fig:extr_compmwz_Z}.

The similarity of Higgsless and SM transverse components can be appreciated in
Fig.~\ref{fig:extr_compzz_noh2}, where we present the $\cos\theta_{e^-}$ distributions for both models.
Transverse distributions (azure curves) differ by an almost constant 5\% factor bin by bin.
However they feature the same shape. This shape similarity is true even for the longitudinal
distributions (red curves), despite a huge difference in terms of total cross sections. This holds
in any of the interesting kinematic regions within the fiducial volume.

Given these premises, we try to extract the polarization fractions from the Higgsless model
unpolarized $\cos\theta_{e^-}$ distribution using SM templates, as done in Ref.~\cite{Ballestrero:2017bxn}.
We compute the SM $\cos\theta_{e^-}$ polarized and
interference normalized distributions,
$F_{ 0}^{\rm SM},\,F_{\rm T}^{\rm SM}$ and $F_{\rm I}^{\rm SM}$. Then we fit the unpolarized
$d\sigma_{\rm full}^{\rm NoH}/d\cos\theta_{e^-}$ distribution of the
Higgsless model  with a superposition of the three SM templates,

\beq
f(\{\mathcal{C}_{\lambda}\},\cos\theta_{e^-})\,=\,\sum_{\lambda = 0, T, I} \mathcal{C}_{\lambda}
\,F_{\lambda}^{\rm SM} (\cos\theta_{e^-})\,.
\eeq
We estimate the best parameters $\overline{\mathcal{C}}_{\lambda}$ by means of a simple $\chi^2$
minimization. In order to evaluate the fit goodness we check:
\begin{enumerate}
\item that $f(\{\overline{\mathcal{C}}_{\lambda}\}, \cos\theta_{e^-})$ reproduces correctly
$d\sigma_{\rm full}^{\rm NoH}/d\cos\theta_{e^-}$, and
\item that each term of the sum after minimization
($\overline{\mathcal{C}}_{\lambda}\,F_{\lambda}^{\rm SM}$)
reproduces the correspondent Higgsless polarized distribution $d\sigma_{\lambda}^{\rm NoH}/d\cos\theta_{e^-}$.
\end{enumerate}

Taking advantage of the strong similarity of the SM and Higgsless transverse differential
cross sections, we have also examined an alternative procedure to extract the longitudinal component, which
assumes that the transverse components and the interference are identical to the SM ones.
This allows to subtract them from the full Higgsless distribution,
in order to deduce the Higgsless longitudinal differential cross section.
This {subtraction} procedure is characterized by a strong bias on the transverse and interference terms,
which are assumed to be
independent of the underlying dynamics, which is supposed, in other words, to have a significant
effect only on the longitudinal polarization of vector bosons.

\begin{figure}[!htb]
\centering
\subfigure[$M_{Z\!Z} >200\,\GeV$\label{fig:extr_fit1_Z_noh}]{\includegraphics[scale=0.37]{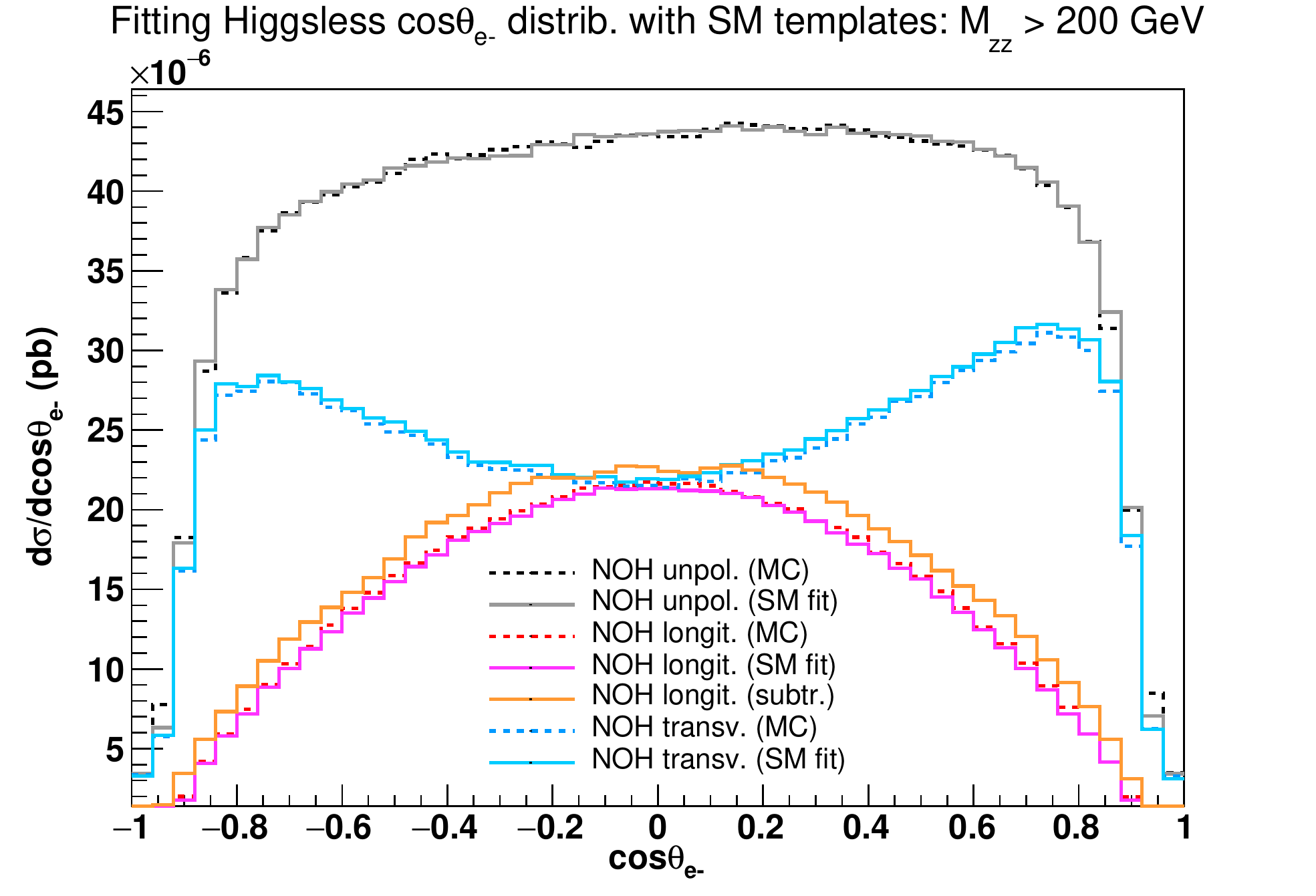}}
\subfigure[$M_{Z\!Z} >500\,\GeV$\label{fig:extr_fit2_Z_noh}]{\includegraphics[scale=0.37]{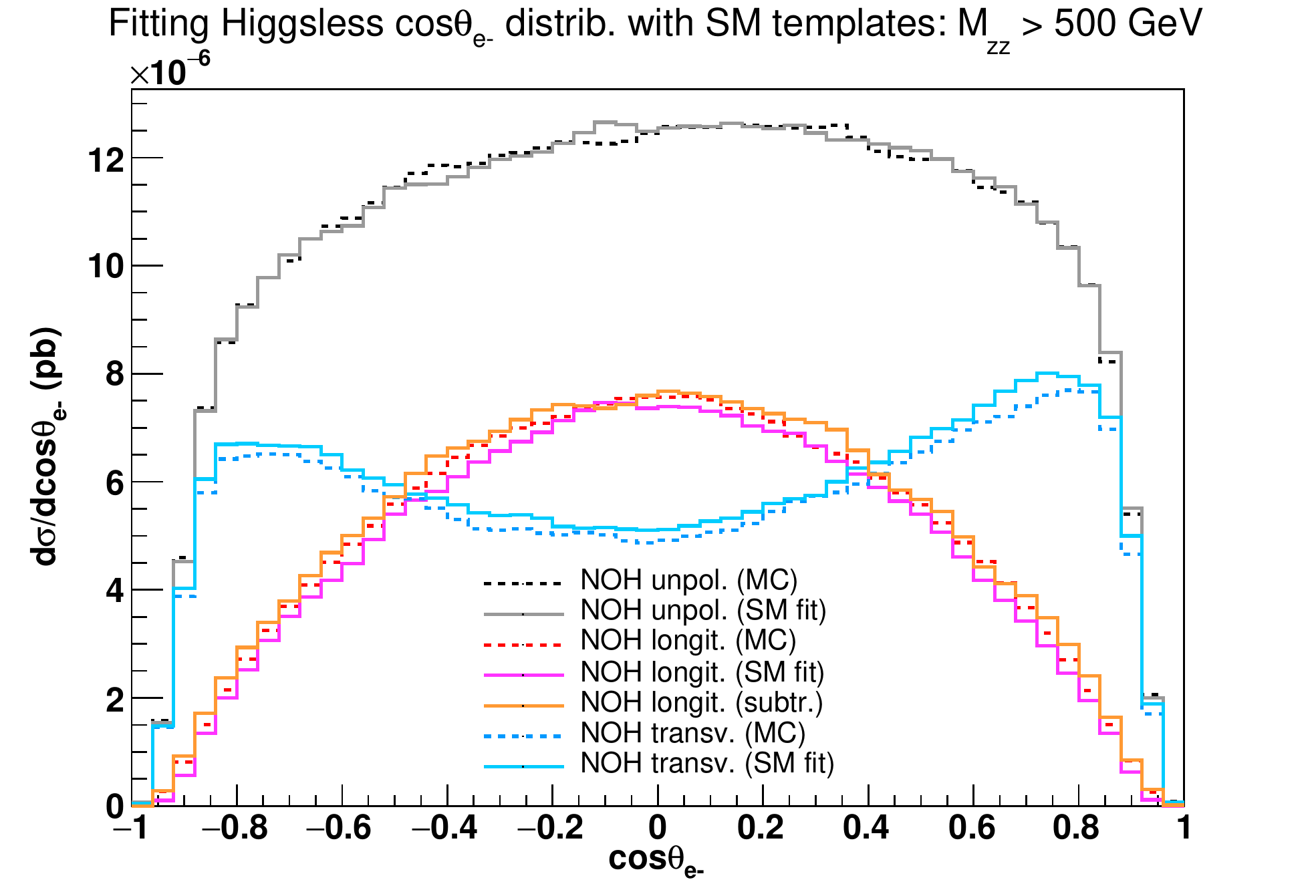}}
\caption{$Z\!Z$ scattering: fit of the Higgsless unpolarized $\cos\theta_{e^-}$ distribution with SM templates.
Fitted and expected
differential cross sections in two $Z\!Z$ invariant mass regions.
For the longitudinal component the result of the
fit (magenta) and the one of the subtraction technique (orange) are compared with the Monte Carlo
expectations
(dashed red).}\label{fig:fitsubtrzz_z_noh}
\end{figure}

We have performed the fit and the subtraction procedure to extract the Higgsless longitudinal component
in the full fiducial region and in the large $M_{Z\!Z}$ region. The results are shown in Fig.~\ref{fig:fitsubtrzz_z_noh}.
When including small values of $M_{Z\!Z}$, the difference between SM and Higgsless transverse cross sections
propagates to the subtraction procedure, leading to a 9.5\% overestimate of the Higgsless longitudinal
cross section. When the analysis is restricted to the region $M_{Z\!Z}>500\,\GeV$,
the longitudinal component is estimated much better, with the total cross section just 2.8\% larger than the
expected value. When looking at
even larger masses, the subtraction procedure improves again (+1.5\% for $M_{Z\!Z}>1000\GeV$).
On the contrary, the fit procedure slightly underestimates the longitudinal component in almost all invariant
mass regions. However, it works better than the subtraction approach when considering the full $Z\!Z$ invariant mass range.
The longitudinal contribution is estimated with at maximum 3\%
discrepancy with respect to the expected value. For $M_{Z\!Z}>200\,\GeV$, the fit underestimates the
longitudinal cross section by only 1.4\%.

These results suggest that, given the  model (quasi)independence of transverse and longitudinal
$\cos\theta_{\ell}$ shapes, a more refined fitting method should enable the extraction of polarized cross section from LHC data
with satisfactory accuracy.
The discrepancy in the transverse cross section between the two models, despite being small and under
control, hampers an accurate extraction of the longitudinal cross section of the Higgsless model by subtracting SM
distributions.

\begin{figure}[!htb]
\centering
\subfigure[$M_{Z\!Z}$ differential cross sections\label{fig:extr_compzz_singl1}]{\includegraphics[scale=0.37]{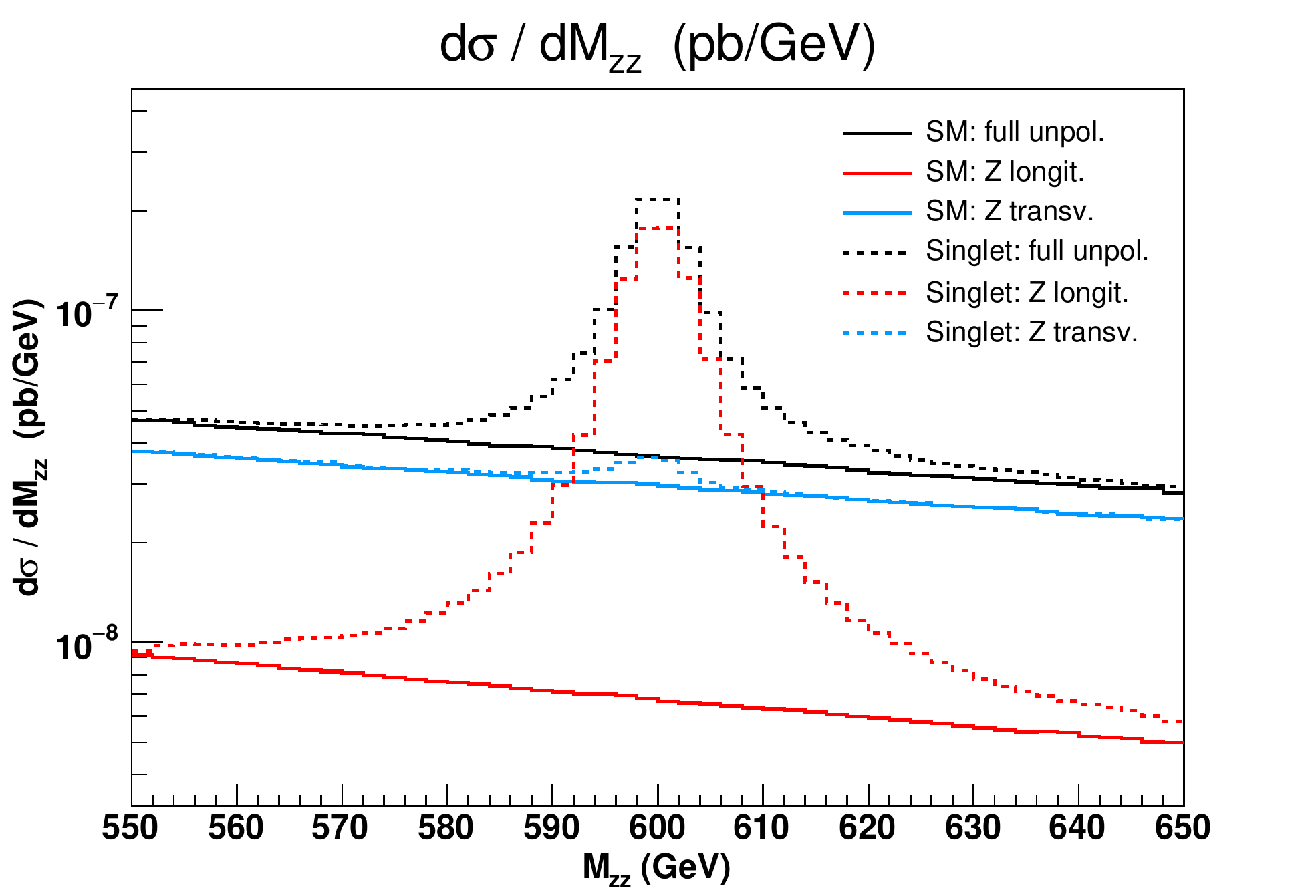}}
\subfigure[$\cos\theta_{e^-}$ differential cross sections\label{fig:extr_compzz_singl2}]{\includegraphics[scale=0.37]{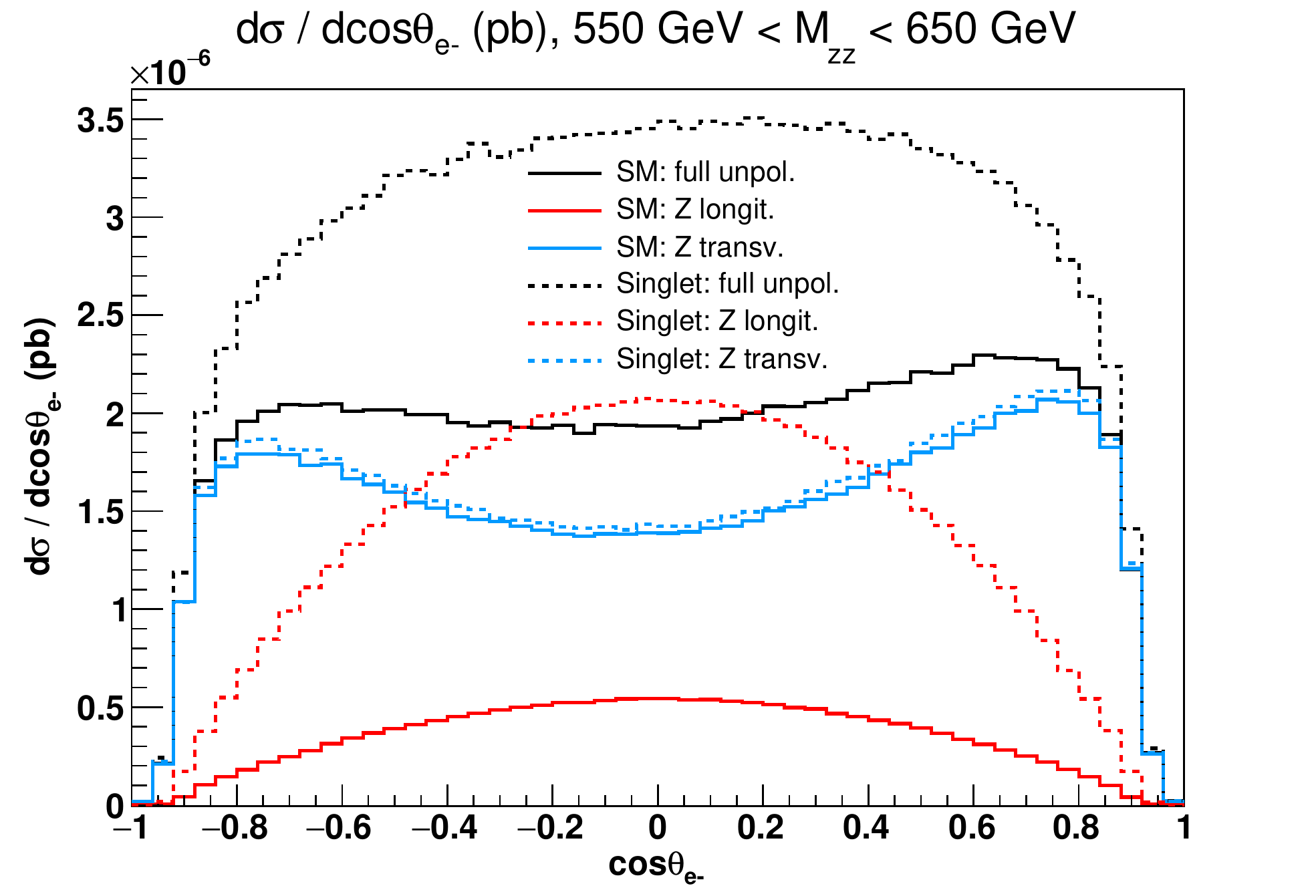}}\\
\subfigure[$\cos\theta_{e^-}$ normalized shapes\label{fig:extr_compzz_singl3}]
{\includegraphics[scale=0.37]{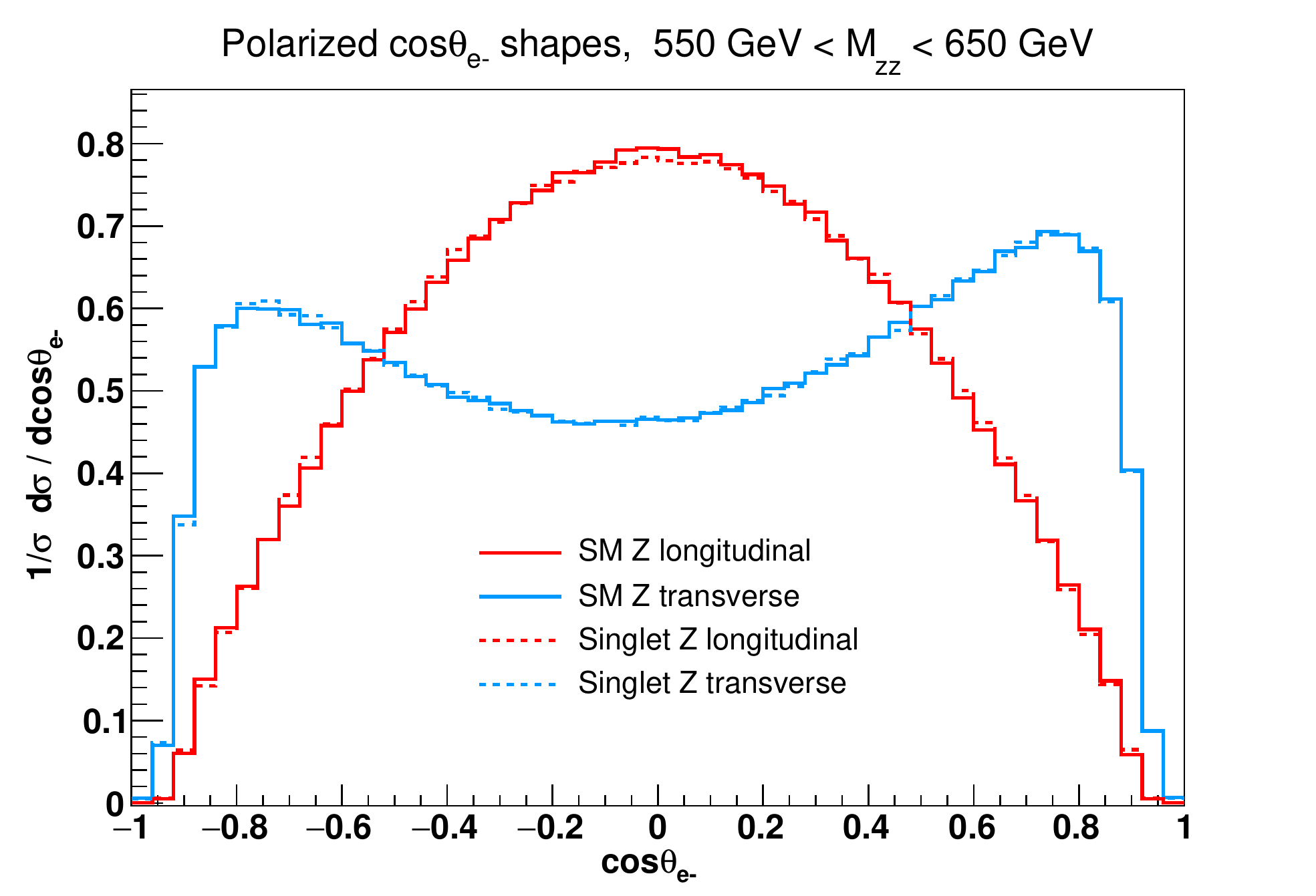}}
\caption{$Z\!Z$ scattering: comparison of Standard Model (solid) and Singlet (dashed) distributions in $M_{Z\!Z}$ and
$\cos\theta_{e^-}$, in the heavy Higgs resonance region ($M_H=600\,\GeV$).
The full set of kinematic cuts (see Sect.~\ref{subsec:setupzz}) is understood.}\label{fig:extr_comp_singlet}
\end{figure}

We now present a few results on the comparison between the SM and its Singlet extension in the invariant
mass region of the heavy Higgs resonance.
The polarized and unpolarized $M_{Z\!Z}$ distributions for both models are shown in
Fig.~\ref{fig:extr_compzz_singl1}.
The Singlet longitudinal distribution (dashed red curve) features a Breit-Wigner resonance on top of the
decreasing SM distribution (solid red curve). Even the transverse component is partially affected by the
additional scalar particle, as can be seen from the small bump of the dashed blue curve around $600\,\GeV$.

The $\cos\theta_{e^-}$ distributions in the region $550\,\GeV<M_{Z\!Z}<650\,\GeV$ are shown in
Fig.~\ref{fig:extr_compzz_singl2}. In Fig.~\ref{fig:extr_compzz_singl3} we show the longitudinal
and transverse normalized $\cos\theta_{e^-}$ shapes for the two models.
The transverse component is essentially insensitive to the presence of the additional resonance,
as the similarity of the blue curves demonstrates, both in shape and total cross section
in the resonance region with a 2.6\% discrepancy.
The shape of the longitudinal component is impressively similar for the two models,
despite a large difference in the total cross section. This holds even when considering a narrower
invariant mass region about the heavy Higgs pole mass, \emph{e.g.} $590\,\GeV<M_{Z\!Z}<610\,\GeV$.

\begin{figure}[!htb]
\centering
\subfigure[$550\,\GeV < M_{Z\!Z} <650\,\GeV$\label{fig:extr_fit1_Z_singl}]{\includegraphics[scale=0.37]{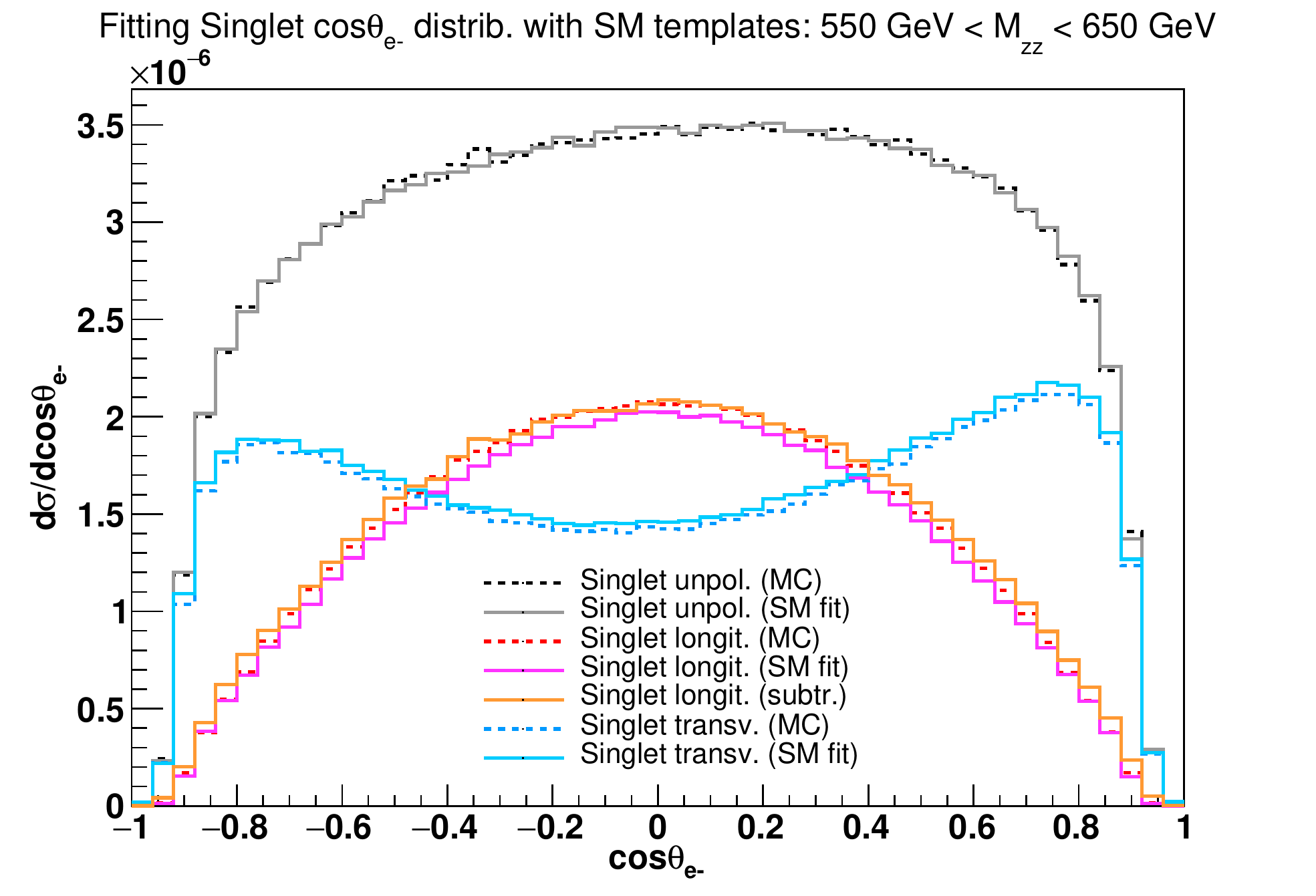}}
\subfigure[$590\,\GeV < M_{Z\!Z} <610\,\GeV$\label{fig:extr_fit2_Z_singl}]{\includegraphics[scale=0.37]{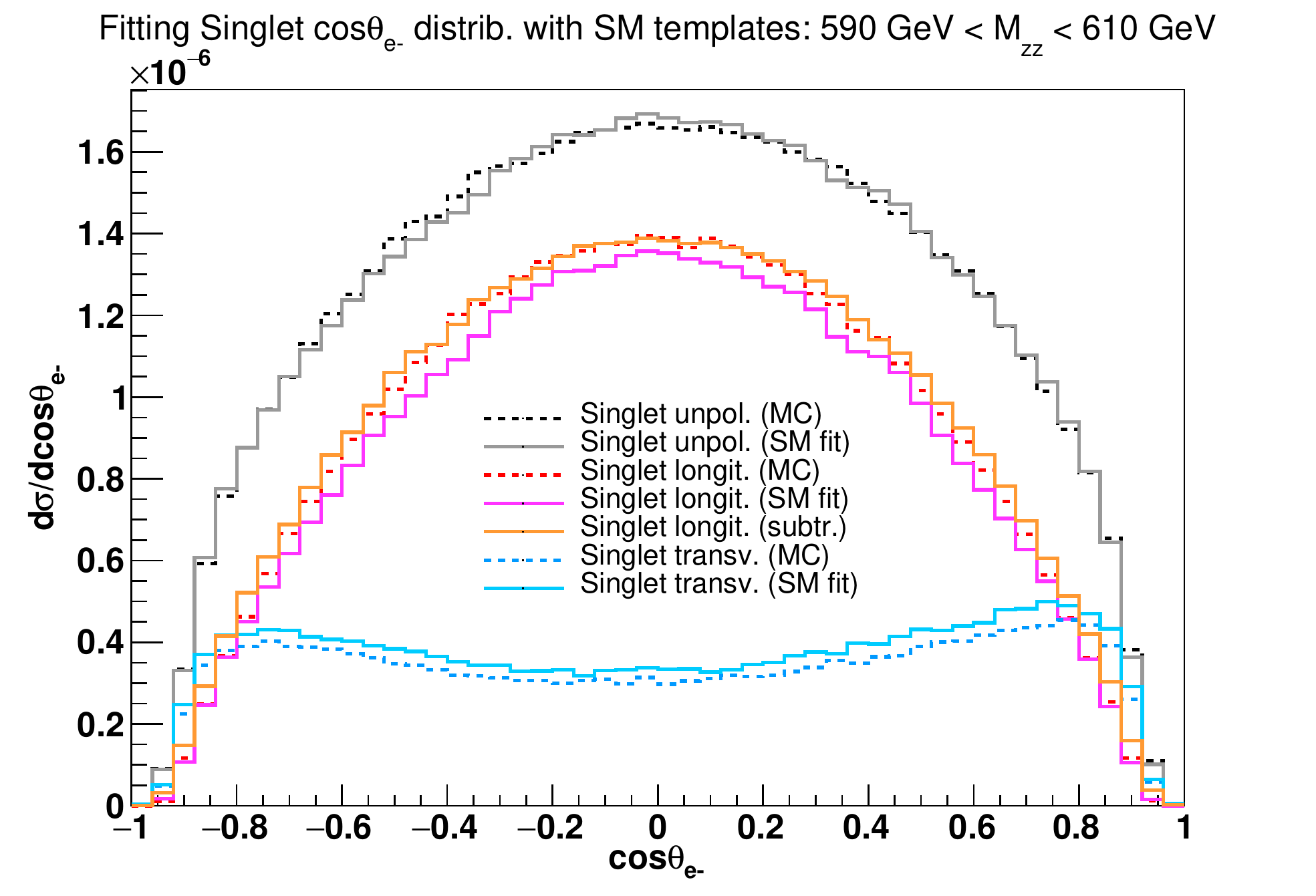}}
\caption{$Z\!Z$ scattering: fit of Singlet unpolarized $\cos\theta_{e^-}$ distribution with SM templates.
Fitted and expected differential cross sections in two $Z\!Z$ invariant mass windows about the heavy Higgs
pole mass. For the longitudinal component the result of the fit (magenta) and the subtraction technique one
(orange) are compared with the Monte Carlo expectation (dashed red).}\label{fig:fitsubtrzz_z_singlet}
\end{figure}

We have performed the fit for a $\pm 50$ GeV and $\pm 10$ GeV mass window around $M_H$.
The result of the fit, shown in Fig.~\ref{fig:fitsubtrzz_z_singlet}, underestimates by less than 4\% the
longitudinal cross section obtained directly with the Monte Carlo.
Conversely, the transverse component is slightly overestimated. Even better results can be obtained via the
subtraction procedure. The longitudinal cross section in this case is reproduced  with a +2.5\% error.
These last results give us confidence that even in the presence of additional resonances interfering with the SM,
it is possible to extract the longitudinal component from LHC data with a few percent accuracy.

\subsection{Polarized $W$ in the $W^+\!Z$ channel}\label{sub:wzextr1}
In a similar fashion as in Sect.~\ref{sub:zzextr}, we investigate how different dynamics affect
vector boson polarizations in $W^+\!Z$ scattering. Since the Higgs contributes to $W^+\!Z$ VBS production only in the $t/u$ channels,
an additional heavy Higgs is expected to produce a rather small enhancement of the total cross section.
Therefore, we present only results for the Higgsless model.

\begin{figure}[!htb]
\centering
\subfigure[$M_{W\!Z}$\label{fig:extr_compmwz}]{\includegraphics[scale=0.37]{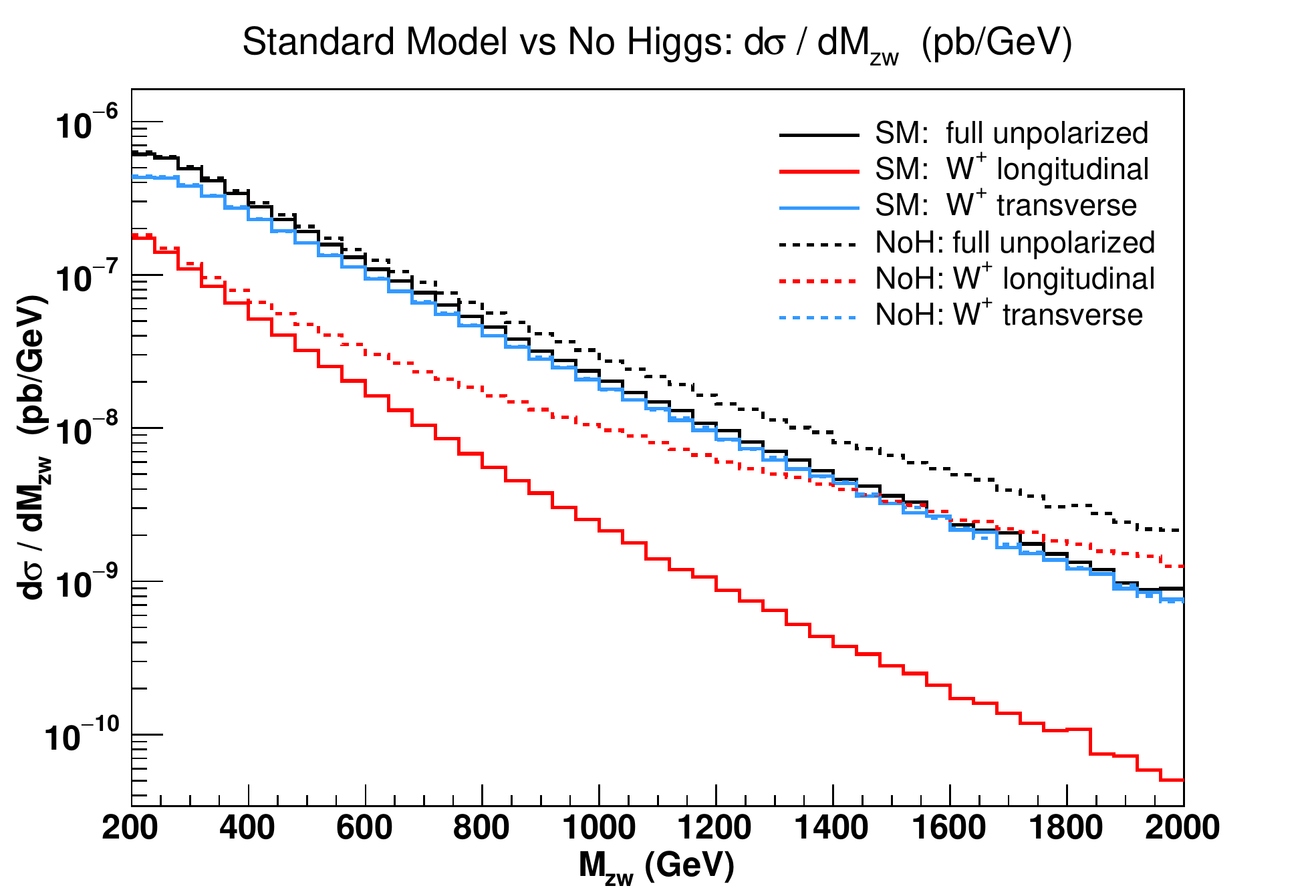}}
\subfigure[$\cos\theta_{\mu^+}$\label{fig:extr_compcth}]{\includegraphics[scale=0.37]{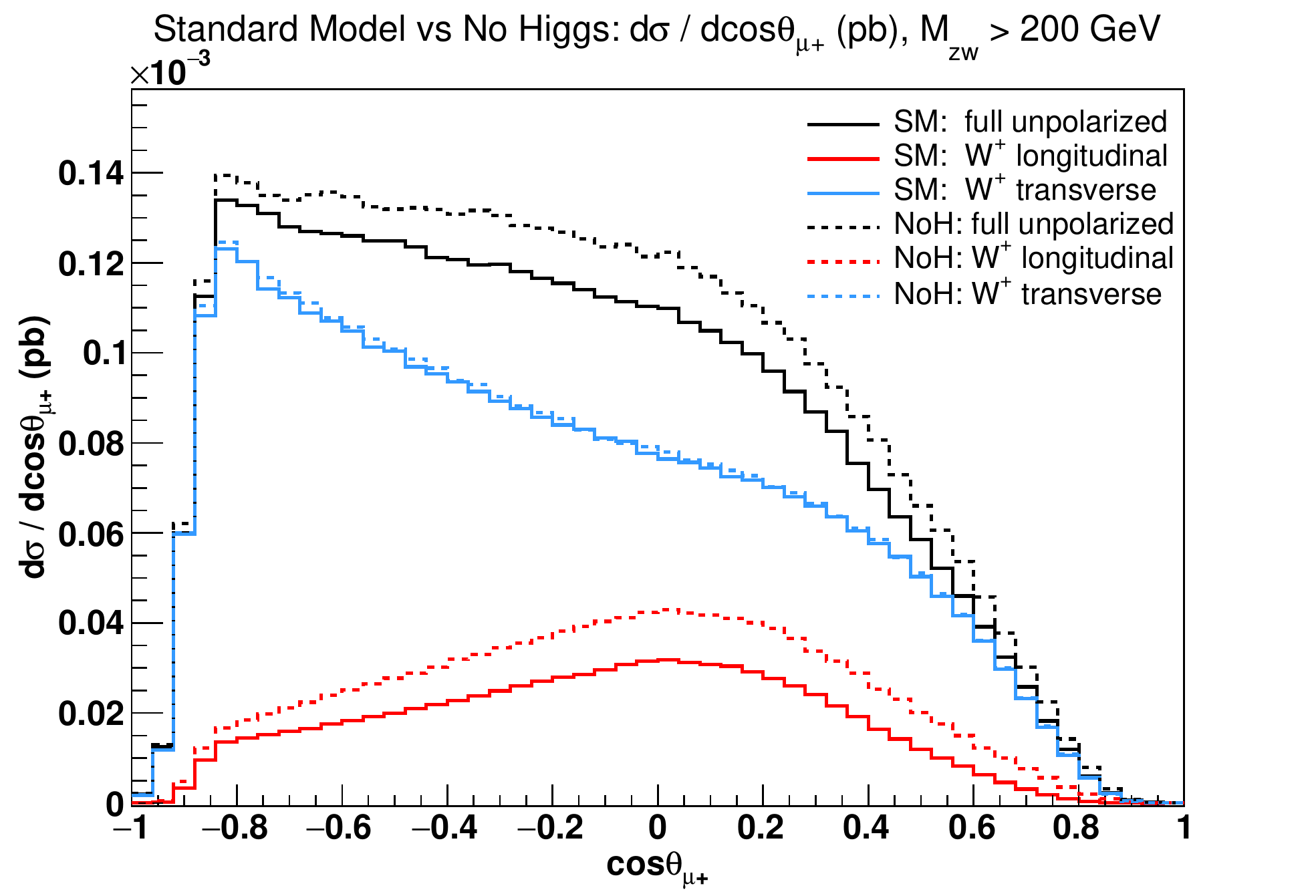}}
\caption{$W^+\!Z$ scattering: comparison of Standard Model (solid) and Higgsless model (dashed) distributions in $M_{W\!Z}$ and
$\cos\theta_{\mu^+}$. Polarized distributions concern the $W^+$ boson. The full set of kinematic cuts
(see Sect.~\ref{subsec:setupwz}) is understood, including lepton and missing transverse momentum cuts,
as well as neutrino reconstruction.}\label{fig:extr_comp}
\end{figure}

Let's consider first a $W^+$ with given polarization and an unpolarized $Z$.
The difference in the total cross section between the SM and the Higgsless model
is due to the longitudinal contribution, while the transverse result is even less sensitive to the underlying
dynamics than in $Z\!Z$ scattering. The two transverse cross sections differ by only 1\%, while the SM
longitudinal cross section is 40\% smaller than the Higgsless one. This is confirmed for the differential
cross sections, as shown in Fig.~\ref{fig:extr_comp}.
The $M_{W\!Z}$ transverse distributions shown in Fig.~\ref{fig:extr_compmwz} are almost identical for the
SM and the Higgsless model over the full invariant mass range.
The same holds for the $\cos\theta_{\mu^+}$ distributions, shown in Fig.~\ref{fig:extr_compcth}.

In the full fiducial region, the longitudinal contribution features a similar shape in the two models, which
proves to be promising for an (almost) model independent fit to extract polarization fractions from the BSM
unpolarized distribution. However we will see that the similarity of longitudinal shapes is not true anymore
at large $M_{W\!Z}$ and large $p_t^{W}$.

As in Sect.~\ref{sub:trans},  using the transverse distributions rather than the
left and right ones separately reduces the interferences among polarizations to
0.1\% of the full unpolarized cross section. The difference between the sum of polarized distributions and the
full results also decreases. Moreover, the interference shape in the two models is very similar.

As we have done for $Z\!Z$ scattering, we try to extract
the cross section for a polarized  $Z$ in the Higgsless model using SM
polarized templates, either through a fit procedure or through
direct subtraction of SM distributions.
We then compare these two different predictions with the result
obtained with Monte Carlo polarized amplitudes.

\begin{figure}[!htb]
\centering
\subfigure[$M_{W\!Z}>200\,\GeV$\label{fig:extr_fit1}]{\includegraphics[scale=0.37]{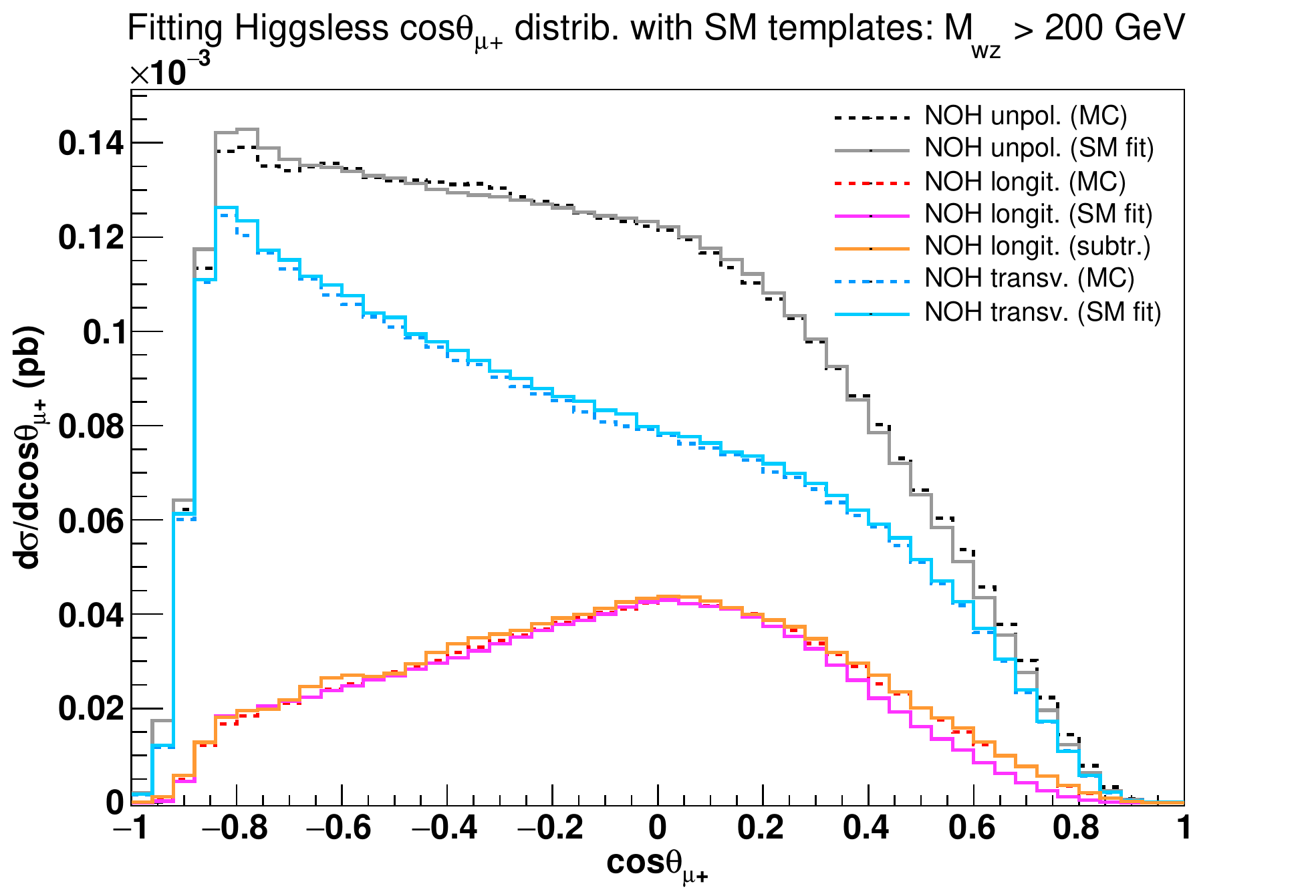}}
\subfigure[$M_{W\!Z}> 500\,\GeV$\label{fig:extr_fit2}]{\includegraphics[scale=0.37]{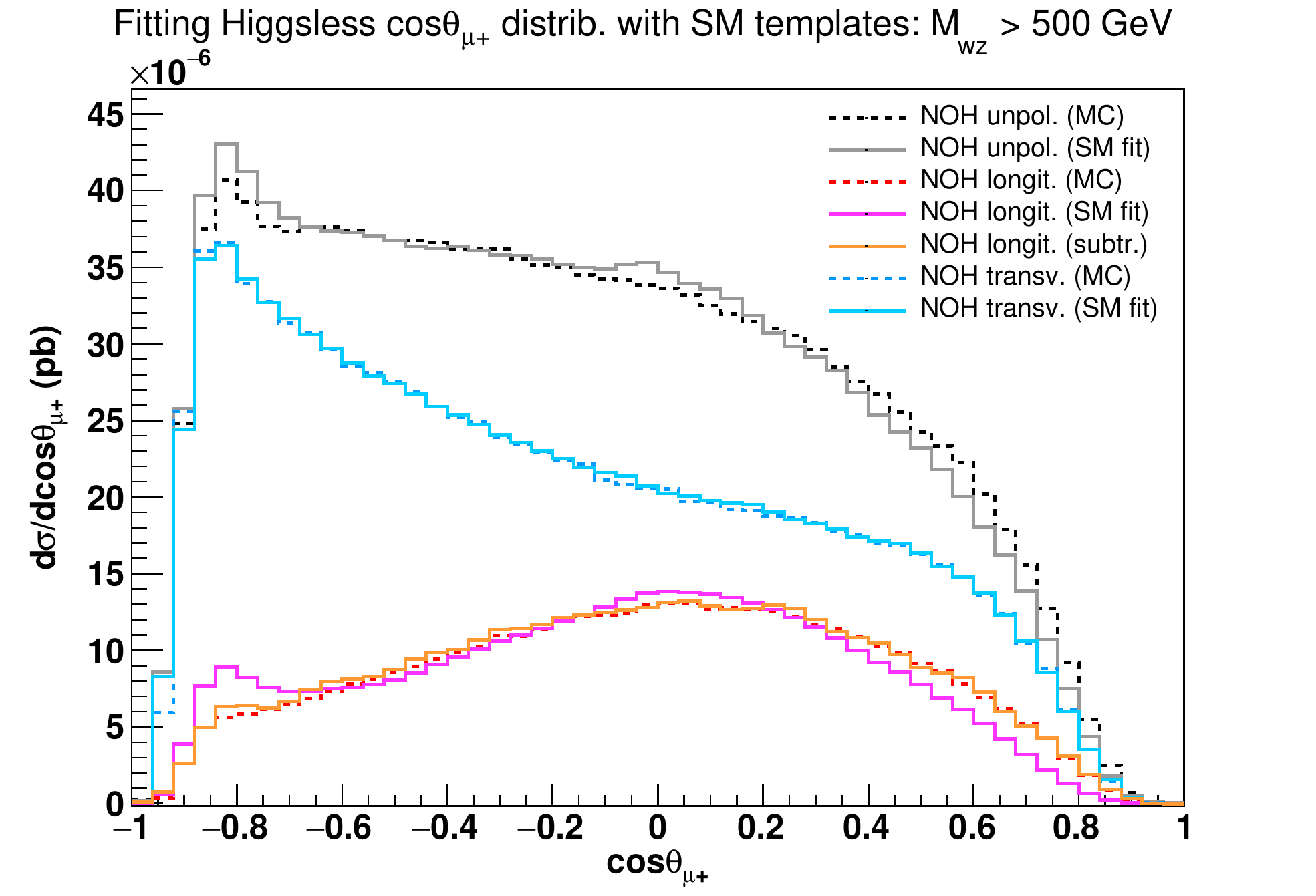}}
\caption{$W^+\!Z$ scattering: fit of Higgsless unpolarized $\cos\theta_{\mu^+}$ distribution with SM templates.
Fitted and expected
distributions, for $M_{W\!Z}>200\,\GeV$ and $M_{W\!Z}>500\,\GeV$.}\label{fig:extr_fit0}
\end{figure}

We have analyzed polarized contributions in a number
of kinematic regions. We show in Fig.~\ref{fig:extr_fit0} the results for the
$\cos\theta_{\mu^+}$ differential distributions, both in the whole fiducial
region ($M_{W\!Z}>200$ GeV) and for $M_{W\!Z}>500\,\GeV$.
In Tab.~\ref{tab:extr_fit} we show the
numerical results of the fit and subtraction procedure for the longitudinal
and transverse cross sections in each of the analyzed kinematic regions.
\begin{table}[hbt]
\begin{tabular}{|c||c|c|c||c|c|c|}
\hline
& \multicolumn{6}{|c|}{\cellcolor{blue!9} Polarized cross sections [$\ab $]} \\
\hline
& \multicolumn{3}{|c||}{ Longitudinal} & \multicolumn{3}{|c|}{ Transverse  }\\
\hline
kinematic region  & MC & Fit & Subtr. & MC & Fit & Subtr. \\
\hline
\hline
$M_{W\!Z}>200\,\GeV$  & 46.90 & 44.93 & 48.37 & 133.10 & 135.16 & 131.73\\
\hline
$M_{W\!Z}>500\,\GeV$  & 16.06 & 15.89 & 16.42 & 38.14 & 38.23 & 37.83\\
\hline
$M_{W\!Z}>1000\,\GeV$  & 4.71 & 5.20& 4.73 & 5.50 & 4.79 & 5.47\\
\hline
$M_{W\!Z}>200\,\GeV$,  $p_t^{W}>200 \,\GeV$ & 13.49 & 13.09 & 13.78 & 43.90 & 44.26 & 43.51\\
\hline
$M_{W\!Z}>200\,\GeV$,  $p_t^{W}>300 \,\GeV$ & 7.89& 7.81& 7.93 & 19.61 & 19.66 & 19.40\\
\hline
$M_{W\!Z}>200\,\GeV$,  $p_t^{W}>400 \,\GeV$ & 4.81& 4.79& 4.84 & 9.12 & 9.26 & 9.03    \\
\hline
$M_{W\!Z}>200\,\GeV$,  $|\eta_{W}|<1 $ & 17.65 & 15.07 & 18.41 & 62.61 & 65.16 & 61.83\\
\hline
$M_{W\!Z}>200\,\GeV$,  $1<|\eta_{W}|<2 $ & 19.42    & 19.36 & 19.95 & 55.91 & 55.70 & 55.35\\
\hline
$M_{W\!Z}>200\,\GeV$,  $2<|\eta_{W}|<3 $ & 8.09 & 8.17 & 8.27 & 13.76 & 13.88 & 13.72\\
\hline
$M_{W\!Z}>200\,\GeV$,  $|\eta_{W}|>3 $ & 1.74 & 1.70 & 1.73 & 0.83 & 0.83 & 0.82\\
\hline
\end{tabular}\quad
\caption{Cross sections ($\ab $), for a longitudinal and transverse $W^+$ in $W^+\!Z$ scattering, in the Higgsless model,
in several kinematic regions: comparison of MC predictions for the Higgsless
model with results obtained via fit and subtraction procedure. The subtraction procedure
results for a transverse $W^+$ coincide with the SM cross sections.} \label{tab:extr_fit}
\end{table}

In the total fiducial region, the Higgsless longitudinal component is reproduced fairly well by the fit,
both in total cross section (-4\%) and in shape (at most 5\% discrepancy, bin by bin). The transverse
cross section is overestimated by 3\%.
Much better results are obtained with the subtraction procedure,
thanks to the very small difference (approximately 1\% in terms of total cross sections)
between the SM and Higgsless transverse component. In this case  both the total cross section (1.5\%
discrepancy) and the
$\cos\theta_{\mu^+}$ distributions (at most 4\% discrepancies, bin by bin) for the longitudinal
component are reproduced accurately, as shown in
Fig.~\ref{fig:extr_fit1}.

When considering less inclusive regions, \emph{e.g.} at large $W\!Z$ invariant mass, the
Higgsless and SM longitudinal $\cos\theta_{\mu^+}$ shapes start to differ, as shown in
Fig.~\ref{fig:extr_fit2}. This results in a poor fit in the region $M_{W\!Z}>500\,\GeV$.
However, for the cross sections, the longitudinal component is only 1\% smaller than the Monte Carlo value.
The full distributions, as well as the transverse ones, are reproduced fairly well by the fit.

When the minimum cut on $M_{W\!Z}$ is  pushed up to 1000 GeV, the fit reproduces the expected polarized
distributions with at most 10\% discrepancies, bin by bin. This shows that a model independent fit
can become inaccurate in some kinematic regimes.

On the contrary, longitudinal $\cos\theta_{\mu^+}$ distributions in the Higgsless case are reproduced by the
subtracted SM distributions within a few percent, in each of the kinematic regions.
In the high energy and forward rapidity regions
($M_{W\!Z}>1000\,\GeV$, $p_t^W>400\,\GeV$, $|\eta_W|>3$), which are
the mostly interesting regions for new physics effects in VBS,
the subtraction procedure reproduces very accurately
the Monte Carlo longitudinal cross sections for the Higgsless model, since the transverse components in the
two models differ by less than 1\%.

The fit results presented in this section show that a model independent extraction of the $W$ polarization
fractions is problematic, due to the lack of universality of the longitudinal $\cos\theta_{\mu^+}$ shapes in this case.
This is not only due to the details of neutrino reconstruction: we have checked that the fit
procedure provides inaccurate results both using a different neutrino reconstruction scheme and
using the true neutrino momentum. Even in these cases the longitudinal
shapes are noticeably model dependent. A factor which contributes to these shape differences
is the strong missing transverse momentum cut ($p_t^{\rm miss} > 40\,\GeV$), which is much harder than
the $p_t$ cut on the anti muon ($p_t^{\ell}>20\,\GeV$).
This introduces a strong asymmetry when boosting into the $W$ rest frame to compute $\cos\theta_{\mu^+}$,
even without neutrino reconstruction. The differences in the $p_t^{\rm miss}$ distribution between the SM
and the Higgsless case are larger than for the $p_t^{\mu^+}$ one and, as a consequence, have a larger effect.
On the contrary, the subtraction procedure has proved promising, despite the strong assumptions it relies on.
The similarity of the transverse cross sections in the SM and in the strong coupling regime is remarkable and
should be investigated further, both in a general EFT framework, and assuming other specific BSM dynamics.

\subsection{Polarized $Z$ in the $W^+\!Z$ channel}\label{sub:wzextr2}

We now consider a polarized $Z$ boson produced via VBS in association
with an unpolarized $W^+$. Differently
from the $W$, the $Z$ boson can be entirely reconstructed.
We then focus on the distributions of the cosine of the electron
angle in the $Z$ CM frame ($\cos\theta_{e^-}$).

As observed previously for the $W^+$, the transverse polarizations of
the $Z$ boson give the same contribution, within 1\%, to the total cross section in
the SM and in the Higgsless model.
Furthermore, both in the SM and in the Higgsless model the adoption of the transverse
component (coherent sum) allows to minimize the interferences, reproducing
at the percent level the full result, when summed to the longitudinal contribution.
{The Higgsless longitudinal component is 30\% larger than the Standard Model
one.}

\begin{figure}[!htb]
\centering
\subfigure[$M_{W\!Z}$\label{fig:extr_compmwz_Z}]{\includegraphics[scale=0.37]{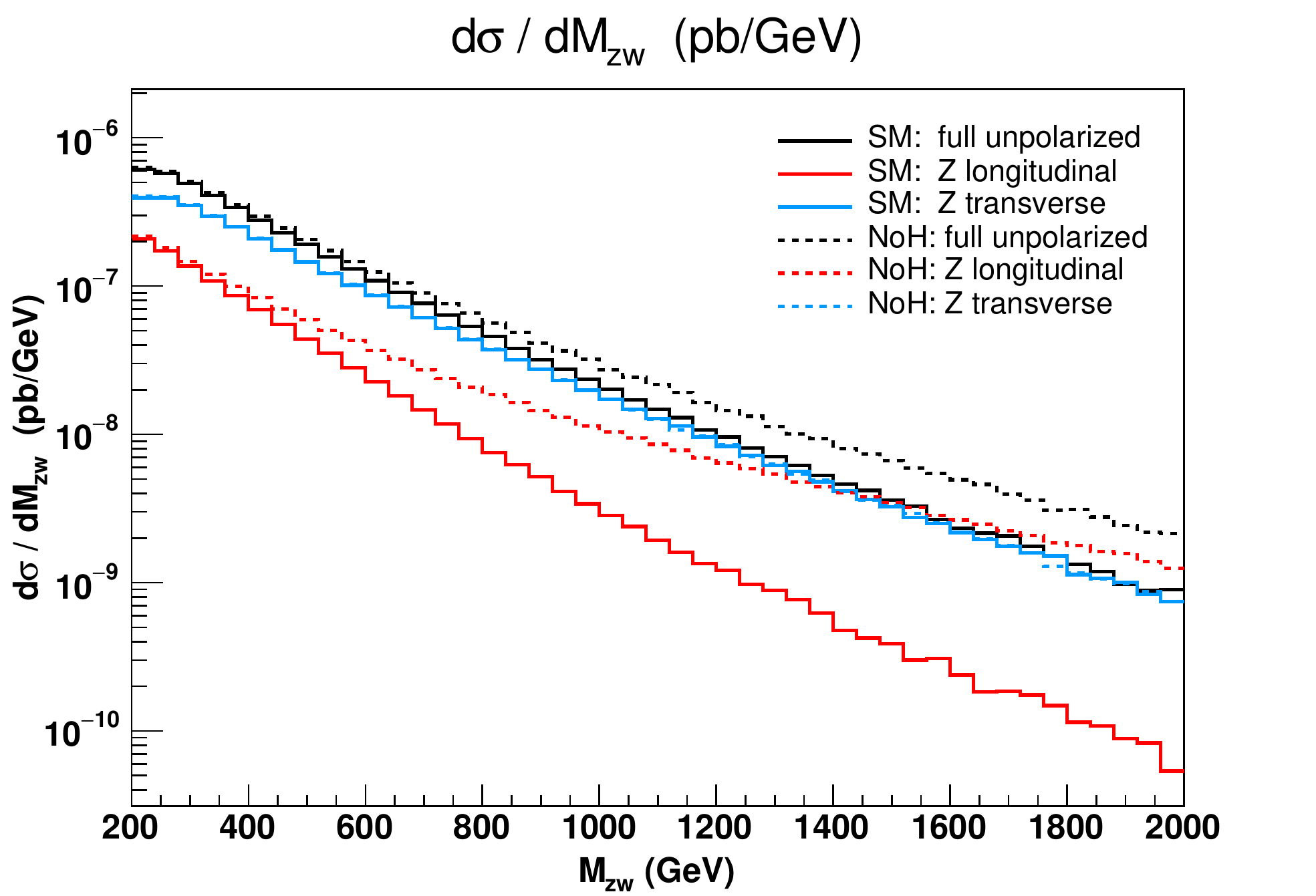}}
\subfigure[$\cos\theta_{e^-}$\label{fig:extr_compcth_Z}]{\includegraphics[scale=0.37]{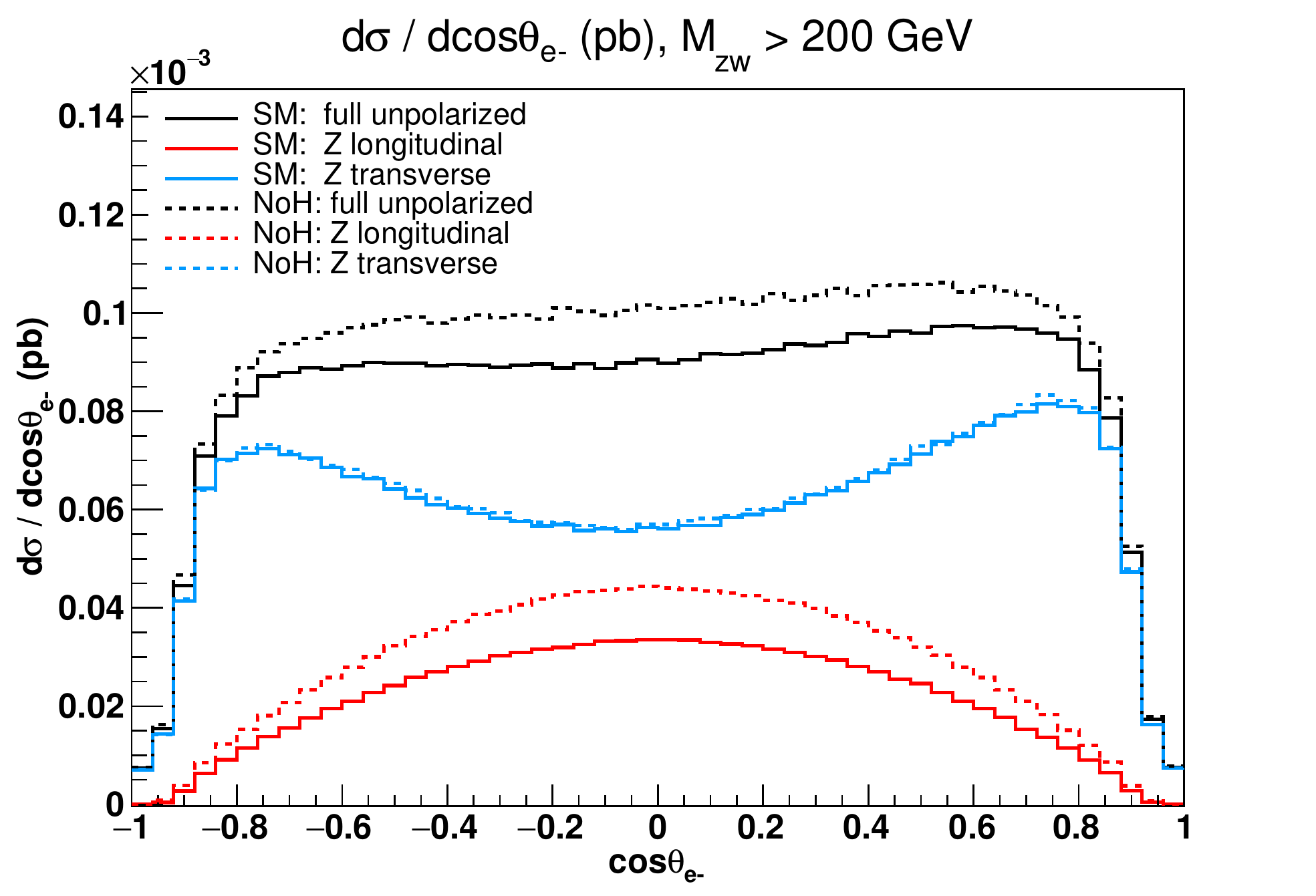}}
\caption{$W^+\!Z$ scattering: comparison of Standard Model (solid) and Higgsless model (dashed) distributions in $M_{W\!Z}$ and
$\cos\theta_{e^-}$. Polarized distributions concern the $Z$ boson. The full set of kinematic cuts
(see Sect.~\ref{subsec:setupwz}) is understood, including lepton and missing transverse momentum cuts, as
well as neutrino reconstruction.}\label{fig:extr_comp_Z}
\end{figure}

As for the $W$, at large boson boson invariant mass the longitudinal
component in the Higgsless model dominates. This effect can be observed
in Fig.~\ref{fig:extr_compmwz_Z}.
The transverse differential distributions are almost identical, even at very large four lepton
invariant masses.

In Fig.~\ref{fig:extr_compcth_Z}, we present the $\cos\theta_{e^-}$ distributions for a polarized $Z$ boson.
The transverse components are almost identical, both in shape and cross section.
The longitudinal component features a very similar shape in the two models.

\begin{figure}[!htb]
\centering
\subfigure[$M_{W\!Z}>500\,\GeV$\label{fig:extr_fit1_Z}]{\includegraphics[scale=0.37]{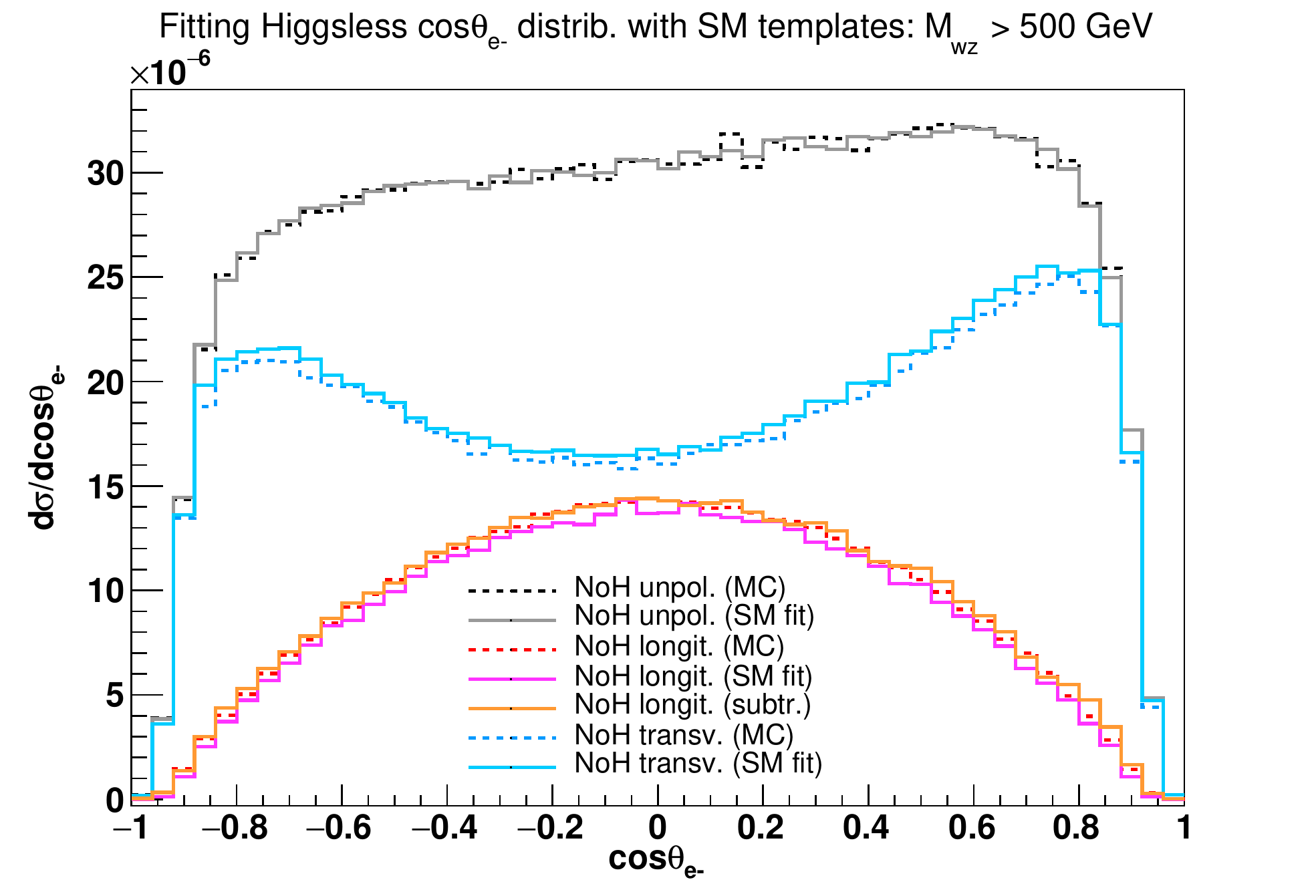}}
\subfigure[$M_{W\!Z}> 200\,\GeV,\,p_t^{Z}> 300\,\GeV$\label{fig:extr_fit2_Z}]{\includegraphics[scale=0.37]{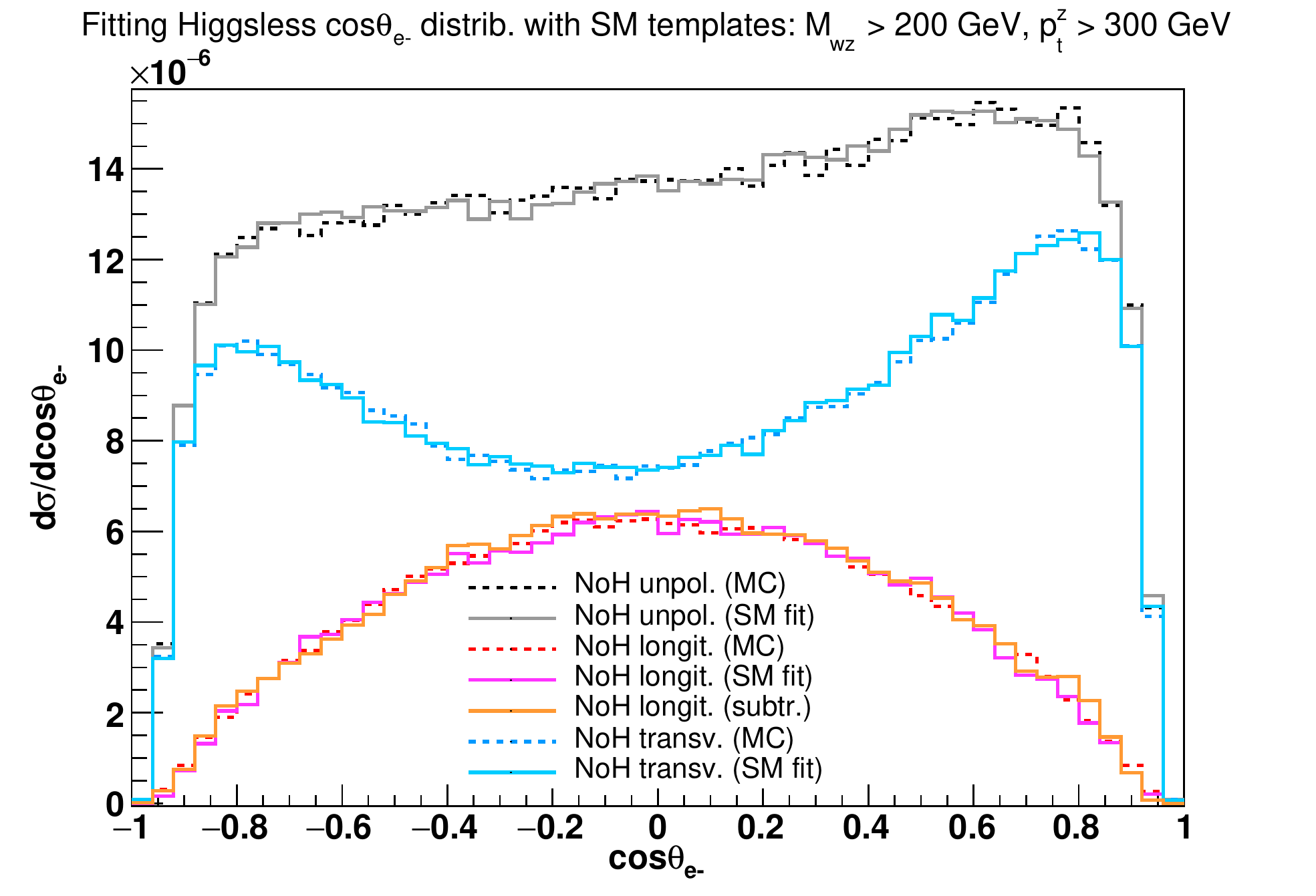}}
\caption{$W^+\!Z$ scattering: fit of Higgsless unpolarized $\cos\theta_{e^-}$ distributions with SM templates, in two different
kinematic regions (large mass and large $p_t$). For the longitudinal component the result of the fit (magenta)
and the one of the subtraction technique (orange) are compared with the Monte Carlo expectation (dashed
red).}\label{fig:fitsubtrwz_z}
\end{figure}

We have determined the longitudinal cross section for the Higgsless model, both through a fit and with the
subtraction technique. Both procedures provide longitudinal cross sections
which differ from the Monte Carlo expectations by less than 5\%, in all the studied kinematic regions.
Numerical results for extracted longitudinal and transverse cross sections are shown in
Tab.~\ref{tab:extr_fitsubtra_Z}. Fitted and expected distributions are
shown in Fig.~\ref{fig:fitsubtrwz_z}, in two specific kinematic regions.

\begin{table}[hbt]
\begin{tabular}{|c||c|c|c||c|c|c|}
\hline
& \multicolumn{6}{|c|}{\cellcolor{blue!9} Polarized cross sections [$\ab $]} \\
\hline
& \multicolumn{3}{|c||}{ Longitudinal} & \multicolumn{3}{|c|}{ Transverse  }\\
\hline
kinematic region  & MC & Fit & Subtr. & MC & Fit & Subtr. \\
\hline
\hline
$M_{W\!Z}>200 \,\GeV$ & 56.27 & 54.88&  57.75 & 122.24 & 124.46 & 120.96\\
\hline
$M_{W\!Z}>500 \,\GeV$ & 18.35 & 17.59& 18.63 & 35.46 & 36.30 & 35.26\\
\hline
$M_{W\!Z}>1000 \,\GeV$ & 4.90 & 4.73& 4.91 & 5.37 & 5.54 & 5.39\\
\hline
$M_{W\!Z}>200 \,\GeV$, $p_t^{Z}>200 \,\GeV$  & 13.97 & 13.58& 14.30 & 37.91 & 38.31 & 37.59\\
\hline
$M_{W\!Z}>200 \,\GeV$, $p_t^{Z}>300 \,\GeV$ & 8.16 & 8.13 & 8.29 & 17.05 & 17.11 & 16.93\\
\hline
$M_{W\!Z}>200 \,\GeV$, $p_t^{Z}>400 \,\GeV$ & 4.94  & 4.84 & 4.99 & 7.92 & 8.05 & 7.92\\
\hline
$M_{W\!Z}>200\,\GeV$,  $|\eta_{Z}|<1 $ & 19.22& 18.32& 19.99 & 62.76 & 63.69 & 61.95\\
\hline
$M_{W\!Z}>200\,\GeV$,  $1<|\eta_{Z}|<2 $ & 22.41& 22.42& 23.03 & 45.42 & 45.59 & 45.08\\
\hline
$M_{W\!Z}>200\,\GeV$,  $2<|\eta_{Z}|<3 $ & 11.72& 11.51& 11.76 & 13.31 & 13.72 & 13.22\\
\hline
$M_{W\!Z}>200\,\GeV$,  $|\eta_{Z}|>3 $ & 2.92 & 2.83 & 2.94 & 0.71 & 0.89 & 0.71\\
\hline
\end{tabular}
\caption{Cross sections ($\ab $) for a longitudinal and transverse $Z$ in $W^+\!Z$ scattering, in the Higgsless model,
in several kinematic regions: comparison of MC predictions for the Higgsless
model with results obtained via fit and subtraction procedure. The subtraction procedure
results for a transverse $Z$ coincide with the SM cross sections.} \label{tab:extr_fitsubtra_Z}
\end{table}

As a general trend, the subtraction procedure overestimates by a few percent the expected values. This is due to
the assumption that the transverse component in the Higgsless model coincides with the SM one. Actually, the
Higgsless transverse component is slightly larger than the SM one, and this
discrepancy propagates in the extraction of the longitudinal component, giving the main contribution to the
few percent discrepancy with respect to the expected value.

On the contrary, the fit procedure underestimates by few percent the expected longitudinal cross section in
the various kinematic regions. This results in a very mild enhancement of the transverse component (see azure
and cyan curve in Fig.~\ref{fig:fitsubtrwz_z}). Differently from the $W$ case, the fit benefits from the strong
similarity between SM and Higgsless longitudinal shapes in each of the considered kinematic regions.

As already observed for the polarizations of the $W^+$, in the large invariant mass
($M_{W\!Z}>1000\,\GeV$), large $p_t$ ($p_t^Z>400\,\GeV$), and forward rapidity ($|\eta_Z|>2$) region the
subtraction procedure reproduces very well the Monte Carlo expected longitudinal cross sections, thanks to
a strong similarity of transverse cross sections in the two models.

These results seem very promising, as they suggest that a model independent extraction of polarization
fractions of the $Z$ boson is viable.
Very good results have been obtained in those regions which are more interesting for new physics in VBS,
\emph{i.e.} large invariant mass of the four-lepton system, large transverse momentum and forward rapidity
of the vector bosons.

\section{Conclusions}
\label{sec:conclusions}
In this paper we have presented a procedure to separate polarization states
of massive weak bosons in VBS processes which involve $Z$ bosons.
We have focused on pure electroweak tree level amplitudes which give
contribution to $W^+\!Z$ and $Z\!Z$ scattering in the fully leptonic
channel at the LHC.
Differently from $W\!W$ scattering processes, that have been investigated
in a previous work \cite{Ballestrero:2017bxn}, separating polarizations
of $Z$ bosons has proved more involved, due to $\gamma$ effects in the
amplitudes.

In both scattering channels, we have checked that a sufficiently tight
cut on the invariant mass of charged lepton pairs around the $Z$ pole
mass is required, to allow for the separation of resonant contributions.
For $W$ bosons in $W\!Z$ scattering, we propose a single On Shell
projection on $W$ resonant diagrams, to avoid unphysical cuts on the
lepton-neutrino system.
We have verified
that the signal for a polarized $Z$ in $Z\!Z$ and $W\!Z$, as well as the
signal for a polarized $W$ in $W\!Z$, reproduce accurately the results which
can be extracted from full $\cos\theta_{\ell}$ distributions by means of
projections onto the first three Legendre polynomials, in the absence of
lepton cuts.
After applying a realistic set of leptonic cuts, the sum of polarized
signals reproduces the full unpolarized results within a few percent.
In $W\!Z$ scattering, the reconstruction of the final state neutrino
generates additional effects on relevant kinematic observables.

The proposed method to separate polarizations at the level of amplitudes
represents a coherent theoretical tool which can be used for LHC data
analyses, and is expected to provide reliable results if the underlying
theory is the Standard Model.
Compared against the results of our proposal, the reweighting
method, which has been widely used to determine approximate polarized signals
in presence of lepton cuts, provides inaccurate predictions particularly at
high diboson invariant mass.

For the extraction of polarization fractions from LHC data,
we have investigated how polarized distributions change with a different
realization of the EWSB, in particular in the presence of
a strongly interacting Higgs sector and an additional heavy Higgs resonance.

Both the approximate independence of the polarized distribution shapes, and
the remarkable similarity of the transverse component in the Standard Model, the
Higgsless model, and the Singlet Extension give us confidence that it will be
possible to estimate polarization fractions with reasonable accuracy by using
Standard Model angular distributions, even in the presence of new physics.

\section*{Acknowledgements}
\noindent This work has been supported by VBSCan COST Action (CA16108) and SPIF INFN project.

\appendix
\section{Neutrino reconstruction}
\label{subsec:vreco}
In this appendix we present several reconstruction schemes we have tried and motivate our choice of a particular procedure.

The presence of a neutrino in $W\!Z$ scattering with fully leptonic decays inhibits the complete reconstruction
of the final state kinematics. To avoid this difficulty,
experiments often measure related, directly observable quantities, as proxies to the decay
angle distribution. Examples are $L_P$ \cite{Chatrchyan:2011ig},
$\cos\theta_{2D}$ \cite{ATLAS:2012au} and  $R_{p_T}$ \cite{Doroba:2012pd}, which is mostly useful for the
$W^+W^+$ channel.
Alternatively, one can attempt to reconstruct the missing component constraining the 
$\ell^+\nu_\ell$ system invariant mass to be equal to the $W$ pole mass ($M_W$) \cite{Chatrchyan:2012kk, ATLAS:2018ogj}.

Identifying the missing transverse momentum with the transverse neutrino momentum, $p^\nu_t$,
only the component along the beam axis, ${p_z^\nu}$, is unknown. The on shell 
condition leads to a quadratic equation in the unknown variable $p_z^\nu$, whose two solutions are
\beq\label{eq:quadraticv}
p_{z\,1,2}^{\nu} = \frac{p_z^\ell\,\xi\,\pm\,\sqrt{\Delta}}{{p^t_\ell}^2}\,\,,
\eeq
where\vspace{-0.3cm}
\beq\label{eq:quadraticv2}
\Delta =  {p_z^\ell}^2\xi^2 - {p^\ell_t}^2\left[{E^\ell}^2{p^\nu_t}^2-\xi^2\right]\,,\qquad  \xi = 
\frac{M_W^2}{2}+{\bf p}_t^\ell\cdot {\bf p}_t^\nu \,\,.
\eeq

The two solutions can be either real or complex, depending on the sign of $\Delta$. In 
particular, $\Delta < 0$ if the transverse mass of the $\ell^+\nu_\ell$ system ($M^{\ell\nu}_{t}$) is 
larger than $M_{W}$. In this case, we need a procedure to determine an approximate real value.
If the transverse mass is smaller than 
$M_{W}$, then $\Delta > 0$: in this case we need a criterion to select one of the two real solutions. The two 
solutions have opposite sign if  $\xi^2 > (E^\ell p^\nu_t)^2$, same sign otherwise.

Several criteria have been used in experimental analyses to get rid of the ambiguity in determining the
unknown longitudinal momentum in processes which involve one neutrino.
We have investigated how different reconstruction schemes fare for unpolarized VBS 
events. The setup is the one of Sect.~\ref{subsec:setupwz}, including lepton and missing $p_t$ cuts.

We first focus on the events with positive $\Delta$, which represent more than $80\%$ of our VBS 
sample. Afterwards, we analyse two procedures for events with $\Delta <0$.

\subsection{Positive $\Delta$}
For $\Delta >0$, we compare five different reconstruction criteria, which we describe briefly in the 
following.

\paragraph{[\texttt{DeltaR}]}
If ${p_{z\,1}^\nu} \cdot {p_{z\,2}^\nu} < 0$, choose the solution with the same sign as $p_z^\ell$. 
Otherwise, choose 
the solution corresponding to the minimum $\Delta R_{\ell\nu}$. This procedure is detailed in 
Ref.~\cite{AmapaneCMS} and was employed for semileptonic VBS in Ref.~\cite{Ballestrero:2008gf}. 
Actually, $\Delta R_{\ell\nu}$ has no discriminating power, since the two solutions give the same
$\Delta\eta_{\ell\nu}$, as can be easily shown in light cone coordinates. 
Thus, we have decided to discard this scheme.

\paragraph{[\texttt{CoM}]}
If ${p_{z\,1}^\nu} \cdot {p_{z\,2}^\nu} < 0$, choose the longitudinal momentum
with the same sign as $p_z^\ell$.
If ${p_{z\,1}^\nu} \cdot {p_{z\,2}^\nu} > 0$, choose the solution which gives the minimum partonic
center of mass invariant mass ($M_{\rm CoM}$), which requires softer initial state partons. 

\paragraph{[\texttt{CoMmod}]}
Choose the solution which reconstructs the minimum $M_{\rm CoM}$, independently of the 
sign of ${p_{z\,1}^\nu} \cdot {p_{z\,2}^\nu}$. 

\paragraph{[\texttt{CMS}]}
Choose the solution with minimum $|p_z^{\nu}|$.
This procedure has been employed by CMS and ATLAS collaborations for analyses of $W\!Z$ 
production \cite{Chatrchyan:2012kk, ATLAS:2018ogj}. 

\paragraph{[\texttt{CMSbis}]}
If ${p_{z\,1}^\nu} \cdot {p_{z\,2}^\nu} < 0$, choose the solution with the same sign as $p_z^\ell$.
Otherwise, choose the solution with minimum $|p^z_{\nu}|$.\\

In order to evaluate the goodness of each reconstruction scheme, we compute the distribution of the relative 
difference between the reconstructed and true value of the neutrino longitudinal momentum, 
$\delta_{p_z}$, defined as
\beq
\delta_{p_z} = \frac{p_z^{\nu, \,\rm{reco}}-p_z^{\nu, \,\rm{true}}}{|p_z^{\nu, \,\rm{true}}|}\,.\nnb
\eeq 
\begin{figure}[!tb]
\centering
\subfigure[$\delta_{p_z}$ \label{fig:Dgt0delta}]{\includegraphics[scale=0.37]{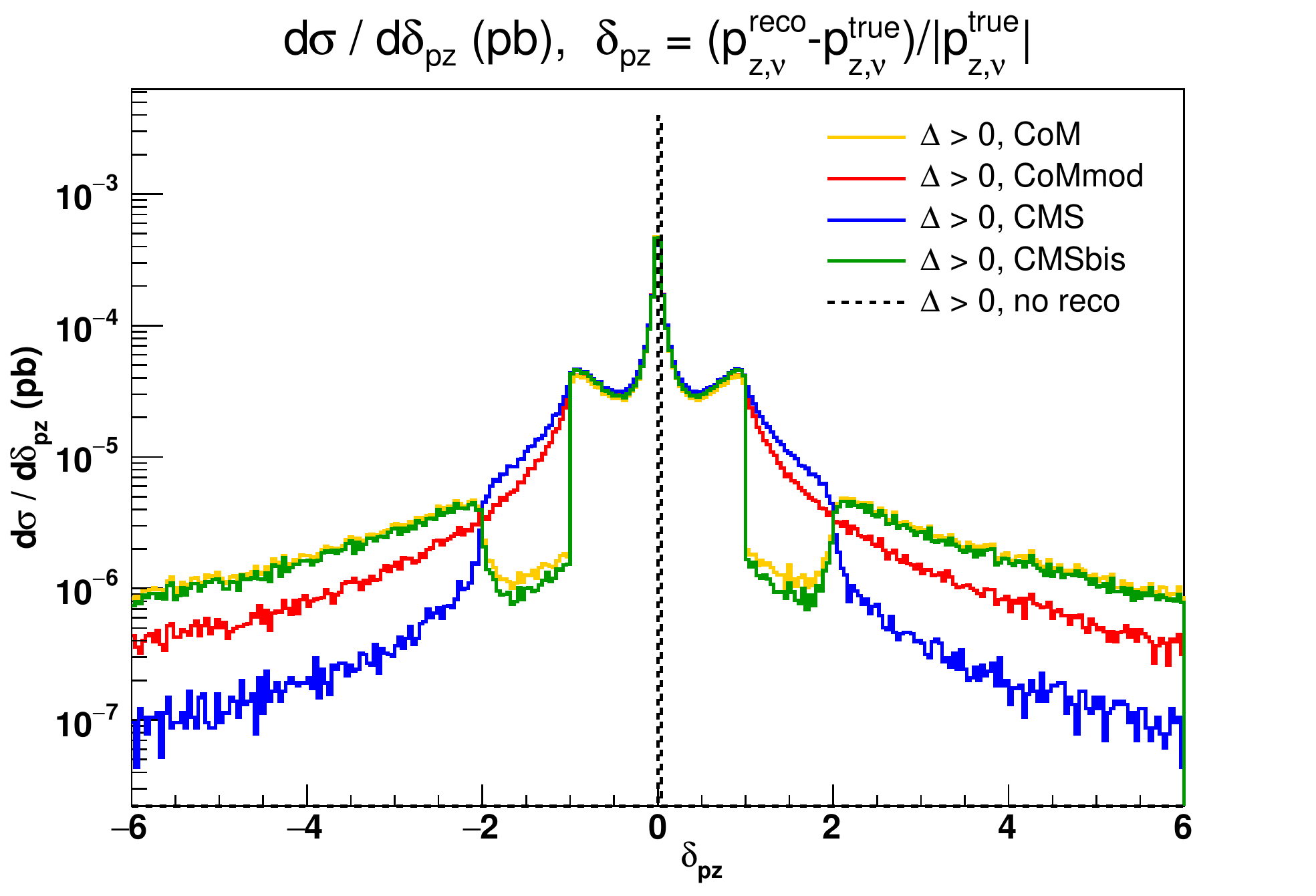}}
\subfigure[$\cos\theta_{\mu^+}$ \label{fig:Dgt0costh}]{\includegraphics[scale=0.37]
{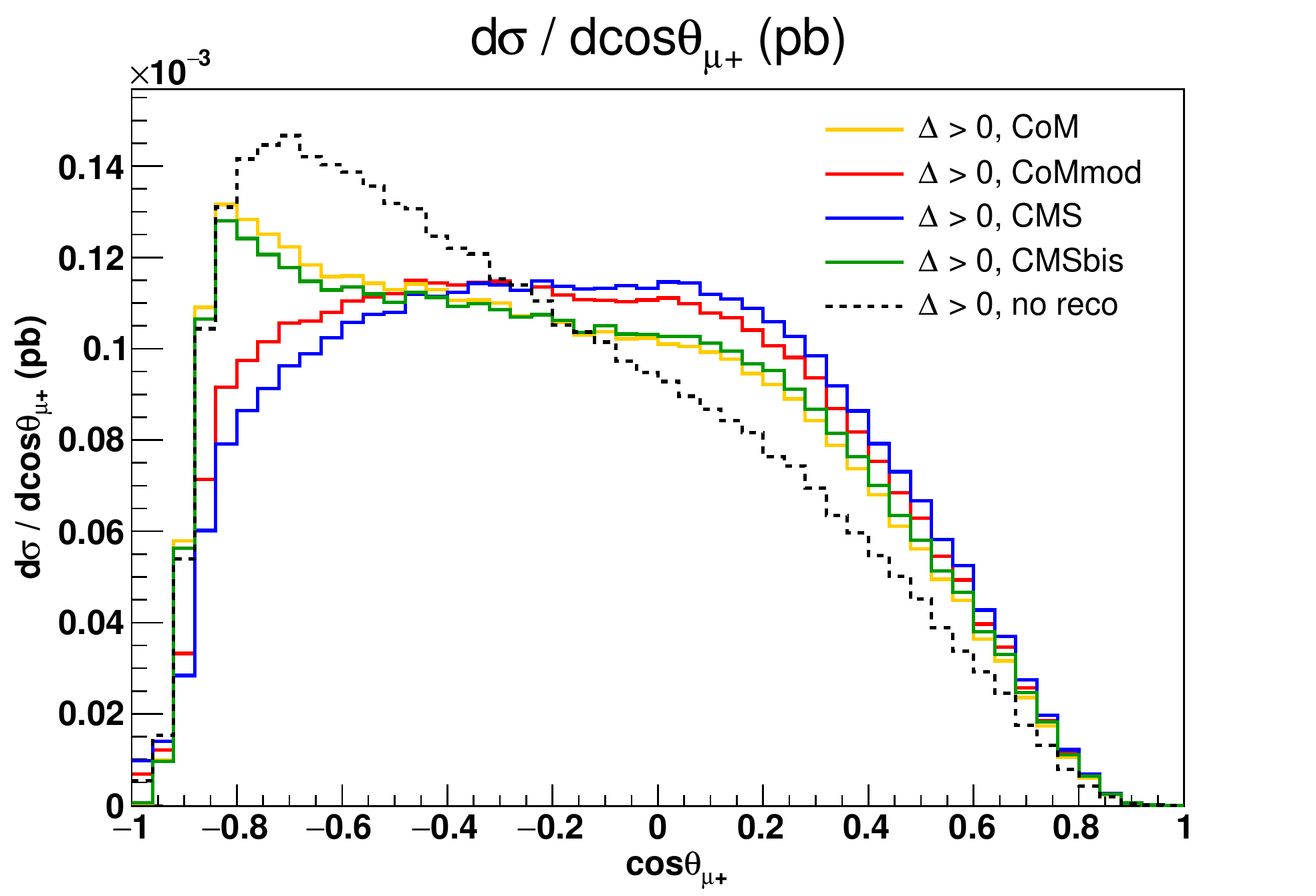}}
\caption{$W^+\!Z$ scattering: neutrino reconstruction, $\Delta>0$. Distributions in 
$\delta_{p_z} = ({p_z^{\nu, \,\rm{reco}}-p_z^{\nu, \,\rm{true}}})/{|p_z^{\nu, \,\rm{true}}|}$ and 
$\cos\theta_{\mu^+}$, obtained with several reconstruction procedures (solid curves), compared with the
generated ones (dashed curve). The following set of cuts is understood: $p_t^j>20$ GeV, 
$|\eta_j|<5$, $M_{jj}>500$ GeV, $\Delta\eta_{jj}> 2.5$ $p_t^\ell>20$ GeV, 
$|\eta_\ell|<2.5$, $p_t^{\rm miss}>40$ GeV, $|M_{e^+e^-}-M_Z|<15$ GeV.}
\end{figure}
We note that total final state invariant mass and the longitudinal momentum of the neutrino are strongly
and positively correlated, therefore minimizing $M_{\rm CoM}$ (as in \texttt{CoMmod}) 
or directly $|p_z^{\nu}|$ (as in \texttt{CMS}) are roughly equivalent procedures.
This results in small differences between the \texttt{CoMmod} and \texttt{CMS} distributions, as well as between the
\texttt{CoM} and the \texttt{CMSbis} distributions, as shown in Fig.~\ref{fig:Dgt0delta}. A crucial role is played by the events with 
${p_{z\,1}^\nu} \cdot {p_{z\,2}^\nu} < 0$.
If we select the solution with the same sign as $p_z^\ell$ (\texttt{CoM, CMSbis}), 
the $\delta_{p_z}$ distribution develops a discontinuity in the region $1<|\delta_{p_z}|<2$, which brings the
reconstructed distribution closer to the true one.
Otherwise, the distributions are smooth (\texttt{CoMmod, CMS}).

Therefore, in addition to the relative shift in $p^\nu_z$, we evaluate how $\cos\theta_{\mu^+}$ distributions 
are affected by reconstruction schemes, since this angular variable has the most relevant role in the phenomenology of 
polarized bosons.
In \Fig{fig:Dgt0costh} we show the reconstructed, unpolarized $\cos\theta_{\mu^+}$ distributions
(solid curves), compared with the true distribution (dashed curve). The peak around -0.7 is not reproduced at all by 
\texttt{CMS, CoMmod} schemes. On the contrary, the other two schemes describe better the correct 
shape of the angular distribution over the whole range, even though not very precisely.

In this paper we have then adopted the \texttt{CoM} prescription to reconstruct events 
with $\Delta>0$ since it reproduces better the $\cos\theta_{\ell}$ distribution.

\subsection{Negative $\Delta$}
For $\Delta <0$, we consider two options to extract a real solution from Eq.~\ref{eq:quadraticv}.
\paragraph{[\texttt{poleMw}]}
$p_z^{\nu}$ is set equal to the real part of the two solutions \cite{AmapaneCMS} :
\beq
p_{z}^{\nu \,\rm{reco}} =  \,\frac{ p^z_\ell\,\xi }{{p^t_\ell}^2}
\eeq
\paragraph{[\texttt{transvMlv}]}
The $W$ pole mass $M_W$ in \eqns{eq:quadraticv}{eq:quadraticv2} is substituted with the transverse mass 
of the lepton neutrino system \cite{Chatrchyan:2012kk}. This forces $\Delta = 0$, and leads to:
\beq
p_z^{\nu \,\rm{reco}} = 
p^z_\ell \,\frac{ {{M_t^{\ell\nu}}^2}+2{\bf p}_t^\ell\cdot {\bf p}_t^\nu }{2{p_t^\ell}^2} \,=
\, p_z^\ell \,\frac{ ({2p_t^\ell p_t^\nu-2{\bf p}_t^\ell\cdot {\bf p}_t^\nu })+2{\bf p}_t^\ell\cdot {\bf p}_t^\nu }
{2{p_t^\ell}^2} \,=\,p_z^\ell \,\frac{ {p_t^\nu} }{{p_t^\ell}}
\eeq
For unpolarized VBS, it turns out that events with negative $\Delta$ account for less than 20\% of the total events.
The two reconstruction procedures discussed here differ by a few percent bin by bin in the $\delta_{p_z}$ 
distribution, both in the central peak region and in the tails. The standard deviation of the distribution suggests 
that the transverse mass method works slightly better, thus we have adopted it.


\bibliographystyle{JHEP}
\bibliography{pol2}


\end{document}